\documentclass[12pt,preprint]{aastex}
\begin{document}

\title{A Comprehensive Study of Gamma-Ray Burst Optical Emission: I. Flares and Early Shallow Decay Component}
\author{Liang Li\altaffilmark{1}, En-Wei Liang\altaffilmark{1,2,3}, Qing-Wen Tang\altaffilmark{1}, Jie-Min Chen\altaffilmark{1}, Shao-Qiang Xi\altaffilmark{1}, Hou-Jun L\"{u}\altaffilmark{3}, He Gao\altaffilmark{3}, Bing Zhang\altaffilmark{3,1}, Jin Zhang\altaffilmark{2,4}, Shuang-Xi Yi\altaffilmark{5}, Rui-Jing Lu\altaffilmark{1}, Lian-Zhong L\"{u}\altaffilmark{1}, and Jian-Yan Wei\altaffilmark{2}}
\altaffiltext{1}{Department of Physics and GXU-NAOC Center for Astrophysics
and Space Sciences, Guangxi University, Nanning 530004,
China; lew@gxu.edu.cn}
\altaffiltext{2}{National Astronomical Observatories, Chinese Academy of Sciences, Beijing, 100012, China}
\altaffiltext{3}{Department of Physics and Astronomy, University of
Nevada, Las Vegas, NV 89154; zhang@physics.unlv.edu}
\altaffiltext{4}{College of Physics and Electronic Engineering, Guangxi Teachers Education
University, Nanning, 530001, China}
\altaffiltext{5}{College of Astronomy and Space Sciences, Nanning University, Nanjing, 210093, China}

\begin{abstract}
Well-sampled optical lightcurves of 146 gamma-ray bursts (GRBs) are compiled from the literature. By empirical fitting we identify eight possible emission components and summarize the results in a ``synthetic'' lightcurve. Both optical flare and early shallow-decay components are likely related to long-term central engine activities. We focus on their statistical properties in this paper. Twenty-four optical flares are obtained from 19 GRBs. The isotropic R-band energy is smaller than $1\%$ of $E_{\gamma, \rm iso}$. The relation between isotropic luminosities of the flares and gamma-rays follows $L^{\rm F}_{\rm R, iso}\propto L_{{\gamma}, \rm iso}^{1.11\pm 0.27}$. Later flares tend to be wider and dimmer, i.e., $w^{\rm F}\sim t^{\rm F}_{\rm p}/2$ and $L^{\rm F}_{\rm R, iso}\propto  [t^{\rm F}_{\rm p}/(1+z)]^{-1.15\pm0.15}$. The detection probability of the optical flares is much smaller than that of X-ray flares. An optical shallow decay segment is observed in 39 GRBs. The relation between the break time and break luminosity is a power-law, with an index of $-0.78\pm 0.08$, similar to that derived from X-ray flares. The X-ray and optical breaks are usually chromatic, but a tentative correlation is found. We suggest that similar to the prompt optical emission that tracks $\gamma$-rays, the optical flares are also related to the erratic behavior of the central engine. The shallow decay component is likely related to a long-lasting spinning-down central engine or piling up of flare materials onto the blastwave. Mixing of different emission components may be the reason of the diverse chromatic afterglow behaviors.
\end{abstract}
\keywords{radiation mechanisms: non-thermal --- gamma-rays: bursts}

\section{Introduction\label{sec:intro}}
Gamma-Ray Bursts (GRBs) and their broadband afterglows are the most luminous phenomena in the Universe. According to the standard model, the broad-band afterglow is from the external shock as the fireball is decelerated by the ambient medium (M\'esz\'aros \& Rees 1997; Sari et al. 1998). The prompt gamma-ray emission, on the other hand, is due to some internal dissipation processes within the relativistic ejecta, either due to internal shocks (Rees \& M\'esz\'aros 1994; Kobayashi et al. 1997; Daigne \& Mochkovitch 1998) or internal magnetic energy dissipation processes (e.g. Usov 1992; Thompson 1994; Drenkhahn \& Spruit 2004; Giannios \& Spruit 2006; Zhang \& Pe'er 2009; Zhang \& Yan 2011), either near or far above the photosphere. In the pre-{\em Swift} era, afterglow observations were mostly made in the optical bands. The data are well explained by the external shock model (e.g., M\'esz\'aros \& Rees 1997; Sari et al. 1998; Panaitescu et al. 1998; Panaitescu \& Kumar 2001;  Huang et al. 2000; see Zhang \& M\'{e}szaros 2004 for review). The successful launch and operation of the {\em Swift} mission (Gehrels et al. 2004) have significantly improved our understanding on the physical origin of GRBs. In particular, early X-ray afterglow observations revealed erratic flares and early plateaus that are difficult to interpret within the standard theoretical framework (Zhang et al. 2006, Nousek et al. 2006). The flares are believed to be produced by late activity of the GRB central engine (Burrows et al. 2005; Fan \& Wei 2005; Zhang et al. 2006; Dai et al. 2006; Proga \& Zhang. 2006; Perna et al. 2006), and the shallow decay segment likely signals a long-lasting wind powered by the GRB central engine after the prompt gamma-ray phase (Zhang et al. 2006). These features indicate that the GRB central engine does not die quickly. One is obliged to accept a more complicated afterglow picture, namely, the observed afterglow emission is a superposition of the traditional external shock afterglow and an afterglow related to the late central engine activity (Zhang 2011).

The prompt localization of GRBs with the X-Ray Telescope (XRT) on board Swift has significantly increased the number of GRBs with optical afterglow detection and redshift measurement. The simultaneous observations with XRT, UVOT, and ground-based optical telescopes in a much wider time window in the afterglow phase have revolutionized our knowledge about the GRB afterglow, and raised some critical problems with the conventional models (e.g., Zhang 2007, 2011; Liang 2010). For example, the multi-wavelength observations sometimes revealed chromatic behaviors between the optical and X-ray bands (e.g. Panaitescu et al. 2006; Liang et al. 2007), suggesting multiple emission components in the afterglow that dominate different energy bands at different epochs. The existence of engine-powered early X-ray emission also obscures the clear separation of long and short GRB, causing confusion in GRB classification (e.g. Zhang et al. 2007, 2009; L\"{u} et al. 2010 for detailed discussion). In order to unveil the underlying physics, it is essential to decompose the lightcurve into different components that have distinct physical meanings.

There are two approaches to decompose the lightcurves into different components: one through theoretical modeling and the other through empirical fitting. Theoretical modeling gives insights into the physical properties of the emission, including radiation mechanisms, micro-physical parameters, properties of the surrounding medium, etc. Intense modeling of optical afterglow data has been carried out in the pre-{\em Swift} era (e.g., Panaitescu et al. 1998; Panaitescu \& Kumar 2001, 2002;  Huang et al. 2000; Wu et al. 2005). This becomes increasingly difficult in the {\em Swift} era. First, there are many more bursts with high quality data to be modeled, and performing such modeling would be time consuming. More importantly, data suggest that it is essentially impossible to interpret all the data with the standard external shock model. The empirical fitting approach, on the other hand, makes use some empirical functions to fit the data, and is suitable to handle a great amount of data. A morphological study can catch insights of various emission components, which can ease statistical analyses of a large sample of data. This has been done by some authors (e.g., Liang \& Zhang 2006; Panaitescu \& Vestrand 2008, 2011; Kann et al. 2010; 2011). Some interesting features have been revealed. For example, Liang \& Zhang (2006), Nardini et al. (2006), and Kann et al. (2006) found two universal tracks of the late optical luminosity lightcurves. Panaitescu \& Vestrant (2008, 2011) showed some general features of the early optical bumps and plateaus in the optical lightcurves. Kann et al. (2010, 2011) compared the optical lightcurves of different types of GRBs.

Different from these previous statistical studies, in a series of papers, we plan to make empirical fitting to the observed optical lightcurves and to identify multiple emission components. After decomposing the lightcurves, we plan to perform statistical analyses of the parameters of various components, and discuss their physical implications. As the first paper in the series, in this paper we present the sample (\S2), the general results of the decomposition, and a ``synthetic'' lightcurve that shows eight possible components with distinct physical origins (\S3). Since flares and the shallow decay component may be directly related to the late central engine activity, in this paper we carry out a detailed study of these two components: flares in \S4 and the shallow decay component in \S5, respectively. We discuss physical implications in \S6, and summarize the results in \S7.

\section{Data\label{sec:data}}
We include all the GRBs that have optical afterglow detections before November 2011 (from Feb. 28 1997 to Nov. 2011) in our sample. A sample of 225 optical lightcurves are complied with data reported in the literature. We make an extensive search for the optical data from published papers or from GCN Circulars if no published paper is available for some GRBs. Well-sampled lightcurves are available for 146 GRBs. In Table 1, we summarize the following information for each GRB: redshift, spectral properties of the optical afterglow and prompt gamma-ray emission, as well as time intervals of optical observations\footnote{A full version of the GRB sample with references to the observational data are available in the electronic form.}.  We collect the optical spectral index $\beta_O$ (defined with the convention $F_\nu \propto \nu^{-\beta_O}$)\footnote{An optical spectral index $\beta_O=0.75$ is adopted for those GRBs whose $\beta_O$ is not available.} and the extinction $A_{\rm V}$ of the host galaxy for each burst from the same literature to reduce the uncertainties introduced by different authors. Galactic extinction correction is made by using a reddening map presented by Schlegel et al. (1998). Since the $A_{\rm V}$ values are available only for some GRBs and the $A_{\rm V}$ is derived from the spectral fits using different extinction curves, we do not make correction for the GRB host galaxy extinction. The $k$-correction in magnitude is calculated by $k=-2.5(\beta_O-1)\log(1+z)$. For the late epoch data ($\sim 10^6$ seconds after the GRB trigger), possible flux contribution from the host galaxy is subtracted. The isotropic gamma-ray energy ($E_{\rm \gamma, iso}$) is derived in the rest frame $1-10^4$ keV energy band using the spectral parameters.

\section{Lightcurve Fitting and a Synthetic Optical Emission Lightcurve \label{sec:data}}
The optical lightcurves are usually composed of one or more power-law segments along with some flares, humps, or re-brightening features. The mix of different components makes the diverse optical afterglow lightcurves. In order to decompose the rich features, we fit the lightcurves with a model of multiple components. The basic component of our model is either a power-law function
\begin{equation}
 F_1 = F_{01} t^{-\alpha}
\end{equation}
or a smooth broken power-law function
\begin{equation}
F_2 =  F_{02} \left[\left(\frac{t}{t_{\rm b,1}}\right)^{\alpha_1\omega}
+\left(\frac{t}{t_{\rm b,1}}\right)^{\alpha_2\omega}\right]^{-1/\omega},
\label{F2}
\end{equation}
where $\alpha$, $\alpha_1$, $\alpha_2$ are the temporal slopes, $t_{\rm b}$ is the break time, and $\omega$ measures the sharpness of the break. In some afterglow models, a double broken power law lightcurve is expected. For example, it is theoretically expected that the afterglow lightcurve may have a shallow segment early on due to energy injection, and then transits to a normal decay segment when energy injection is over, and finally steepens due to a jet break (e.g. in the canonical X-ray afterglow lightcurve, Zhang et al. 2006). We therefore also consider a smooth triple power-law function to fit some lightcurves. In this case, we extend equation (\ref{F2}) to the following function (Liang et al. 2008)
\begin{equation}\label{STPL}
F_3=(F_{2}^{-\omega_2}+F_j^{-\omega_2})^{-1/\omega_2}
\end{equation}
where $\omega_2$ is the sharpness factor of the jet break at $t_{b,2}$, and
\begin{equation}\label{PL}
F_j=F_2(t_{\rm b,2})\left(\frac{t}{t_{b,2}}\right)^{-\alpha_3}.
\end{equation}

\begin{figure*}
\includegraphics[angle=0,scale=0.350,width=0.3\textwidth,height=0.25\textheight]{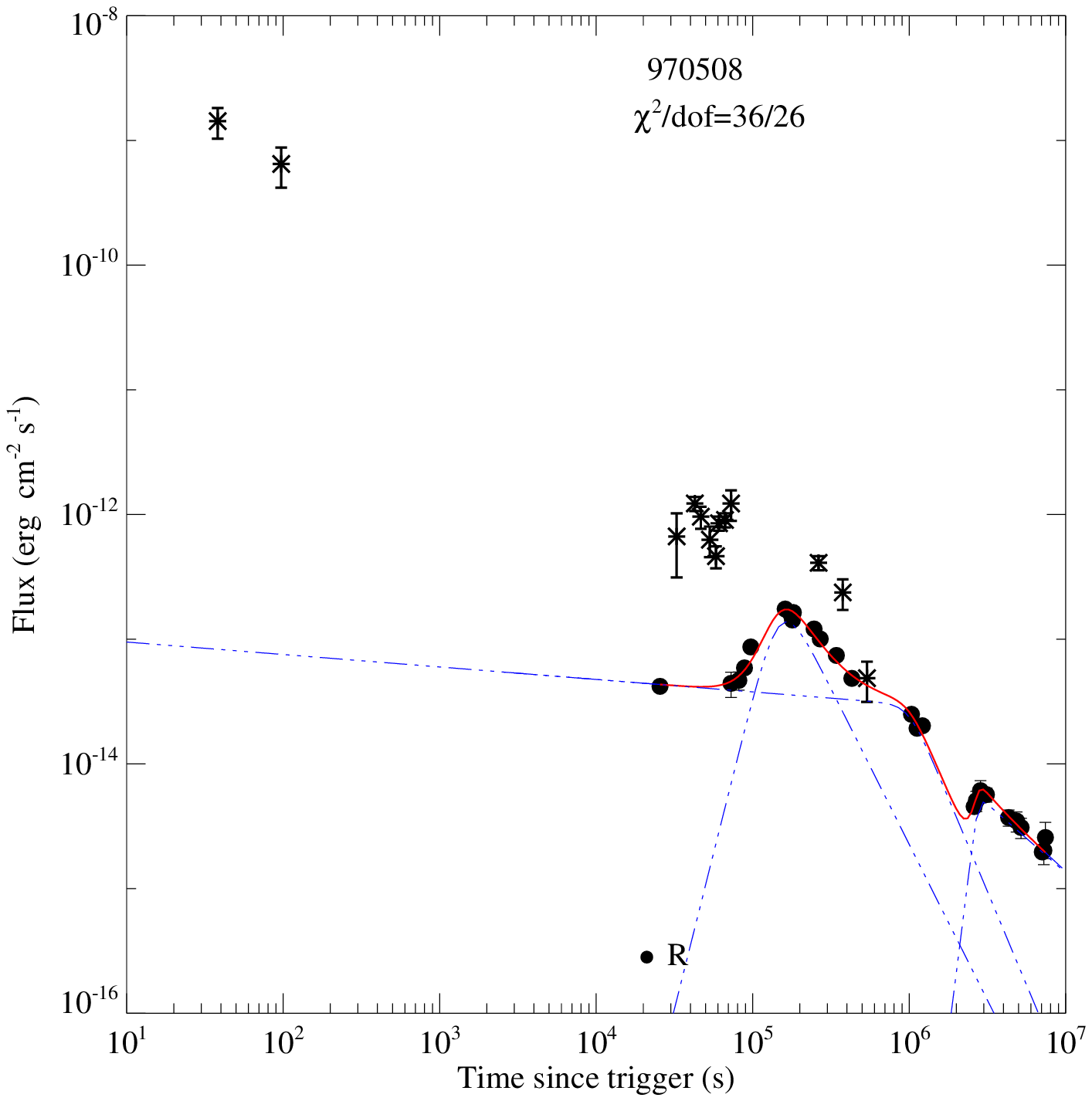}
\includegraphics[angle=0,scale=0.350,width=0.3\textwidth,height=0.25\textheight]{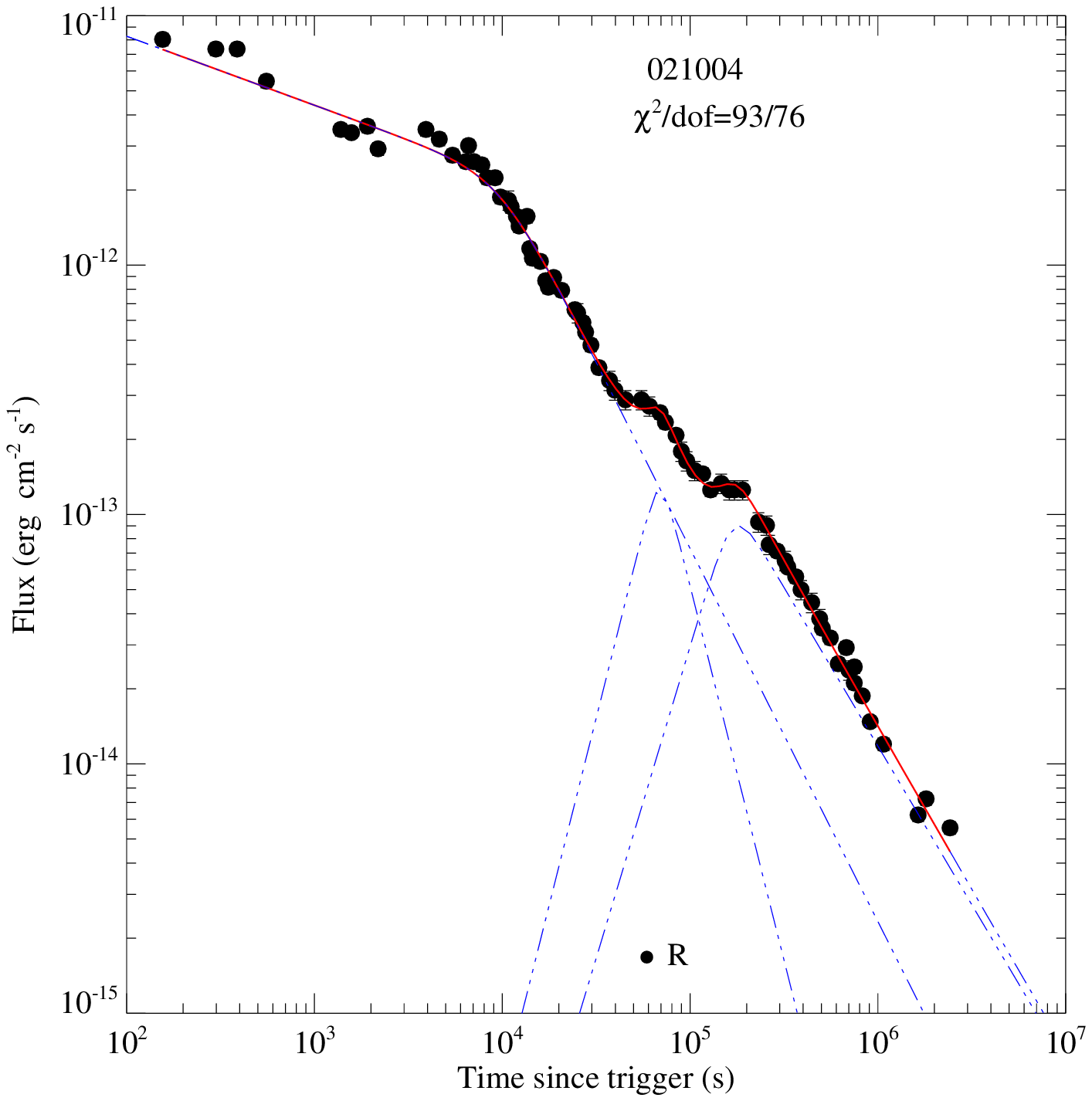}
\includegraphics[angle=0,scale=0.350,width=0.3\textwidth,height=0.25\textheight]{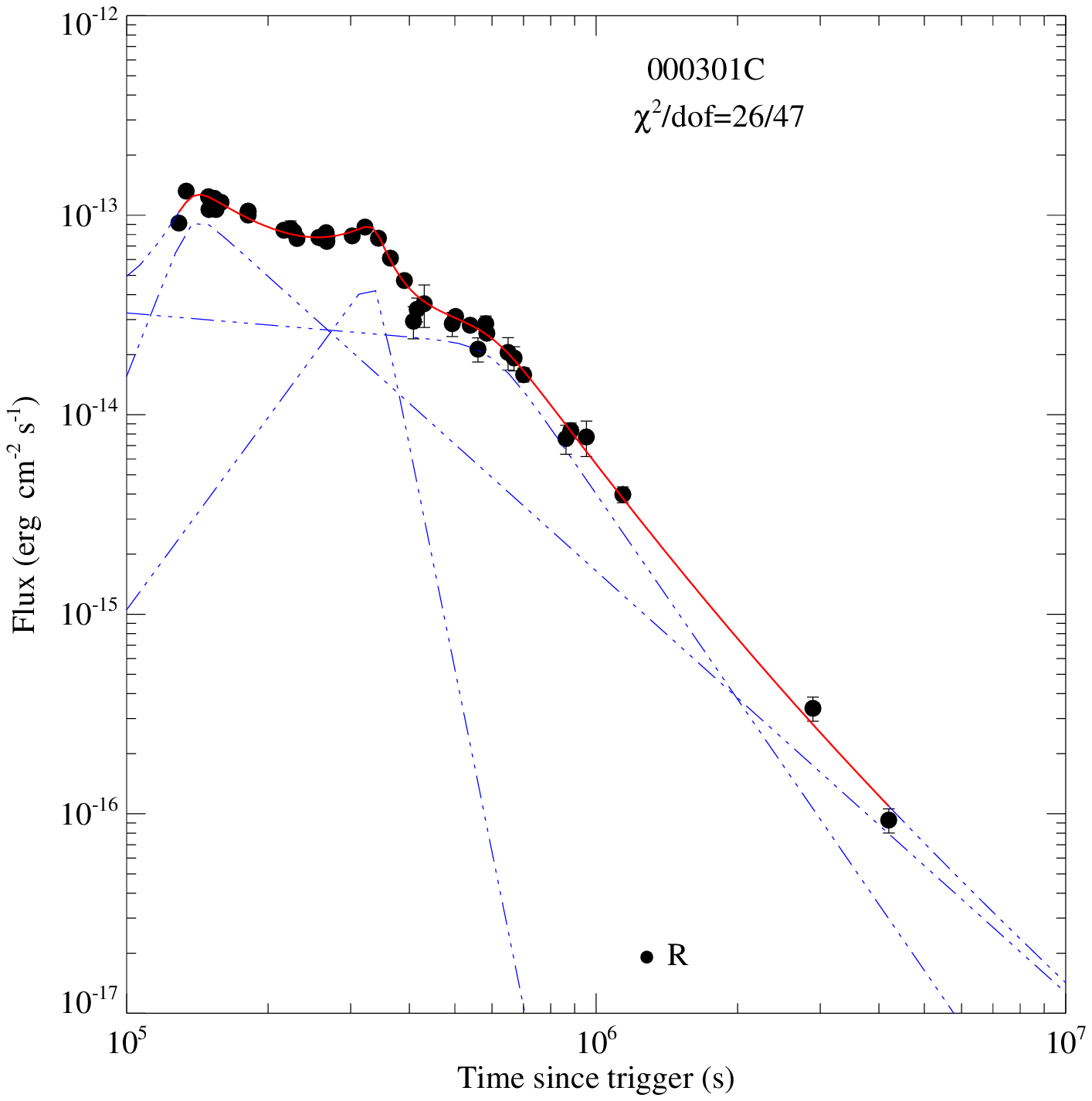}
\includegraphics[angle=0,scale=0.350,width=0.3\textwidth,height=0.25\textheight]{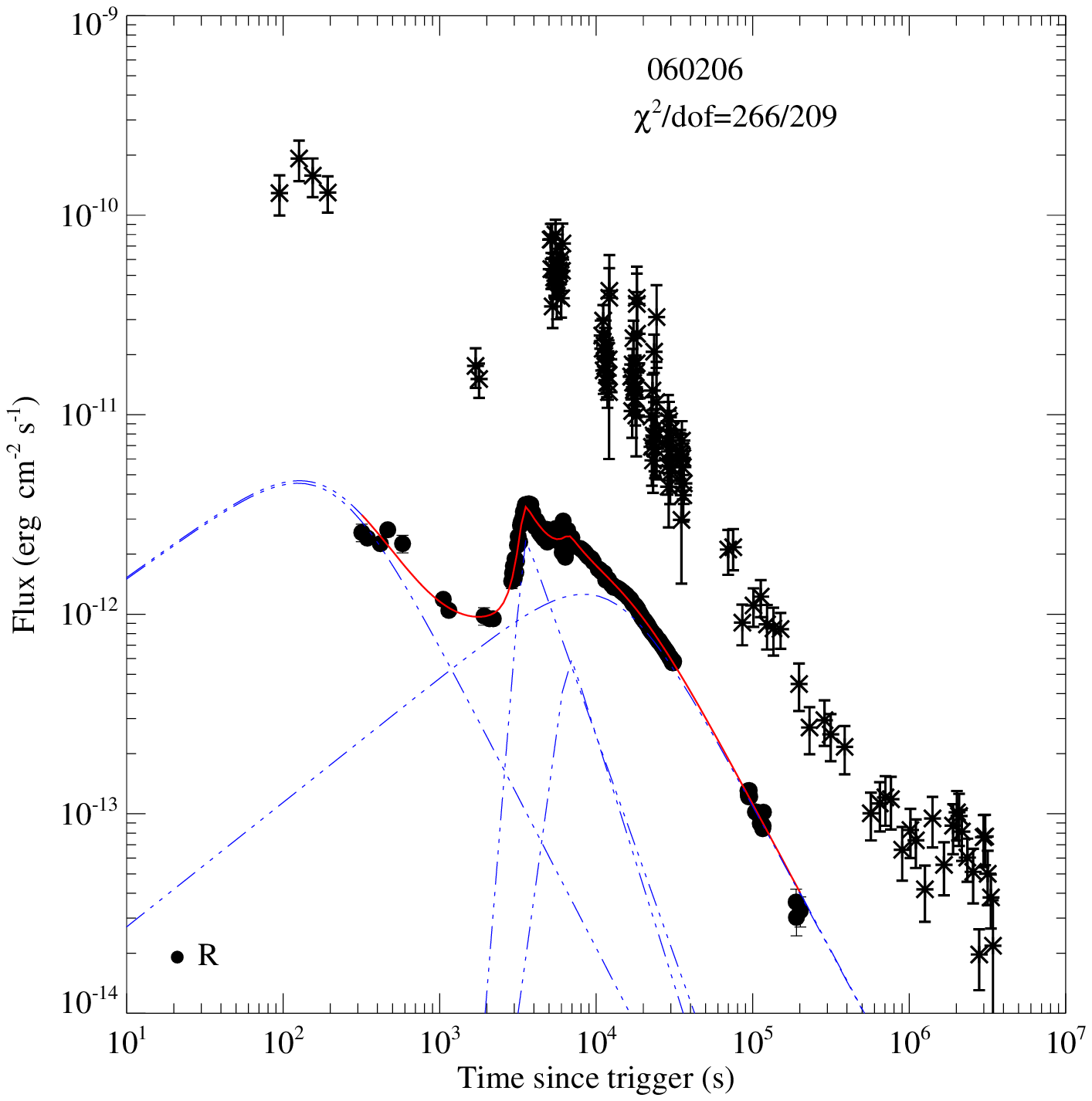}
\includegraphics[angle=0,scale=0.350,width=0.3\textwidth,height=0.25\textheight]{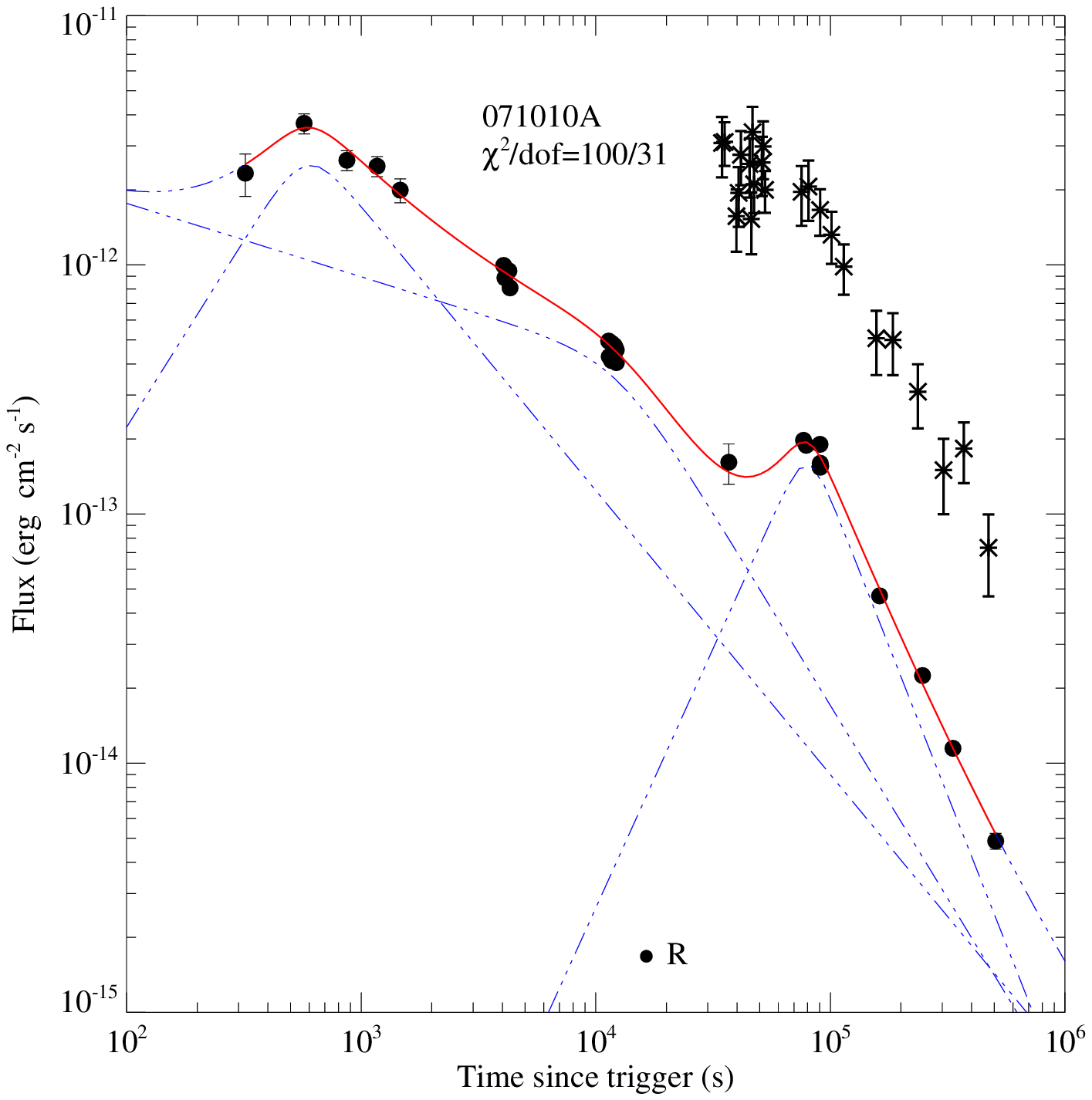}\hfill
\includegraphics[angle=0,scale=0.350,width=0.3\textwidth,height=0.25\textheight]{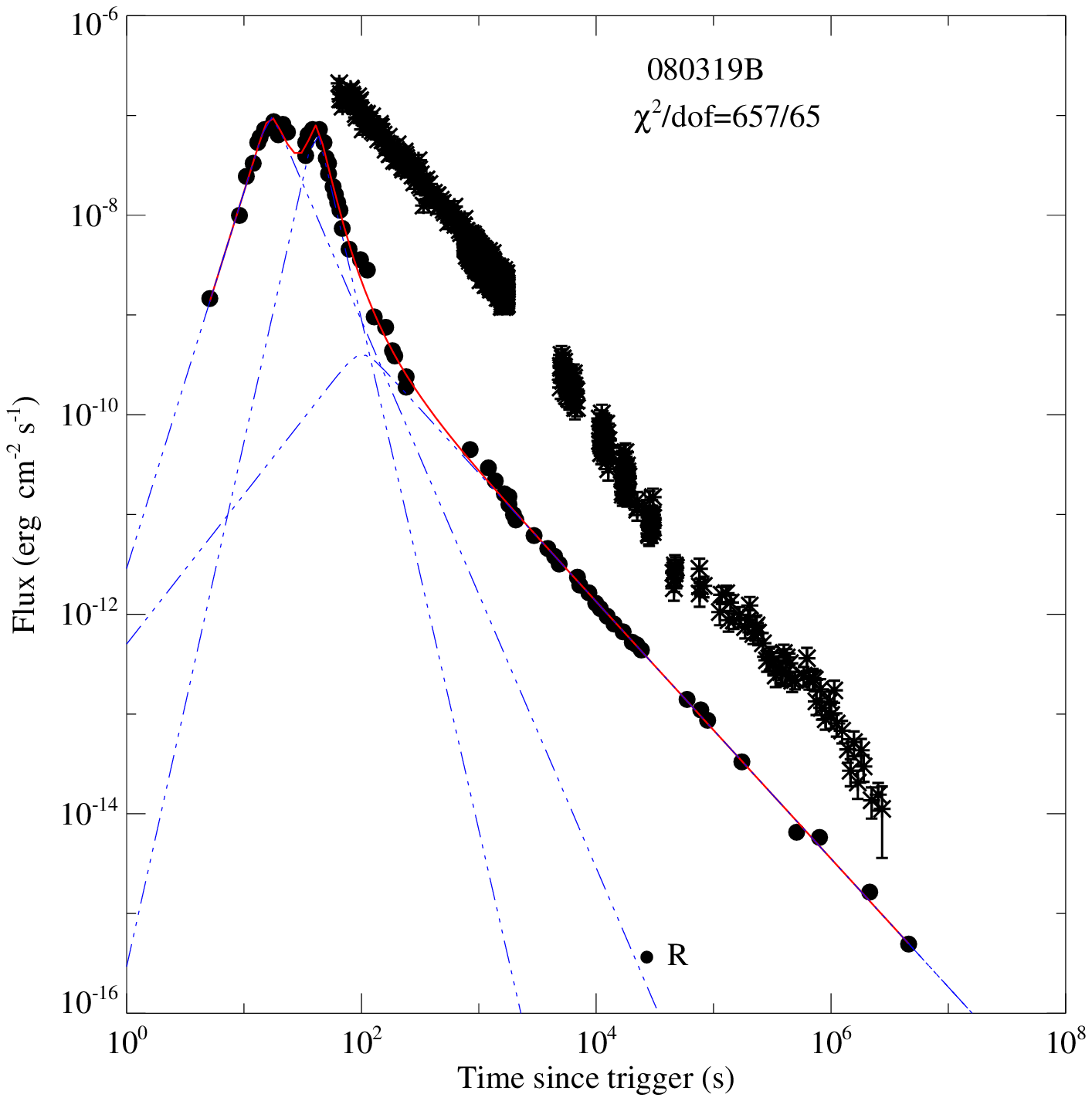}
\caption{Examples of our model fits (solid lines) to the optical lightcurves with multiple components (dashed or dash-dotted lines). The solid lines represent the best fit to the data. Simultaneous X-ray data
observed with {\em Swift}/XRT (crosses with error bars) are also presented.}
\label{Opt_LC_6}
\end{figure*}

We developed an IDL code to make best fits with a subroutine called MPFIT\footnote{http://www.physics.wisc.edu/~craigm/idl/fitting.html.}. The parameter $\omega$ is usually fixed to 3 or 1 in our fitting. The approach of our lightcurve fitting is as follows. Initially, we introduce a minimum number of components based on eye inspection of the global feature of the lightcurve. If the reduced $\chi^2_{\rm r}$ is much larger than 1, we continue to add more components and re-do the fit, until the reduced $\chi^2_{\rm r}$ becomes close to 1. The reduced $\chi^2$ of a simple power-law fits to the lightcurves of some GRBs, such as GRBs 020813, 030328, 050416A , and 070110, are $\sim 1$. However, a smooth broken power-law can significantly improve the fit\footnote{Note that the lightcurves of these GRBs are usually poorly sampled in the initial shallow-decay segment, leading to uncertainty to reveal the initial shallow-decay segment.}, which reduces $>50\%$ of the $\chi^2_{\rm r}$ value. We therefore adopt the smooth broken power-law fit for these GRBs. The $\chi^2_{\rm r}$ values for some lightcurves are much smaller than 1, indicating that some model parameters are poorly constrained. For these cases, we fix some parameters to make the fits. The erratic fluctuations of some data points with small error bars in some GRBs, such as GRB 030329, make $\chi^2_{\rm r}$ much larger than 1. We do not add additional components for these lightcurves, so that their $\chi^2_{\rm r}$ values remain much larger than 1. The most challenging problem in our fit is to extract severely overlapping flares/bumps from the lightcurves. The slopes of these flares/bumps are usually quite uncertain. In our fitting, we first set all the parameters free to get the best fit to the global lightcurve, and then adjust the rising slopes to ensure that the fitting curve crosses the data point around the peak time of each flare/bump. Finally, we fix the rising slopes and perform the best fits again. As an example, Figure \ref{Opt_LC_6} shows six lightcurves with the best-fit multiple components decomposed.

\begin{figure*}
\includegraphics[angle=0,scale=1]{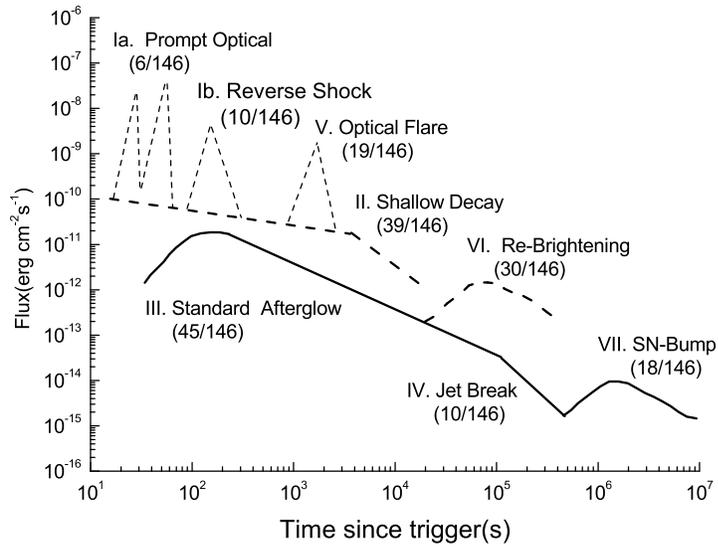}
\caption{A synthetic cartoon lightcurve of multiple optical emission components based on our analysis.} \label{Cartoon}
\end{figure*}

Even though individual optical lightcurves can differ significantly, after synthesizing many lightcurves, one can come up with a synthetic afterglow lightcurve, as shown in Fig.\ref{Cartoon}. In contrast to the five-component canonical X-ray lightcurve (Zhang et al. 2006), the synthetic optical lightcurve includes eight components that may have distinct physical origins. These components are: Ia. prompt optical flares; Ib. an early optical flare from the reverse shock; II. early shallow decay segment; III. the standard afterglow component (an onset hump followed by a normal decay segment); IV. the post jet break phase; V. optical flares; VI. rebrightening humps; VII. late supernova (SN) bumps. The components II-V can find their counterparts in the canonical X-ray lightcurve. These components can be distinguished based on the parameters of our multi-component fits. For example, flares usually have rapid rise and fall. We define a flare if the absolute value of its rising and decaying slopes are steeper than 2. If flares occur during the prompt emission phase, they are grouped as Ia (prompt optical flares). A reverse shock flare (Ib) is a huge flare (or optical flash) that peaks slightly after the end of prompt emission (the decay slope can be somewhat shallower than 2). All the optical flares (V) afterwards are considered ``late'' (with respect Ia and Ib), even though most of them actually happen in the early afterglow phase.

We define different components based on theoretical guidance. For example, flares are defined as features with both steep rising ($\alpha_1 <- 2$ with the convention $F_{\nu} \propto t^{-\alpha}$) and decaying ($\alpha_2 > 2$) slopes. A shallow decay component, on the other hand, is defined as a segment whose decay is shallower than what is predicted in the constant energy afterglow model. As shown in Table 1, the spectral indices of the optical afterglows are usually smaller than 1 within error. This is consistent with the standard external shock synchrotron radiation model in the spectral regime of $\nu_{\rm m}<\nu_{\rm O}<\nu_{\rm c}$  (with the electron power-law index $p>2$), where $\nu_{\rm m}$ and $\nu_{\rm c}$ are the injection and cooling frequencies, respectively. In this spectral regime, one has the closure relation $\alpha_{\rm O}=3\beta_{\rm O}/2$.  A shallow decay component (II) is then defined by the condition $\alpha_{\rm O} < 3\beta_{\rm O}/2$ (with the convention $F_\nu \propto t^{-\alpha_{\rm O}}$) within error. Similarly, a post-jet-break decay segment IV is defined as the segment beyond a steepening break after the normal decay segment (with $\alpha_{\rm O}=3\beta_{\rm O}/2$)\footnote{The radiation physics during the jet-break segment would be the same as the normal decay segment, but the break separates two different dynamical evolution regimes of the GRB fireball. We therefore define the post-jet-break segment as a new component in our analysis, to echo the definition in the canonical X-ray afterglow lightcurve (Zhang et al. 2006).}. An afterglow onset feature is characterized by a smooth hump peaking at less than 1 hour post trigger, which is followed by a normal power-law decay component (III). A rebrightening hump (VI) is similar to the early afterglow onset hump but is much later. It differs from optical flares by much shallower rise and decay as well as a much smoother peak. The supernova bump (VII) is a special late rebrightening peaking at around 1-2 weeks after GRB trigger, which usually shows a red color.

After decomposing the lightcurves, we are able to group all the identified components to one of these eight components. The early optical afterglow lightcurves ($t<10^3$ s) of about one-third GRBs show a smooth hump. Another one-third lightcurves start with a shallow decay segment. Twenty-four optical flares are observed in 19 GRBs. Late rebrightenning humps are observed in 30 GRBs. A jet like break is detected in 10 GRBs. A clear SN bump is detected in 18 GRBs. The detected fraction of each component is marked in the synthetic lightcurve (Fig.\ref{Cartoon}).

We report our statistical results of various components in a series of papers.  As the first paper of the series, this paper focuses on the optical flares and the shallow decay segment. The reason to discuss them together is because they are both likely related to late central engine activity of GRBs (see \S6 for more discussion). Notice that the prompt optical flares and early reversed shock flares are not included in this paper and we'll discuss them separately. Throughout the paper, we mark the parameters of the flares and the shallow decay segment with the superscripts ``F'' and ``S'', respectively.

\section {Flares}

\begin{figure*}
\includegraphics[angle=0,scale=0.350,width=0.3\textwidth,height=0.25\textheight]{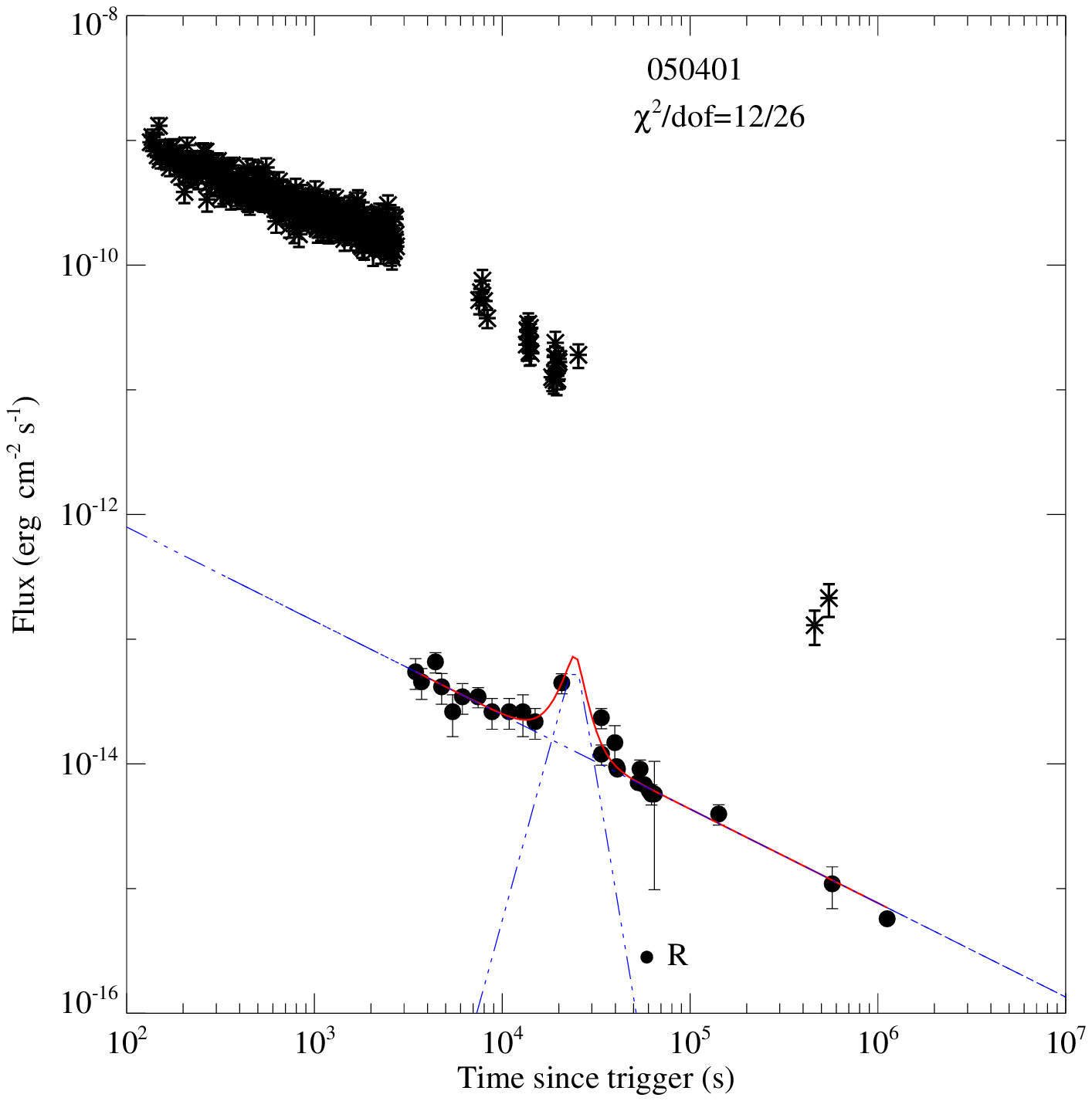}
\includegraphics[angle=0,scale=0.350,width=0.3\textwidth,height=0.25\textheight]{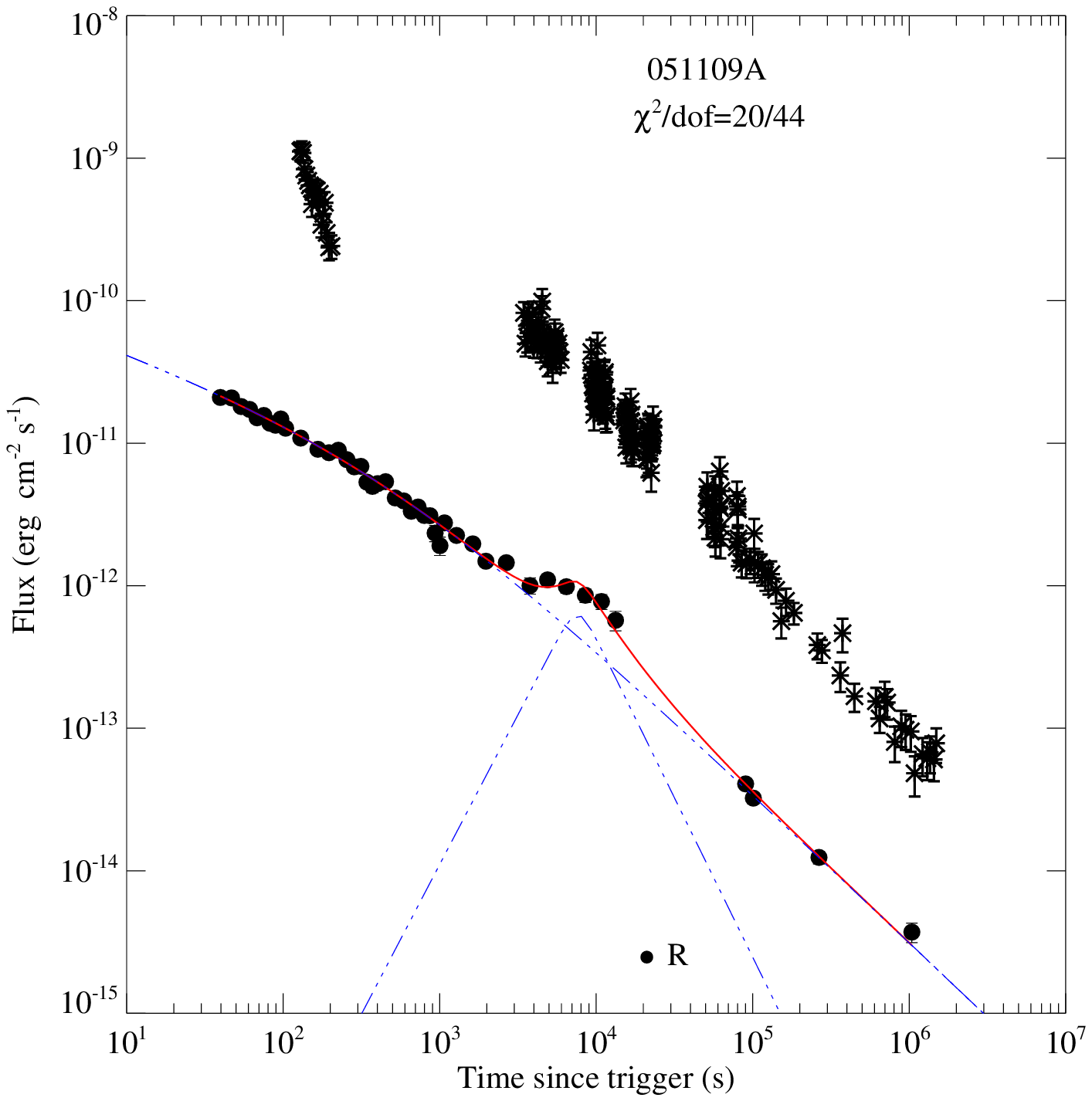}
\includegraphics[angle=0,scale=0.350,width=0.3\textwidth,height=0.25\textheight]{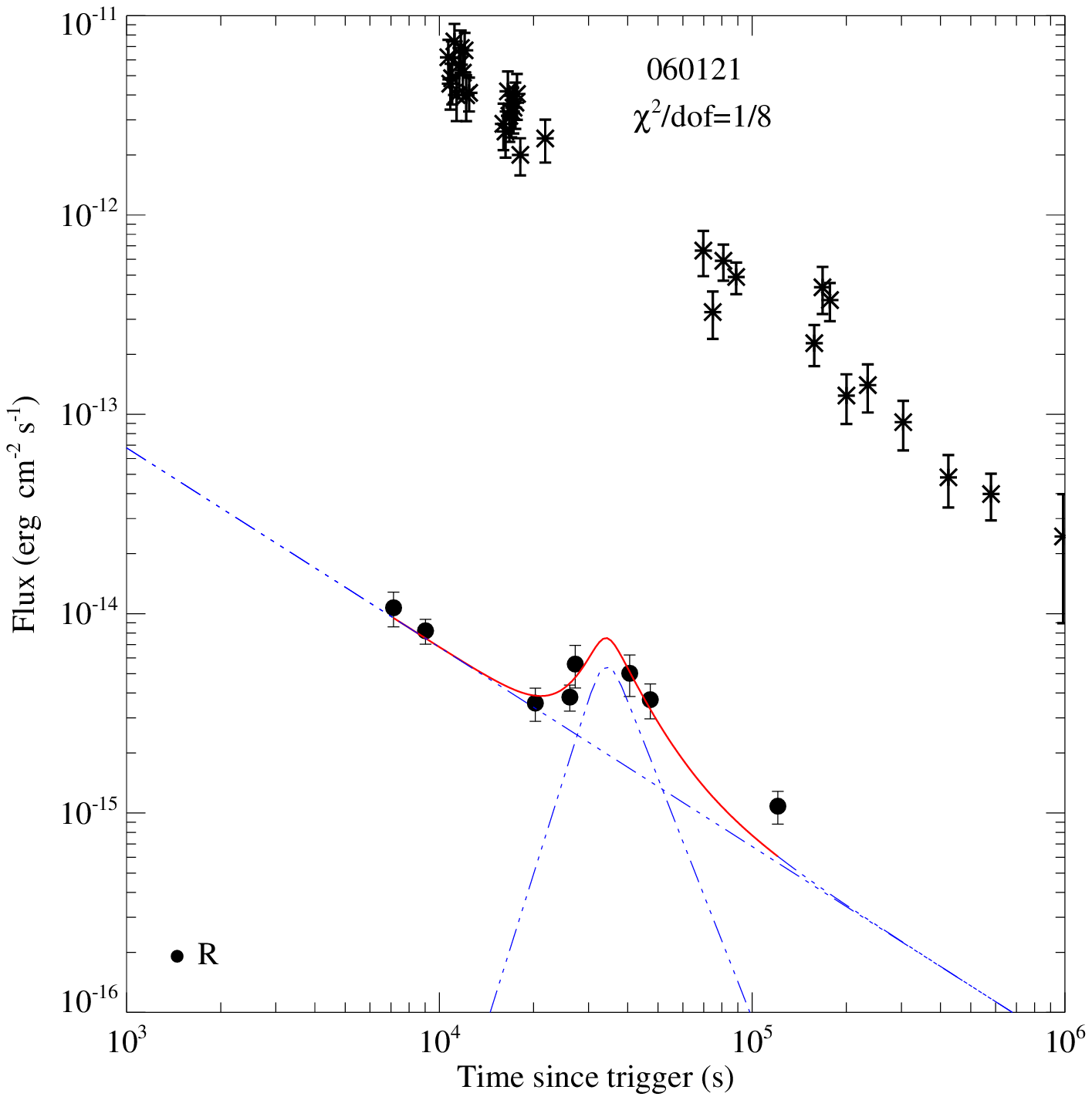}
\includegraphics[angle=0,scale=0.350,width=0.3\textwidth,height=0.25\textheight]{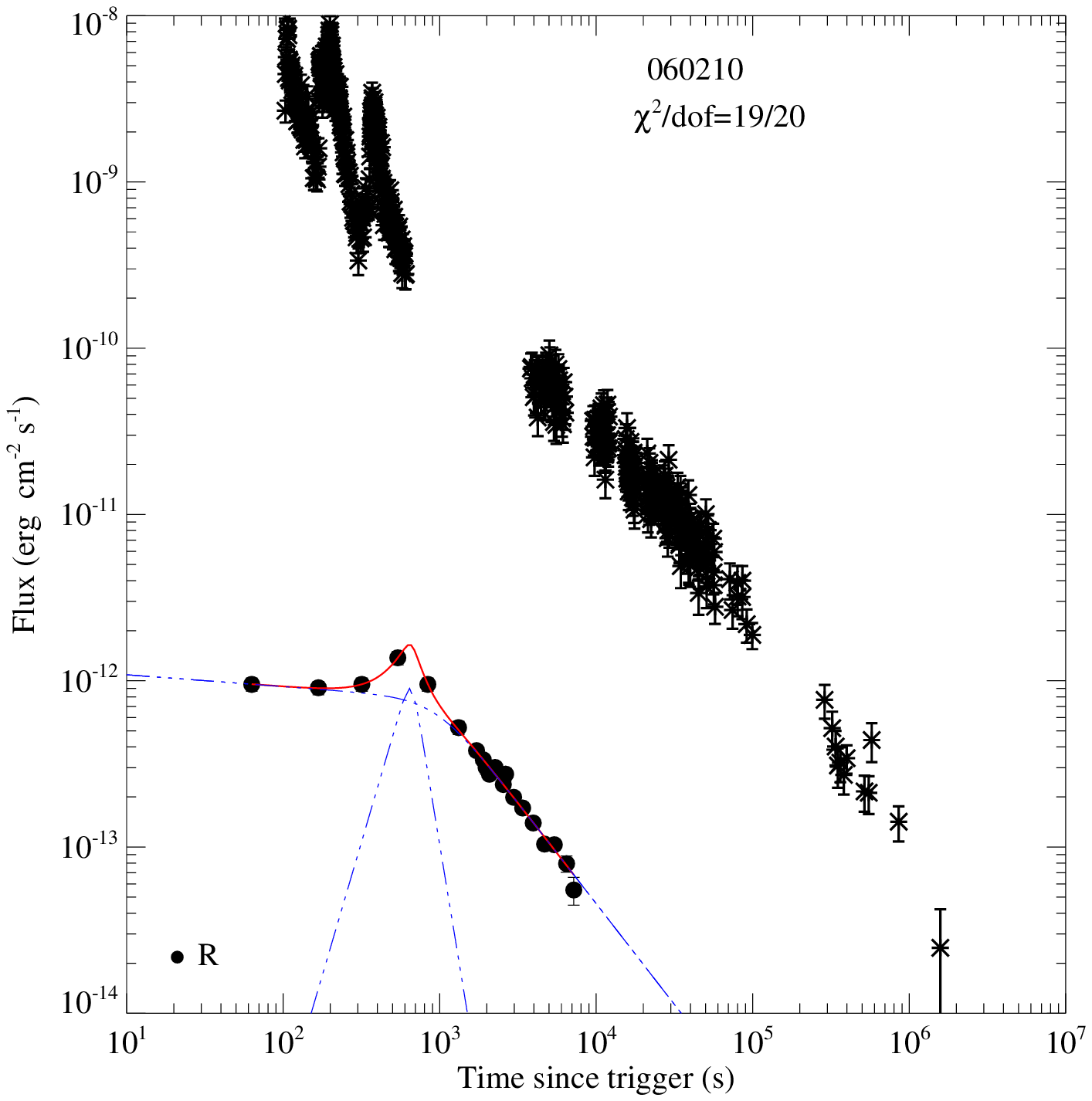}
\includegraphics[angle=0,scale=0.350,width=0.3\textwidth,height=0.25\textheight]{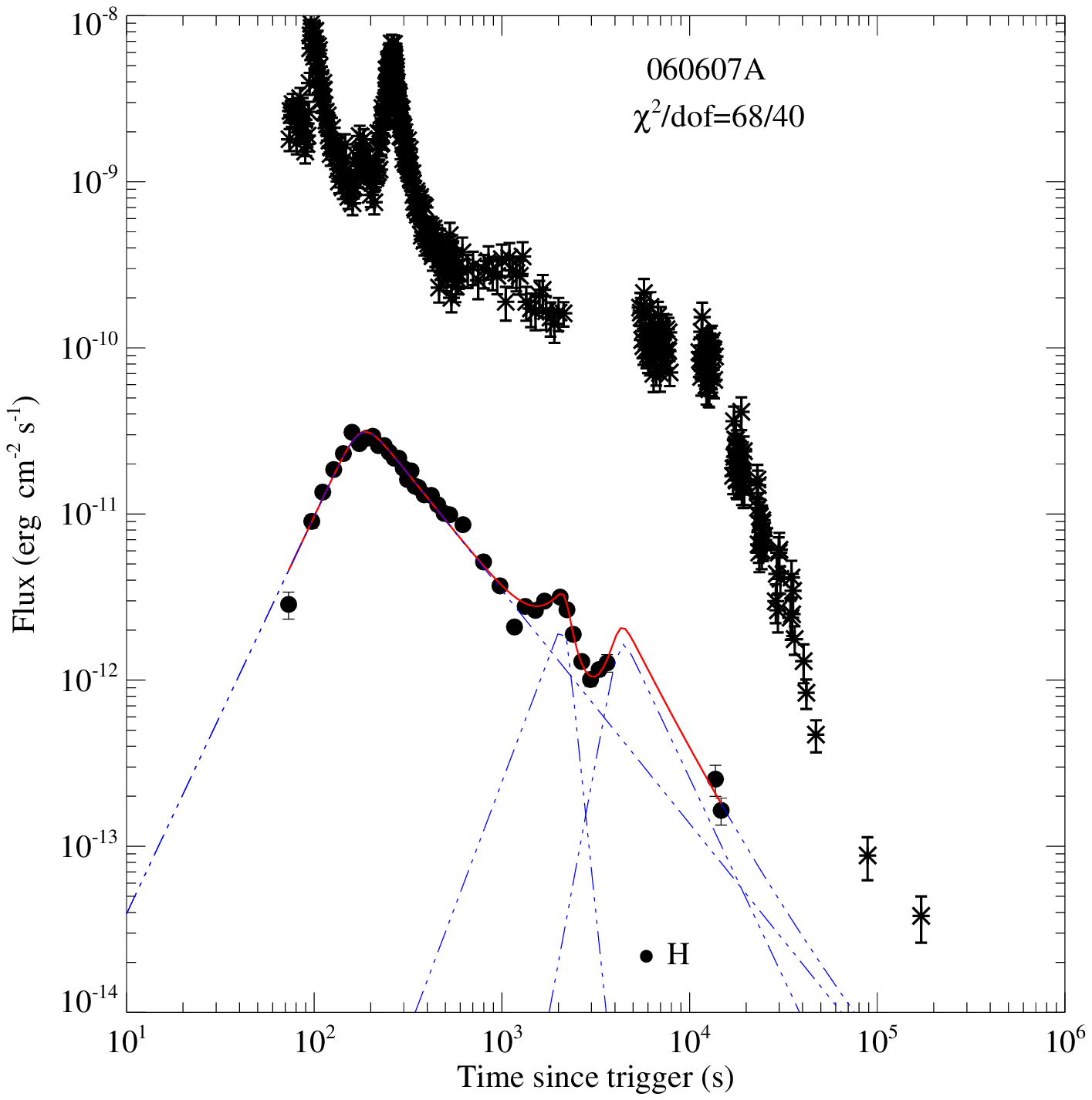}
\includegraphics[angle=0,scale=0.350,width=0.3\textwidth,height=0.25\textheight]{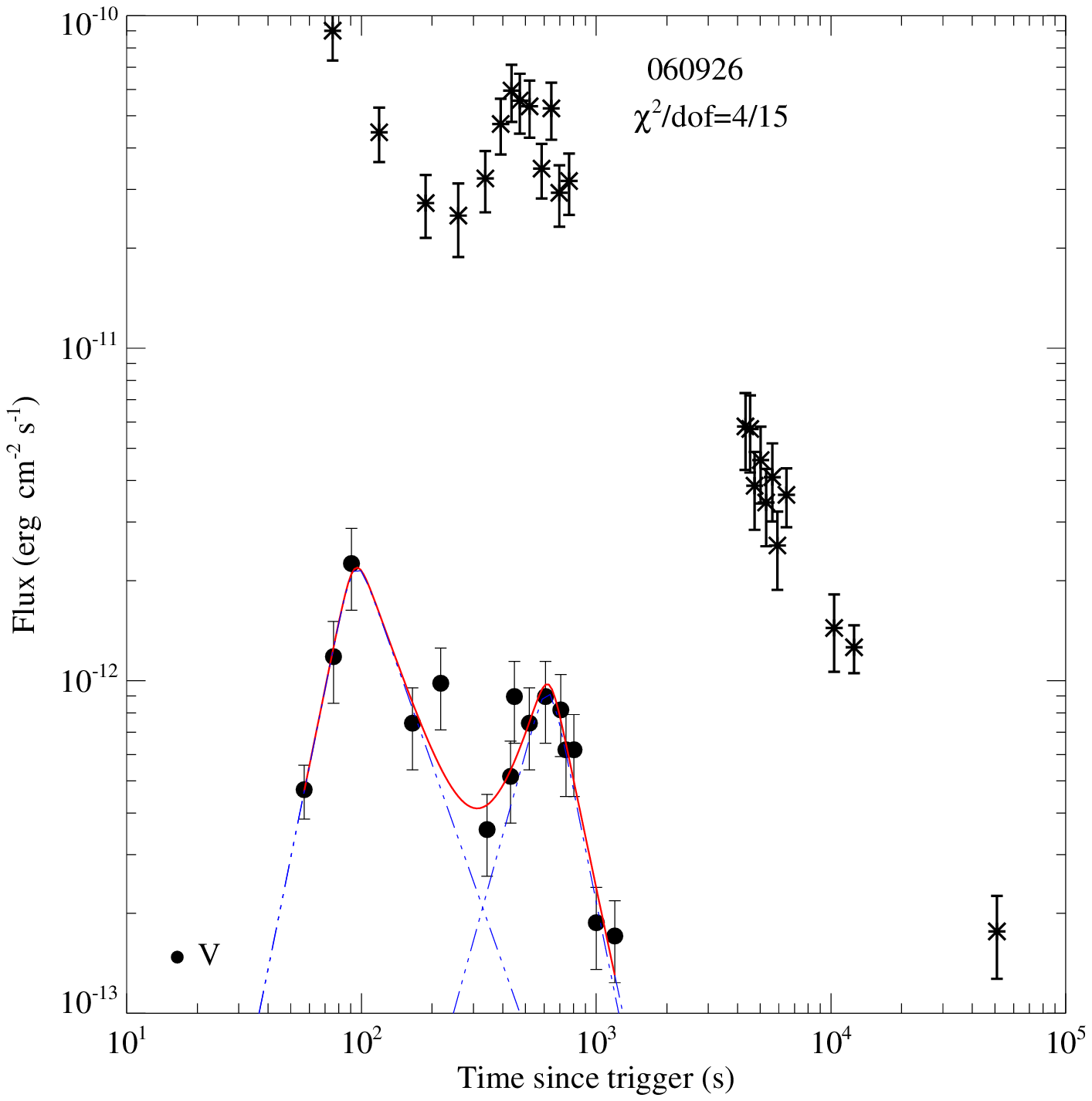}
\includegraphics[angle=0,scale=0.350,width=0.3\textwidth,height=0.25\textheight]{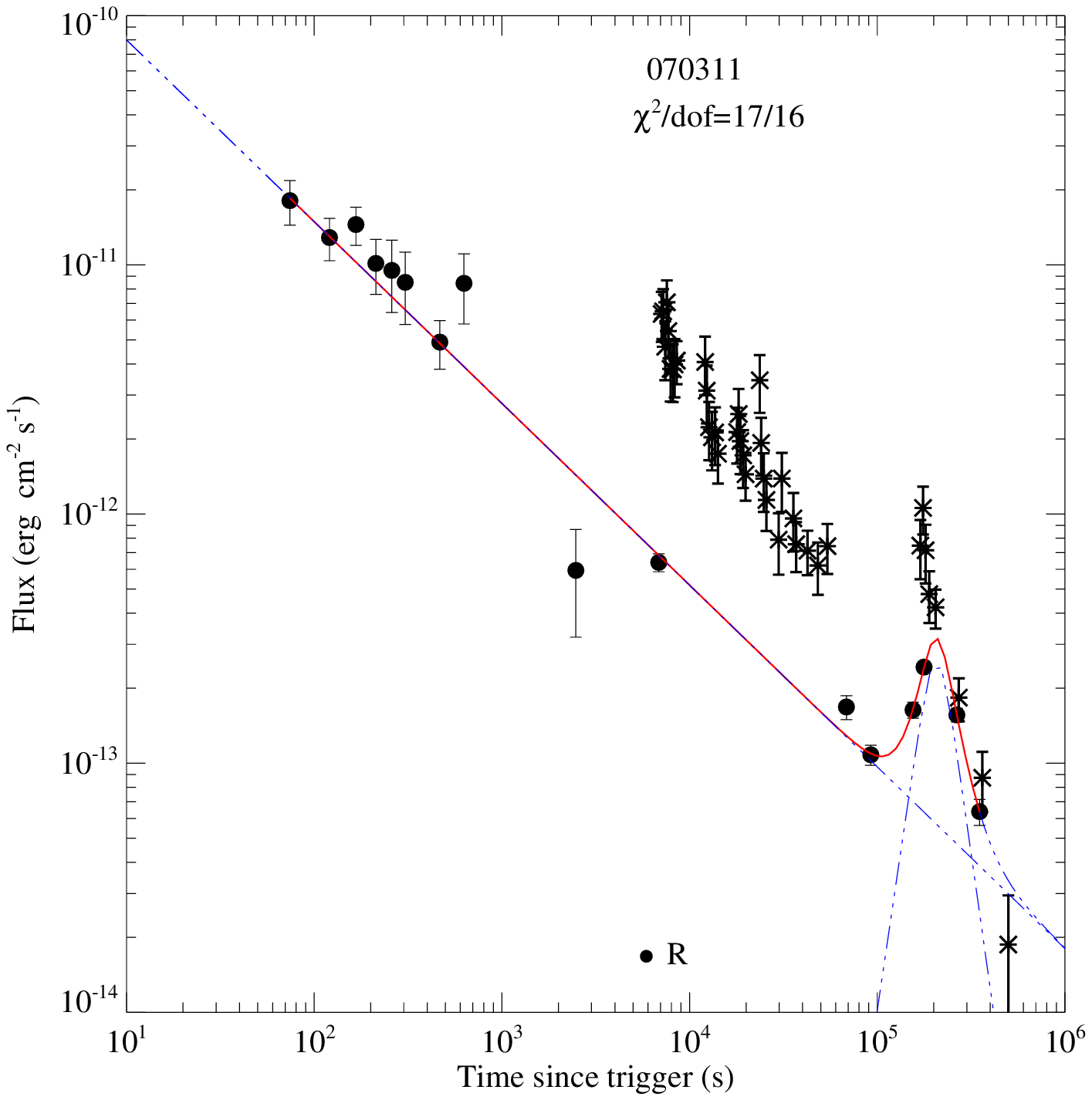}
\includegraphics[angle=0,scale=0.350,width=0.3\textwidth,height=0.25\textheight]{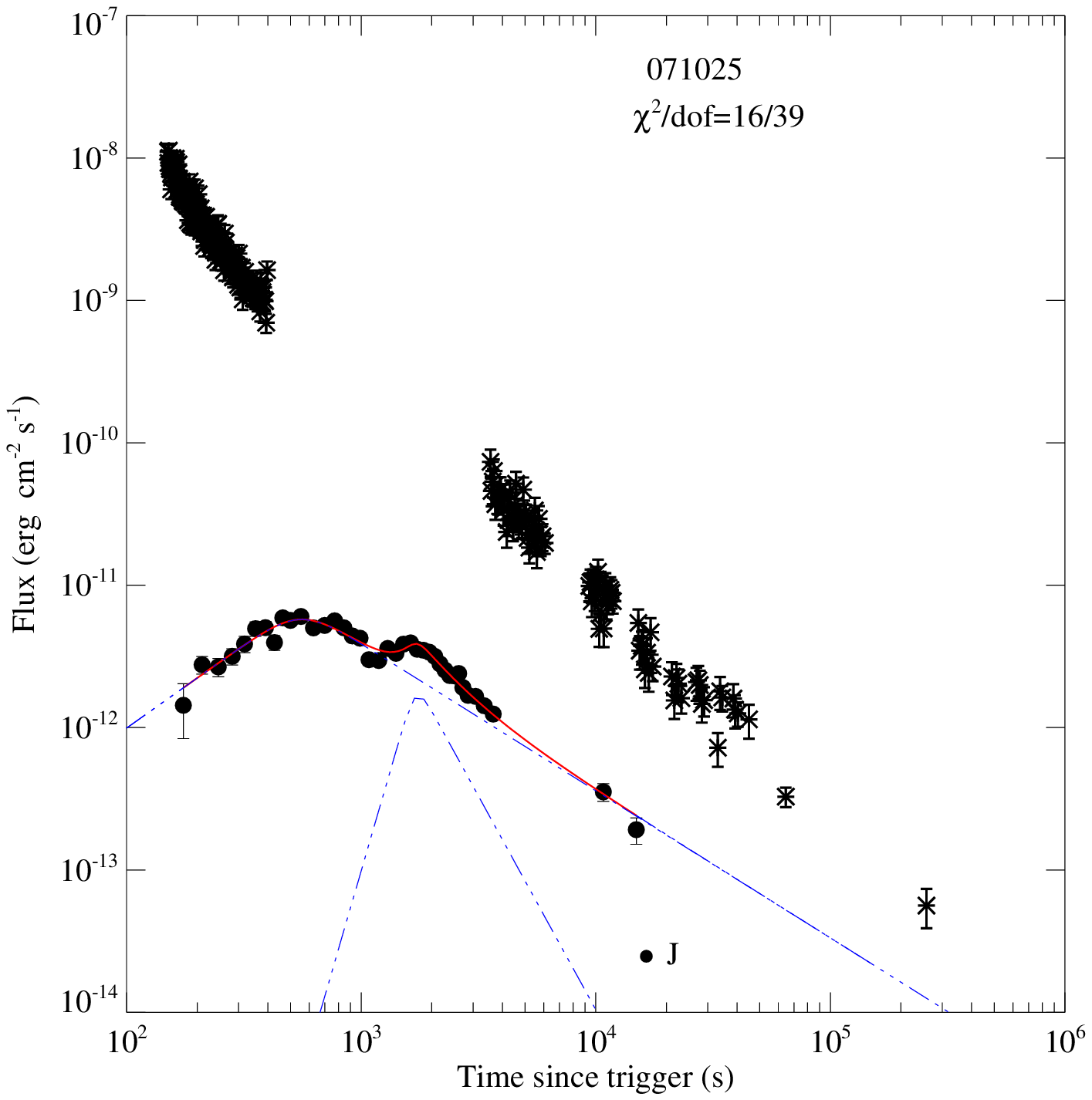}
\includegraphics[angle=0,scale=0.350,width=0.3\textwidth,height=0.25\textheight]{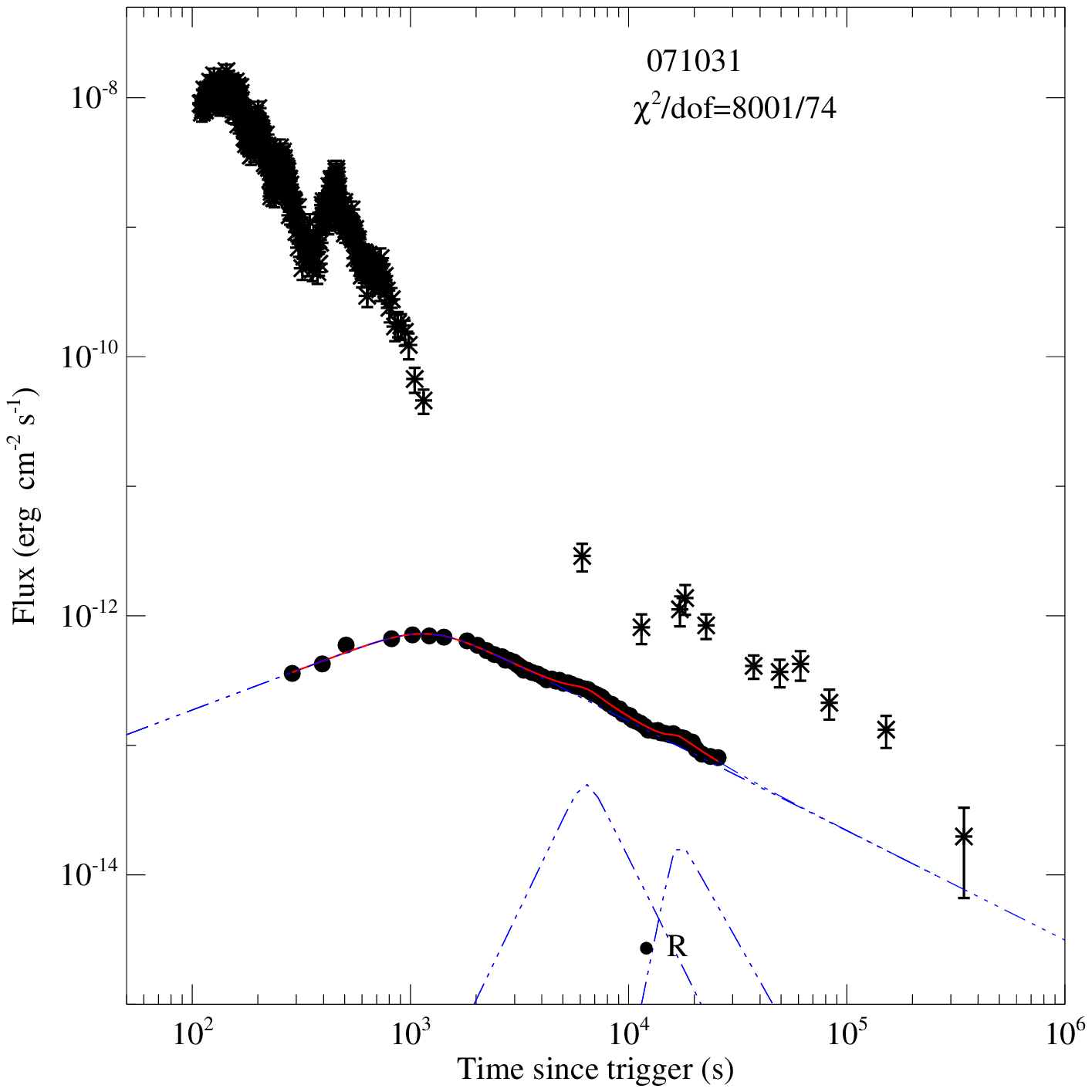}
\includegraphics[angle=0,scale=0.350,width=0.3\textwidth,height=0.25\textheight]{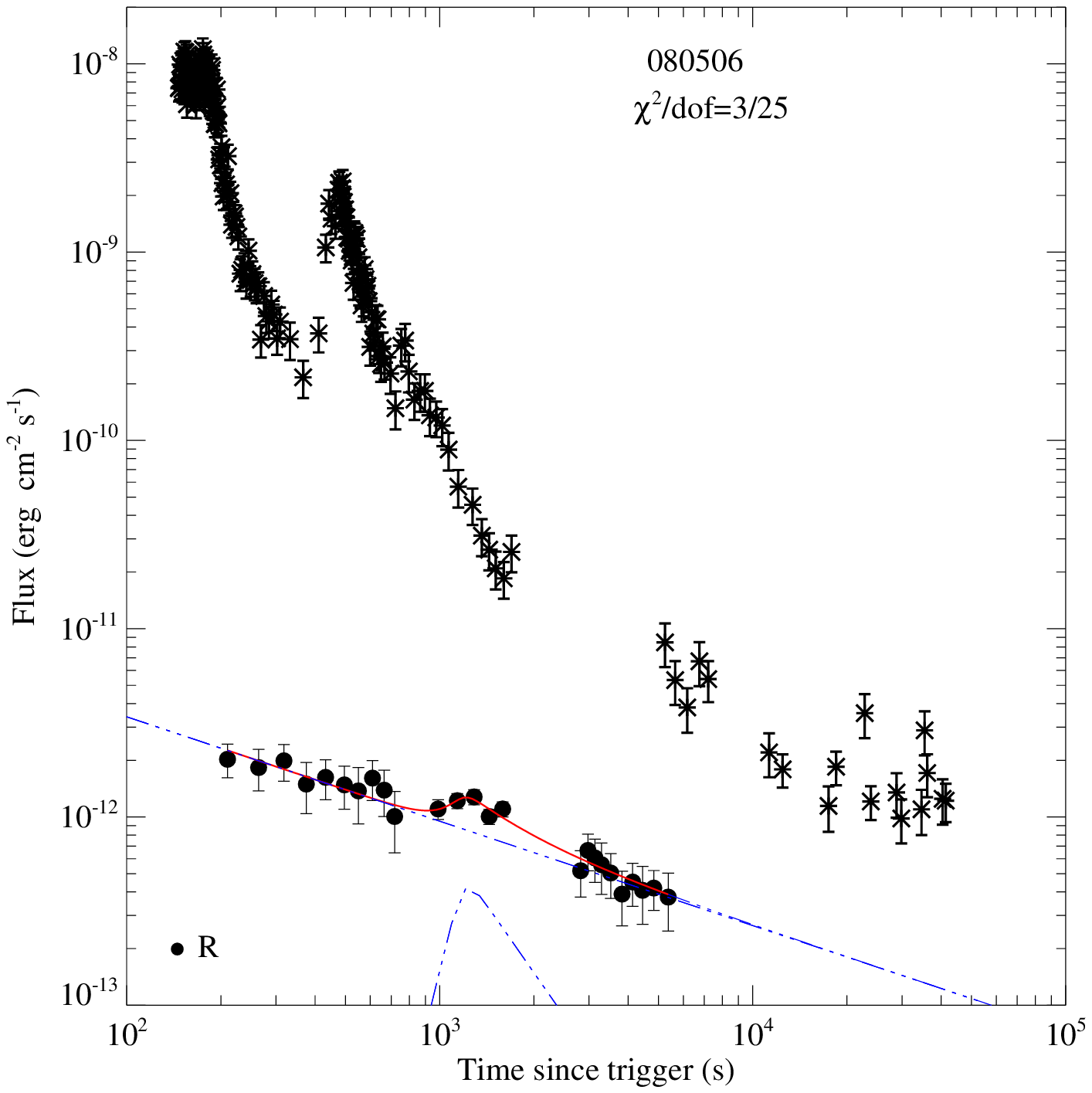}
\includegraphics[angle=0,scale=0.350,width=0.3\textwidth,height=0.25\textheight]{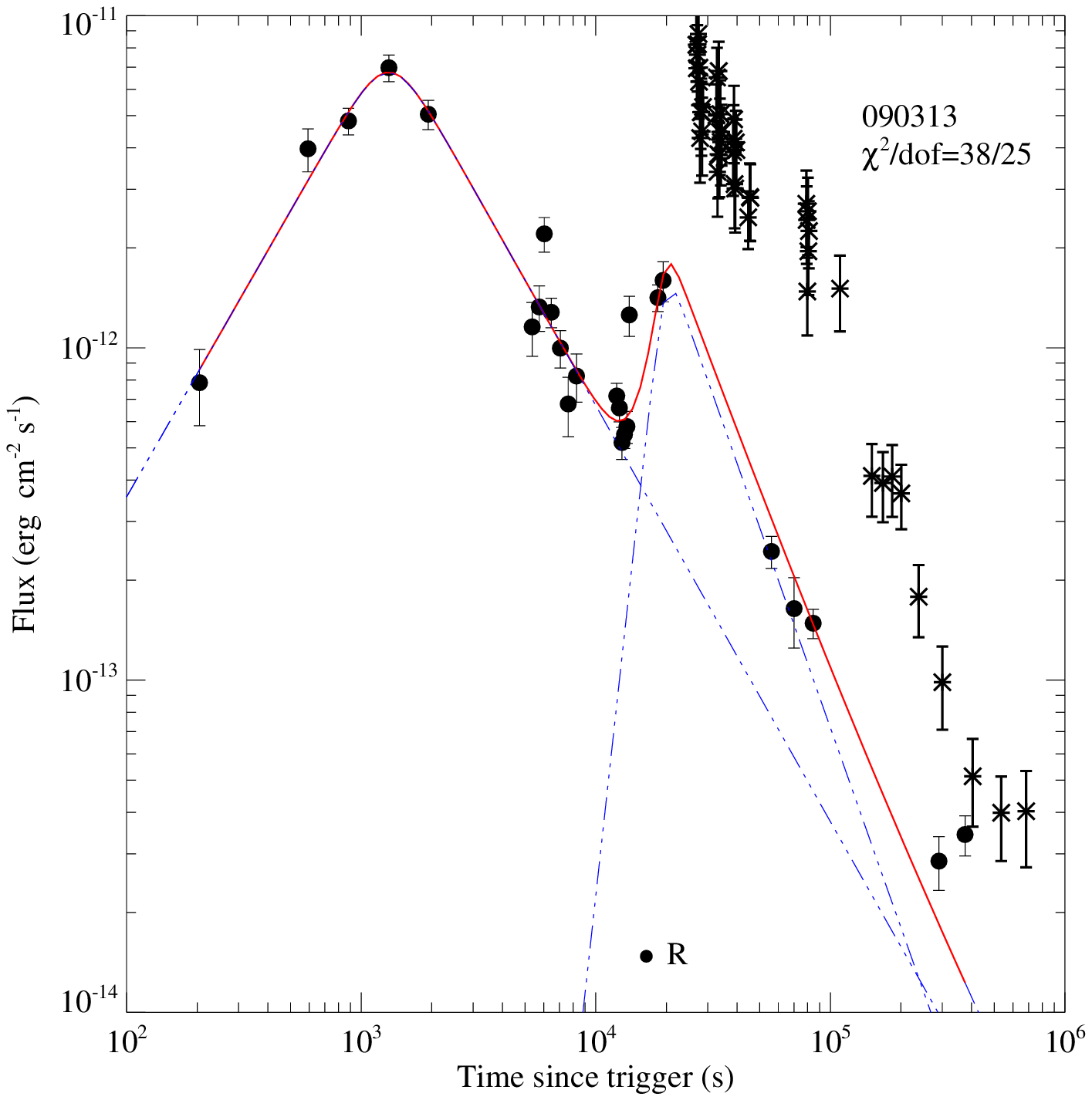}
\includegraphics[angle=0,scale=0.350,width=0.3\textwidth,height=0.25\textheight]{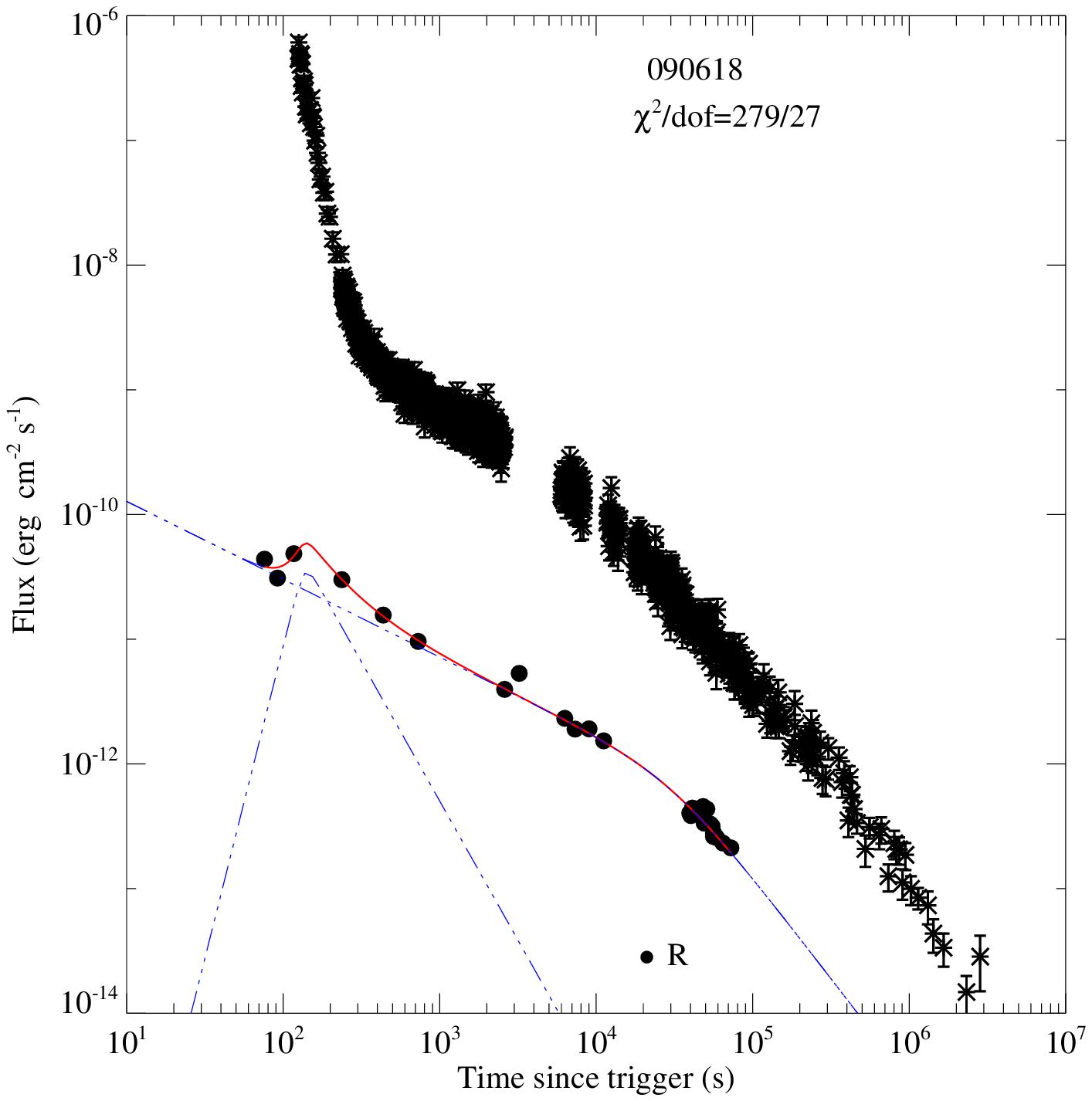}

\caption{Optical afterglow lightcurves with clear detections of at least one optical flare. The line styles and symbols are the same as Figure 1.}
\label{Opt_LC_Flares}
\end{figure*}

We get 24 flares in 19 GRBs. A flare is clearly seen in 14 out of the 19 GRBs, as shown in Figure \ref{Opt_LC_Flares} along with our best fit results. Some flares may be embedded in the lightcurves as shown in Figure \ref{Opt_LC_6}. Notice that most of the well-sampled optical lightcurves are in the R band. For a few GRBs, the flares are well-sampled in other bands. We correct these lightcurves to the R band with the optical spectral indices. The fitting parameters (the flux at peak time and the rising and decaying slopes) of the flares and the derived temporal properties, including the peak time ($t_{\rm p}^{\rm F}$), the width ($w^{\rm F}$) measured at the full-width-half-maximum (FWHM), the rising timescale $t_r^{\rm F}$, the decay timescale ($t_d^{\rm F}$), the ratio  of the $t_r^{\rm F}$ to $t^{\rm F}_d$ ($R^{\rm F}_{\rm rd}$) and the ratio  of the $t_r^{\rm F}$ to $t_{\rm p}^{\rm F}$ ($R^{\rm F}_{\rm rp}$) derived from the fitting parameters are summarized in Table 2. With the fitting parameters, we calculate the isotropic peak luminosity $(L^{\rm F}_{\rm R,iso})$ and the total energy release ($E^{\rm F}_{\rm R, iso}$) in the $R$ band.  The $E^{\rm F}_{\rm R, iso}$ is integrated from $t^{\rm F}_{\rm p}/5$ to $5t^{\rm F}_{\rm p}$. Our results are reported in Table 2.

We show the distributions of $R^{\rm F}_{\rm rp}$, $R^{\rm F}_{\rm rd}$, $t^{\rm F}_{\rm p}$, $w^{\rm F}$, $L^{\rm F}_{\rm R,iso}$ in Figure \ref{Flare_Dis}. The distribution of $R^{\rm F}_{\rm rp}$ is clustered around $0.1-0.3$, being consistent with the expectation of an internal origin of the flares. The ratio $R^{\rm F}_{\rm rd}$ is similar to that observed in GRB pulses, but the rising wing of some flares are even longer (in log scale) than the decaying wing. The $t^{\rm F}_{\rm p}$ ranges from $\sim$ tens of seconds to $\sim 10^6$ seconds.  The $w^{\rm F}$ values are in the same range as $t^{\rm F}_{\rm p}$. The $L^{\rm F}_{\rm R, iso}$ ranges from $10^{43}$ to $10^{49}$ erg s$^{-1}$, with a typical value of $10^{46}$ erg s$^{-1}$.

Relations of $w^{\rm F}$ and $L^{\rm F}_{\rm p,iso}$ of the flares as a function of $t^{\rm F}_{\rm p}$ are shown in Figure \ref{Flare_Corr}. A tight correlation between $w^{\rm F}$ and $t^{\rm F}_{\rm p}$ is found. The best fit gives $\log w^{\rm F}=-0.32+1.01\log t^{\rm F}_{\rm p}$, i.e., $w^{\rm F}\sim t^{\rm F}_{\rm p}/2$. The $L^{\rm F}_{R,p}$ is anti-correlated with $t^{\rm F}_{\rm p}$ in the burst frame, i.e., $\log L^{\rm F}_{\rm R, iso,48}=(1.89\pm 0.52)-(1.15\pm0.15) \log [t^{\rm F}_{\rm p}/(1+z)]$ with a Spearman correlation coefficient 0.85 and a chance probability $p<10^{-4}$. Therefore, later flares tends to be dimmer and wider than earlier flares.

The possible relations between the flare properties and $E_{\gamma, \rm iso}$ are shown in Figure \ref{GRB_Flare_Corr}. Since more than one flares are detected in a few GRBs, we select only the brightest one for our analysis. As shown in Figure \ref{GRB_Flare_Corr}, $E^{\rm F}_{\rm R, iso}$ is typically smaller than 1/100 of $E_{\gamma, \rm iso}$. The flare R-band luminosity $L^{\rm F}_{\rm R, iso}$ is correlated with gamma-ray luminosity $L_{\rm \gamma, iso}$, i.e., $\log L^{\rm F}_{\rm R, iso}/10^{48}=(-3.97\pm 0.60)+(1.14\pm 0.27)\log L_{\rm \gamma, iso}/10^{50}$ with a Spearman correlation coefficient of $r=0.75$ and a chance probability $p\sim 10^{-3}$. The flares in GRBs 050401, 060926, and 090726 are out of the $3\sigma$ region of the fit. Without considering the flares in these three GRBs, it is found that the $t^{'\rm F}_{\rm p}$ is also tightly anti-correlated with $E_{\gamma, \rm iso}$, i.e., $\log t^{'\rm F}_{\rm p}=(5.38\pm 0.30)-(0.78\pm 0.09)\log E_{\rm \gamma, iso}/10^{50} $ (with $r=0.92$). Similarly, a tight anti-correlation between $L_{\rm R, p}$ and $t_{\rm p}$ in the burst frame is found without considering the flares in the three GRBs, e.g., $\log [t^{\rm F}_{p}/(1+z)]=(7.57\pm 0.60)-(1.35\pm 0.17)\log E_{\rm \gamma, iso, 50}$, with a Spearman correlation coefficient 0.91. These results indicate that the optical flares of a GRB that has a larger $E_{\gamma, \rm iso}$ tends to peak earlier and brighter.

It is interesting to study whether optical flares are associated with X-ray flares. Early flares are frequently seen in X-ray afterglow lightcurves (Burrows et al. 2005, O'Brien et al. 2006). However, as shown above, early optical flares are only observed in the lightcurves of GRBs 060210, 060926, 090618, and 090726 in our sample. The fraction of GRBs with detected early optical flares is much lower than that for X-ray flares. Among the 19 GRBs with optical flare detections, 16 had early Swift XRT observations. Their X-ray afterglow lightcurves are also shown in Figure 2. Simultaneous observations with XRT during the optical flares are available for GRBs 050401, 060206, 060210, 060607A, 060926, 070311, 071010A, 071031, 080506, 090618, and 100728B.  An X-ray flare that may be associated with the optical flare is only observed in GRBs 060926, 070311, and 071010A. The optical flares of these three GRBs are lagged behind the corresponding X-ray flares. Measuring the lags with the peak time of the flares, we get 196 seconds, $7.7\times 10^4$ seconds, and $2.45\times 10^4$ seconds for the flares in GRBs 060926, 070311, and 071010A, respectively. The lag is potentially  proportional to the peak time of the flares with the three flares.

\section{Shallow decay segment}
A shallow decay segment is defined with a criterion that the initial decay slope of this segment is shallower than $3\beta_{\rm O}/2$ within error. We get a sample of 39 out of 146 GRBs that have such a shallow decay segment. Some examples are shown in Figure \ref{Shallow_Opt_LC}. Figure \ref{Shallow_Dis} shows the distributions of the decay slopes ($\alpha^{\rm S}_{\rm 1}$ and $\alpha^{\rm S}_{\rm 2}$), the break times ($t^{\rm S}_{\rm b}$), and the luminosity at the break ($L^{\rm S}_{b, \rm iso}$) of our sample. Thirty-one shallow decay segments transit to a decay slope of $1\sim 2.5$, and 5 have shallow decay segment followed by a sharp drop with a decay slope steeper than 2.5. About half of the shallow decay segments look like a plateau, with $|\alpha^{\rm S}_{b,1}|\leq 0.3$. The break time ranges from tens of seconds to several days after the GRB trigger, with a typical $t^{\rm S}_{\rm p}$ of $\sim 10^4$ seconds. The $L^{\rm S}_{\rm R,b}$ typically varies from $10^{43}$ to $10^{47}$ erg s$^{-1}$, and even reaches $\sim 10^{49}$ erg s$^{-1}$ in a few GRBs with an early break. The break luminosity $L^{\rm S}_{\rm R,b}$ is anti-correlated with $t^{\rm S}_{\rm b}$, as shown in Figure \ref{Shallow_Corr}. The best fit gives $\log L^{\rm S}_{\rm R,48}=(1.75\pm 0.22)-(0.78\pm 0.08)\log [t^{\rm S}_{\rm b}/(1+z)]$, with a Spearman correlation coefficient $r=0.86$ and a chance probability $\rho<10^{-4}$.  No correlation between $E^{S}_{\rm R,iso}$ and $t^{S}_{\rm b}$ is observed.

A shallow decay segment is commonly seen in the well-sampled of X-ray afterglow lightcurves detected by {\em Swift} XRT (e.g., Liang et al. 2007; Evans et al. 2009), except for a few GRBs whose XRT lightcurves decay as a single power-law (Liang et al. 2009). It was also reported that the X-ray luminosity at the break time is correlated with the break time (Dainotti et al. 2010). We over-plot $L_{\rm b, iso}$ as a function of $t_{\rm b}$ in the burst frame in Figure \ref{Shallow_Corr}. One can observe that optical data share a similar relation to the X-ray data. Since the X-ray luminosity is measured in the 0.3-10 KeV energy band and the optical luminosity is in the R band, the X-ray luminosity lie significantly above the optical luminosity (the $\nu F_\nu$ peak is at $\nu_c$ for slow cooling, which may be close to the X-ray band). The observed photon spectral indices of the X-ray spectra are $\sim 2$ (Liang et al. 2007). Therefore, the X-ray energy spectra are flat and the derived $L_{\rm b, iso}-t_{\rm b}$ relation in the 1 KeV band is roughly consistent with that derived from the entire X-ray band.

We examine the chromaticity of the shallow decay segments in the X-ray and optical bands.  The X-ray observations are available for 17 out of the 34 GRBs. We extract the underlying afterglow components II and III (by removing flares) for the X-ray and opical samples, and compare the parameters $\alpha^{\rm S}_{\rm 1}$, $\alpha^{\rm S}_{\rm 2}$, and $t^{\rm S}_{\rm b}$ of the two samples in Figure \ref{Shallow_Opt_Xray}. It is found that the $t^{\rm S}_{\rm b}$ data points are scattered around the equality line, and a tentative correlation between the break times of the optical and X-ray lightcurves is observed, with a chance probability of the correlation of  $\sim 0.15$. These is no correlation between the decay slopes of the X-ray and optical lightcurves. The decay segment prior to the break times in the optical bands tends to be steeper than that in the X-ray band, but the post-break slopes are roughly consistent, except for those $\alpha^{\rm S}_{2}>2.5$ in the optical bands.

We integrate the total R-band energy release ($E^{\rm S}_{R, \rm iso}$) in the shallow decay phase from 10 seconds to $t^{\rm S}_{\rm b}$ post trigger. We show $E^{S}_{\rm R, iso}$ as a function of $E_{\rm \gamma, iso}$ in Figure \ref{Eiso_ERiso}. A trend of rough proportionality between $E^{S}_{\rm R, iso}$ and $E_{\rm \gamma, iso}$ is observed. The Spearman correlation analysis shows that the chance probability of the correlation between the two quantities is $\sim {6\times 10^{-3}}$. Our robust fit yields $\log E^{\rm S}_{\rm R, iso}=0.40+0.47\log E_{\rm \gamma, iso}$. The correlation between $L^{\rm S}_{\rm b, iso}$ and $E_{\gamma, \rm iso}$ is much worse. We get $\log L^{\rm S}_{\rm R, b}=-5.57+1.13\log E_{\rm \gamma, iso}$, with a chance probability of 0.16, as shown in Figure \ref{Eiso_ERiso}.

Physically, there are two types of shallow-decay segment (plateau), as observed in the X-ray band (Liang et al. 2007). The majority of X-ray plateaus are followed by a normal decay with decay index typically around -1. These plateaus are likely of an external shock origin, with the shallow decay segment caused by continuous energy injection into the blastwave (Rees \& M\'esz\'aros 1998; Dai \& Lu 1998; Sari \& M\'esz\'aros 2000; Zhang \& M\'esz\'aros 2001a). This scenario has been applied to interpret most X-ray plateaus discovered by {\em Swift} (Zhang et al. 2006; Nousek et al. 2006). A small fraction of plateaus, first found by Troja et al. (2007) in GRB 070110 and studied systematically by Liang et al. (2007), are followed by a much steeper decay (index steeper than -3), which cannot be interpreted within the external shock model. These plateaus are called ``internal plateaus'' by Liang et al. (2007), since they have to be powered by internal dissipation of a late outflow. Looking at our optical shallow decay component sample, most lightcurves have a shallow decay component followed by a normal decay segment. Nonetheless, we identify two possible internal plateaus in GRBs 060605 and 080413B, which show superposition of a normal decay segment and a possible internal plateau with sharp drop of slope (see Figure \ref{Internal_Plateau})\footnote{An issue of defining such cases is that the possibility of the sharp drop is caused by a flare is not ruled out.}. As shown in Fig. \ref{Internal_Plateau}, the early optical lightcurves of GRBs 060605 and 080413B are a smooth bump and a normal decay segment, respectively, which are consistent with the standard afterglow model. Their late lightcurve rapidly decays with a slope of $\alpha>2.5$, and then flattens to a level consistent with the normal decay.  This suggests that the plateau is likely internal and is superposed on the external component. We also revealed evidence of such a component from the lightcurves of GRB 970508 (ending at $\sim 10^6$ seconds with a slope 3.0), 050319 (ending at $\sim 490$ seconds and $3.3\times 10^5$ with slopes 3 and 2.5, respectively). The plateau end times of these GRBs range from tens of seconds to several days after the GRB trigger.

\section{Physical implications}
\subsection{From Prompt Gamma-Ray Pulses to Late Optical Flares:  Global Evolution of an Erratic Central Engine}

As shown above, flares are an independent component superimposed onto the afterglow component. The observed relations between $E_{\gamma, \rm iso}$ or $L^{\rm F}_{R, iso}$ and $t^{\rm F}_{\rm p}$ indicate that the prompt gamma-ray emission and late optical flare emission could have the same physical origin. The temporal evolution from the GRB phase to late optical flares may signal global evolution of the erratic GRB central engine.

The most extensively discussed GRB central engine model is a hyper-accreting black hole surrounded by a neutrino-dominated accretion disk or torus (NDAFs, e.g., Popham et al. 1999; Narayan et al. 2001; Kohri \& Mineshige 2002; Di Matteo et al. 2002; Kohri et al. 2005; Gu et al. 2006; Chen \& Beloborodov 2007; Liu et al. 2007, 2008, 2010). The prompt gamma-ray phase and the late X-ray and optical flares in a burst are usually well-separated, indicating that the central engine may intermittently eject a series of shells during these emission episodes. Random collisions of these shells would make internal shocks or magnetic turbulent reconnections, which would result in the observed variability (e.g. Kobayashi et al. 1997; Zhang \& Yan 2011). With gamma-ray data alone, no significant trend of width and intensity evolution was found (Fenimore et al. 1995). However, considering both prompt gamma-ray pulse and late X-ray and optical flares, we find that these episodes are correlated and show clear temporal evolution, as shown in Figs.\ref{Flare_Corr} and \ref{GRB_Flare_Corr}. We also show the $w-t_{\rm p}$ relation for single-pulse GRBs observed with CGRO/BATSE and the X-ray flares observed with {\em Swift}/XRT in comparison with the optical flares. We find that they follow the same relation (see also Chincarini et al. 2007, 2010; Margutti et al. 2010). The general trend is that late flares/pulses tend to be wider and dimmer. This cannot be caused by hydrodynamical spreading of the shells ejected at late times, but demand that the central engine is ejecting thicker, and dimmer shells at late times (Maxham \& Zhang 2009). This may be interpreted as late flares being produced by clumps at larger radii, so that spreading during accretion process would increase the accretion time into the black hole (Perna et al. 2006; Proga \& Zhang 2006).

Liang et al. (2010) discovered a tight correlation between $E_{\gamma, \rm iso}$ and the initial Lorentz factor of the GRB ejecta, i.e., $\Gamma_0=182 (E_{\gamma, \rm iso}/10^{52} {\rm erg})^{0.25}$. Replacing $E_{\gamma, \rm iso}$ with $L_{\gamma, \rm iso}$, L\"{u} et al. (2011) got $\Gamma_0= 264(L_{\gamma, \rm iso}/10^{52} {\rm erg s^{-1}})^{0.27}$, and suggested that the correlation may be interpreted as a natural consequence of the interplay between neutrino-annihilation luminosity and neutrino mass loading from a NDAF. At even lower accretion rate, the neutrino annihilation mechanism would be inefficient to power a jet (Popham et al. 1999; Fan, Zhang \& Proga 2005). It is unclear whether luminosity is always proportional to the accretion rate $\dot m$ (it can be maintained so if the jet is magnetically launched), and whether the $L_{\rm iso}-\Gamma$ relation can be extended to lower luminosities. If one naively extends both correlations to lower luminosities, the general trend of decaying $L$ with time ($L_{R, p}\propto t^{-1.15\pm 0.15}$) is consistent with decreasing $\dot m$ with $t$ as expected in several models (e.g. $\dot m \propto t^{-1.2}$ of Cannizzo et al. 1990 and $\dot m \propto t^{-1.25}$ of Frank et al. 2002). The $L-\Gamma$ relation would predict that the Lorentz factors of optical flares are below 10. In order to maintain a low $E_p$ for these flares, the standard internal shock model would have difficulty (Zhang \& M\'esz\'aros 2002), and one needs to attribute to photosphere emission (Thompson et al. 2007) or magnetic dissipations (e.g. Zhang \& Yan 2011) in order to account for the observations.

\subsection{The Shallow Decay Segment as Probe of Late Energy Injection}
In the framework of the GRB fireball models, the shallow decay segment followed by a normal decay can be interpreted as a blastwave with continuous energy injection. There are two types of energy injection, one related to a long-lasting central engine (Dai \& Lu 1998; Zhang \& M\'esz\'aros 2001a), and another related to distribution of Lorentz factor in the promptly ejected outflow (Rees \& M\'esz\'aros 1998). The two cases have effectively similar predictions and cannot be differentiated from the data. The existence of internal plateaus observed in some GRBs suggest that at least for some GRBs, indeed a long-lasting central engine is at work. This may be related to spindown of the central engine, either a rapidly spinning black hole or a rapidly spinning magnetar. Theoretical modeling also suggests that the energy of flares can pile up onto the blastwave and make a shallow decay segment (Maxham \& Zhang 2009). Overlapping optical flares on the shallow decay segment are observed in some GRBs, such as 970508 and 000301C. This is good evidence of two emission sites: an internal origin of the optical flares and the external shock origin of the shallow decay segment. So an observed shallow decay component can be a probe of the central engine activity and energy injection into the blastwave.

For a long-lasting central engine, one may parameterize the central engine luminosity history as  $L=L_0t^{-q}$. The external shock closure relation between the decay slope and the spectral slope can be written as $\alpha=(q-1)+(2+q)\beta_O/2$ in case of $\nu_m<\nu_O<\nu_c$ (Zhang et al. 2006).
With the observed $\alpha$ and $\beta_O$, we derive the $q$ values for these GRBs in this spectral regime. The typical $q$ value is 0.5, as shown in in Figure \ref{Shallow_q_Dis}. For a black hole - torus system, the wind may be driven by neutrino annihilation or the Blandford-Znajek (1977) mechanism.
For a neutrino-driven wind, the annihilation luminosity can be estimated as $\log L_{\nu \bar{\nu}}=43.6+4.89\log (\dot{m}/0.01 {\rm M_\odot})+3.4 a_*$, where $\dot{m}$ is the accretion rate, $a_*$ is the spin parameter of the central black hole (W. H. Lei \& B. Zhang 2012, in preparation). Assuming that the pre-collapse density profile is $\rho\propto r^{\tau}$, the mass enclosed within $r$ increases with radius as $r^{\tau^{'}}$, where $\tau^{'}=3+\tau$ for $\tau>-3$ and $\tau^{'}\sim 0$ for $\tau<-3$. Then, the mass fall-back rate onto the disk is given as $\dot{m}_{\rm f}\propto t^{(2\tau+\tau^{'}+3)/(3-\tau)}$. Following Kumar et al. (2008), $\tau=-1.8$, we get $\dot{m}_{\rm f}\propto t^{-0.3}$. If the fall-back mass is accreted into the black hole, we obtain $\log L_{\nu \bar{\nu}}\propto t^{-1.5}$. This is inconsistent with $q$ values for most bursts in our sample. For the Blandford-Znajek mechanism (1977), the maximum of the power can be estimated with $L_{\rm BZ}\sim \dot{m}c^2$. If $\dot{m}\propto t^{-0.3}$  as discussed above, the decay slope is consistent with the $q$ values for most GRBs in our sample. Alternatively, if the long-lasting wind is driven by a spinning down magnetar, one has $L(T)=L_{0,\rm em}/(1+T/T_{\rm em})^2$, where $T_{\rm em}$ is the characteristic timescale for dipolar spin-down. The predicted $q$ values are 0 or 2, not consistent with the typical value $q=0.5$. On the other hand, if the shallow decay is caused by injection of flare energies into the blastwave (e.g. Maxham \& Zhang 2009), the predicted decay slope depends on the energetics and temporal distributions of the flares, and can be more flexible.

One issue to explain the shallow decay segment with energy injection onto the blastwave is the chromatic breaks in the optical and X-ray bands (e.g., Panaitescu et al. 2006; Liang et al. 2007), as shown in Figures \ref{Shallow_Opt_LC} and  \ref{Shallow_Opt_Xray}. This raises the concern regarding whether the X-ray and optical emission are from the same emission component (e.g., Liang et al. 2009; Racusin et al. 2008). Notice that the decay slopes after the break in the optical and X-ray bands are usually consistent with the expectation of external shock models. This implies that the radiation from the two energy bands could share a similar origin. Introducing inverse Compton scattering in the X-ray band may cause chromatic behavior between the X-ray and optical bands (e.g. Panaitescu \& Kumar 2000; Zhang \& M\'esz\'aros 2001b). Alternatively, the X-rays may be emission from a long lasting wind. A long-lasting reverse shock may introduce further complications (e.g. Uhm et al. 2012). In general, mixing of different emission components may be the reason for the complex chromatic behaviors observed in different energy bands.

\section{Conclusions}

We have systematically decomposed the optical afterglow lightcurves for 146 GRBs before November 2011 that have good quality optical data. By fitting the lightcurves with multiple components, we get a synthetic optical lightcurve that includes 8 components with distinct physical origins. We plan to study these components in detail in a series of papers, and in this paper we focus on the optical flares and the shallow-decay segment. Our results can be summarized as the following:

\begin{itemize}
 \item We obtained 24 optical flare events in 19 GRBs. The $t^{\rm F}_{\rm p}$ ranges from several tens of seconds to several days post the GRB trigger, and it is tightly correlated with the width and peak luminosity of the flares, i.e., $w^{\rm F}\sim t^{\rm F}_{\rm p}/2$ and $\log L^{\rm F}_{\rm R, iso,48}=(1.89\pm 0.52)-(1.15\pm0.15) \log [t^{\rm F}_{\rm p}/(1+z)]$, suggesting that flares peaking later tend to be dimmer and wider. The parameters $t^{\rm F}_{\rm p}$ and $E^{\rm F}_{\rm R, iso}$ are also corrected with $E_{\gamma, \rm iso}$, suggesting that GRBs with larger $E_{\gamma, \rm iso}$ tend to have optical flares peaking earlier and being brighter.
\item  The fraction of GRBs with detected optical flares and the number of flares in a GRB are much smaller than the case of X-ray flares. Among the 19 GRBs with detected optical flares, 16 have early {\em Swift} XRT observations. Only four cases, i.e. GRBs 970508, 060926, 070311, and 071010A, show detection of associated X-ray flares. The optical flares in the three GRBs are lagged behind the corresponding X-ray flares, similar to the spectral lag observed in prompt gamma-ray emission, but the time lag is much longer what is observed in the prompt phase.

\item  We get a sample of 42 shallow decay segments from 39 GRBs. About half of the shallow decay segments look like a plateau, with a decay slope $\alpha^{\rm S}_{b}$ being smaller than 0.3. Thirty-two out of the 39 shallow decay segments transit to a decay with a slope of $1\sim 2.5$, and 5 of them are followed by a sharp drop with a decay slope steeper than 2.5. The break times range from tens of seconds to several days after the GRB trigger, with a typical $t^{\rm S}_{\rm b}\sim 10^4$ seconds.. No clear correlation between $E^{\rm S}_{R, \rm iso}$ and  $E_{\gamma, \rm iso}$ is found.

\item The break times of the shallow decay segment in the optical and X-ray bands are chromatic for most GRBs, but they are tentative correlated. The decay prior to the break time in the X-ray band tends to be steeper than that in the optical band, and the decay slopes post the break time in the two energy bands are roughly consistent with each other. The $L^{\rm S}_{\rm R, iso}$ is anti-correlated with $t_{b}^{S}$, which is similar to the case for X-ray plateaus.
\end{itemize}

 We discussed the physical implications of the optical flares and the optical shallow decay segment, both are related to late GRB central engine activities. The observations strengthen the trend that the GRB central engine dies out gradually with the decreasing luminosity with time. The late central engine activity can be either erratic (for flares) or steady (for internal plateaus), both could add energy to the blastwave to make a shallow decay segment in the lightcurve. The observed afterglow is a mix of various emission components of the external and internal origins, and the variation of the strengths of different components lead to diverse chromatic afterglow behaviors.

\acknowledgments We acknowledge the use of the public data from the Swift data archive. We thank helpful discussion with Zi-Gao Dai, Xue-Feng Wu and Shuang-Nan Zhang. This work is supported by the National Natural Science Foundation of China (Grants No. 11025313, 10873002, 11078008, 11063001, and 11163001), the ``973" Program of China (Grant 2009CB824800), Special Foundation for Distinguished Expert Program of Guangxi
, the Guangxi SHI-BAI-QIAN project (Grant 2007201), the Guangxi Natural Science Foundation
(2010GXNSFA013112, 2011GXNSFB018063 and 2010GXNSFC013011), the special funding for national outstanding young scientist (Contract No. 2011-135), and the 3th Innovation Projet of Guangxi University. BZ acknowledges support from NASA (NNX10AD48G) and NSF (AST-0908362).

\clearpage
\begin{deluxetable}{lcccccccccccccccccccc}
\rotate
\tablewidth{650pt}
\tabletypesize{\footnotesize}
\tablecaption{Properties of the GRB sample with well-sampled optical lightcurves}
\tablenum{1}
\tablehead{
\colhead{GRB}
&\colhead{$z$}
&\colhead{$\beta_{\rm O}$}
&\colhead{$A_{\rm V}$}
&\colhead{$T_{\rm start}$\tablenotemark{a}}
&\colhead{$T_{\rm end}$\tablenotemark{a}}
&\colhead{$E_{\rm iso}$\tablenotemark{b}}
&\colhead{$E_{\rm p}$\tablenotemark{c}}
&\colhead{$\alpha$}
&\colhead{$\beta$}
&\colhead{Refs.\tablenotemark{d}}
}
\startdata
970228&0.695&0.78$\pm$0.022&0.5&69.98&3180.00&254.9$\pm$23.2&195$\pm$64&-1.54$\pm$0.08&-2.5$\pm$0.4&(1),(1),(1),(2)\\
970508&0.835&1.11&...&25.57&7420.00&42.6$\pm$7.1&145$\pm$43&-1.71$\pm$0.1&-2.2$\pm$0.25&(1),(1),(-),(3)\\
971214&3.42&0.87$\pm$0.13&0.43$\pm$0.08&46.66&304.99&3469.1$\pm$354.8&685$\pm$133&-0.76$\pm$0.1&-2.7$\pm$1.1&(1),(1),(1),(2)\\
980326&1&0.80$\pm$0.4&0&36.46&1850.00&21.8$\pm$4.4&71$\pm$36&-1.23$\pm$0.21&-2.48$\pm$0.31&(4),(1),(1),(3)\\
980425&0.0085&...&1.9$\pm$0.1&564.42&3880.00&$(83\pm8.7)\times10^{-4}$&119$\pm$24&-1$\pm$0.3&-2.1$\pm$0.1&(1),(-),(5),(2)\\
980519&...&1.07$\pm$0.12&...&29.98&231.38&...&...&...&...&(-),(6),(-),(-)\\
980613&1.096&0.6&0.45&59.44&3440.00&34.8$\pm$7.0&194$\pm$89&-1.43$\pm$0.24&-2.7$\pm$0.6&(1),(1),(1),(3)\\
980703&0.966&1.013$\pm$0.016&1.5$\pm$0.11&81.26&343.92&685.6$\pm$59.6&502$\pm$100&-1.31$\pm$0.14&-2.39$\pm$0.26&(1),(1),(1),(2)\\
990123&1.6&0.75$\pm$0.07&0&0.02&0.61&62818.1$\pm$8375.7&2030$\pm$160&-0.89$\pm$0.08&-2.45$\pm$0.97&(1),(1),(1),(2)\\
990510&1.619&0.55&0.22$\pm$0.07&12.44&340.24&1979.4$\pm$208.4&423$\pm$42&-1.23$\pm$0.05&-2.7$\pm$0.4&(1),(1),(7),(2)\\
990712&0.434&0.99$\pm$0.02&0.5$\pm$0.1&15.25&320.24&77.3$\pm$3.6&93$\pm$15&-1.88$\pm$0.07&-2.48$\pm$0.56&(1),(1),(5),(2)\\
991208&0.706&0.75$\pm$0.03&0.05&179.52&613.24&2230.0&313$\pm$31&...&...&(7),(8),(9),(10)\\
991216&1.02&0.57$\pm$0.08&...&38.75&9470.00&6534.7$\pm$687.9&641$\pm$128&-1.234$\pm$0.13&-2.18$\pm$0.39&(4),(6),(-),(2)\\
000301C&2.03&0.7&0.09$\pm$0.04&129.17&4200.00&...&...&...&...&(1),(1),(1),(-)\\
000418&1.12&0.75&0.96&214.27&3930.00&910.0&284$\pm$21&...&...&(1),(1),(1),(-)\\
000630&...&...&...&73.44&336.10&...&216$\pm$56&-0.67$\pm$0.38&-2.18&(-),(-),(-),(11)\\
000926&2.07&1.00$\pm$0.18&0.18$\pm$0.06&74.48&505.21&2710.0$\pm$2964.5&310$\pm$20&...&...&(1),(1),(1),(10)\\
010222&1.48&1.07$\pm$0.09&0&13.09&186.21&8570.9$\pm$125.8&291$\pm$43&-1.05$\pm$0.16&-2.14$\pm$0.58&(1),(1),(1),(11)\\
011121&0.36&0.8$\pm$0.15&0&37.35&1490.00&780.0&$>$700&...&$>$-2&(1),(1),(1),(12)\\
020405&0.69&1.43$\pm$0.08&0&85.04&882.60&1328.9$\pm$125.7&364$\pm$73&0.25&-1.87$\pm$0.23&(1),(1),(1),(12)\\
020813&1.25&0.85$\pm$0.07&0.14$\pm$0.04&6.05&5000.00&7626.5$\pm$762.6&211$\pm$42&-1.05$\pm$0.11&-2.3&(1),(1),(1),(12)\\
020903&0.251&...&...&57.02&4548.00&1.137$\pm$0.858&2.7&-1&$<$-2&(13),(-),(-),(13)\\
021004&2.335&0.39&0.3&21.12&2030.00&499.5$\pm$115.3&267$\pm$117&-1$\pm$0.2&...&(1),(1),(1),(10)\\
021211&1.01&0.69&0&0.13&8.96&111.5$\pm$10.1&47$\pm$9&-0.85$\pm$0.09&-2.37$\pm$0.42&(1),(1),(1),(12)\\
030226&1.98&0.7$\pm$0.03&0.53&17.34&353.70&1053.4$\pm$98.8&108$\pm$22&-0.95$\pm$0.1&-2.3&(1),(1),(1),(12)\\
030323&3.37&0.89$\pm$0.04&0.13$\pm$0.09&34.68&895.75&280.0&...&...&...&(1),(1),(7),(-)\\
030328&1.52&0.36$\pm$0.45&0.05$\pm$0.15&4.90&227.47&2705.8$\pm$208.1&110$\pm$22&-1$\pm$0.11&-2.3&(1),(14),(14),(12)\\
030329&0.17&0.5&0.3$\pm$0.03&11.17&2860.00&154.6$\pm$14.1&68$\pm$2&-1.26$\pm$0.02&-2.28$\pm$0.05&(1),(1),(1),(12)\\
030418&...&...&...&0.29&7.19&...&...&...&...&(-),(-),(-),(-)\\
030429&2.65&0.75&0.34&12.53&574.04&216.0&...&...&...&(1),(1),(1),(-)\\
030723&0.4&0.66$\pm$0.21&0.32$\pm$0.22&15.00&6050.00&4.5&4.8&-1&-2&(-),(14),(14),(15)\\
030725&...&2.9$\pm$0.6&...&335.23&1280.00&...&...&...&...&(-),(16),(-),(-)\\
040924&0.859&0.7&0&0.95&62.99&141.3$\pm$11.5&102$\pm$35&-1.17$\pm$0.05&...&(1),(1),(5),(17)\\
041006&0.716&0.55&0&0.23&5580.00&1082.1$\pm$282.3&108$\pm$22&-1.37$\pm$0.14&...&(1),(1),(1),(17)\\
041218&...&...&...&8.77&18.62&...&...&...&...&(-),(-),(-),(-)\\
041219A&0.31&...&6.8$\pm$1.6&0.44&186.56&...&...&...&...&(18),(-),(18),(-)\\
050319&3.24&0.74$\pm$0.42&0.05$\pm$0.09&0.04&994.12&1186.6$\pm$187.4&45$\pm$43&-2$\pm$0.2&...&(1),(7),(19),(20)\\
050401&2.9&0.39$\pm$0.05&0.65$\pm$0.04&3.46&1120.00&6469.4$\pm$1362.0&119$\pm$16&-0.83$\pm$0.13&-2.37$\pm$0.09&(1),(21),(21),(22)\\
050408&1.2357&0.28$\pm$0.33&0.73$\pm$0.18&3.35&3670.00&...&...&...&...&(7),(23),(23),(-)\\
050416A&0.65&1.3&0.7&4.09&32.61&16.2$\pm$1.7&28.6$\pm$8.3&-1.01&-3.4&(5),(24),(5,(17)\\
050502A&3.793&0.76$\pm$0.16&0&0.05&17.85&...&...&...&...&(1),(5),(5),(-)\\
050525A&0.606&0.97$\pm$0.1&0.25$\pm$0.16&0.07&35.64&379.5$\pm$73.0&127$\pm$5.5&-1.01$\pm$0.11&...&(1),(1),(1),(17)\\
050603&2.821&0.2$\pm$0.1&...&34.09&219.71&5665.8$\pm$323.8&1333$\pm$107&-0.79$\pm$0.06&-2.15$\pm$0.09&(25),(26),(-),(17)\\
050721&...&1.16$\pm$0.35&...&1.48&248.60&460.0$\pm$90.0&63$\pm$21&-1.8$\pm$0.2&...&(-),(27),(-),(20)\\
050730&3.969&0.52$\pm$0.05&0.12$\pm$0.02&0.07&72.70&900.0$\pm$300.0&196$\pm$87&-1.4$\pm$0.1&...&(1),(5),(21),(20)\\
050801&1.56&1$\pm$0.16&0.3$\pm$0.18&0.02&21.65&46.1$\pm$8.4&44$\pm$42&-1.9$\pm$0.2&...&(25),(28),(5),(20)\\
050820A&2.612&0.72$\pm$0.03&0.07$\pm$0.01&0.23&663.30&15923.6$\pm$1244.0&1325$\pm$277&-1.12$\pm$0.14&...&(1),(5),(19),(17)\\
050824&0.83&0.4$\pm$0.04&0.15$\pm$0.03&0.63&8990.00&15.0$\pm$4.0&13$\pm$12&...&-2.9$\pm$0.4&(21),(21),(21),(20)\\
050904&6.29&1.31$\pm$1.2&1&11.05&459.36&83951.5$\pm$8395.2&3178$\pm$1094&-1.11$\pm$0.06&-2.2$\pm$0.4&(29),(29),(30),(17)\\
050922C&2.198&0.51$\pm$0.05&0&0.74&606.01&574.9$\pm$195.7&417$\pm$118&-0.83$\pm$0.26&...&(31),(5),(5),(17)\\
051021&...&...&...&1.61&35.82&...&99$\pm$32&-0.4$\pm$0.8&...&(-),(-),(-),(20)\\
051028&3.7&0.6&0.7&8.21&58.26&1868.2$\pm$168.1&298$\pm$73&-0.73$\pm$0.22&...&(-),(32),(33),(33)\\
051109A&2.346&0.7&...&0.04&13.30&1198.5$\pm$117.5&539$\pm$381&-1.25$\pm$0.5&...&(25),(34),(-),(17)\\
051111&1.55&0.76$\pm$0.07&0.2$\pm$0.1&0.03&7.59&1507.3$\pm$150.7&447$\pm$175&-1.22$\pm$0.09&-2.1$\pm$0.27&(25),(53),(-),(22)\\
051221A&0.5459&0.64$\pm$0.05&...&11.12&445.12&35.5$\pm$1.1&390$\pm$190&-1.34$\pm$0.06&...&(20),(35),(-),(20)\\
060110&$<$5&0.8&...&0.03&4.78&320.0$\pm$60.0&135$\pm$47&-1.58$\pm$0.08&...&(25),(36),(-),(20)\\
060111B&...&0.7&3.6$\pm$0.5&0.03&13700.00&1100.0$\pm$500.0&540$\pm$280&-0.9$\pm$0.2&...&(-),(37),(38),(20)\\
060117&...&...&...&0.13&0.50&3600.0$\pm$200.0&72$\pm$5&-1.4$\pm$0.1&...&(-),(-),(-),(20)\\
060121&4.5&...&...&7.14&3120.00&249.6$\pm$23.3&134$\pm$32&0.82$\pm$0.38&...&(39),(-),(-),(39)\\
060124&2.296&0.73$\pm$0.08&0.05$\pm$0.26&3.34&1980.00&4987.8$\pm$733.5&636$\pm$162&-1.48$\pm$0.02&...&(25),(40),(7),(17)\\
060206&4.048&0.73$\pm$0.05&0.01$\pm$0.02&2.89&201.58&1568.8$\pm$336.2&381$\pm$98&-1.06$\pm$0.34&...&(37),(5),(37),(17)\\
060210&3.91&0.37&1.18$\pm$0.1&0.06&7.19&4150.0$\pm$570.0&575$\pm$186&-1.12$\pm$0.26&...&(1),(37),(19),(17)\\
060218&0.0331&...&0.5&0.25&2850.00&0.5$\pm$0.0&4.9$\pm$0.3&-1.622$\pm$0.16&...&(25),(-),(5),(17)\\
060418&1.489&0.78$\pm$0.09&0.12$\pm$0.05&0.08&7.66&4859.4$\pm$1056.4&572$\pm$114&-1.5$\pm$0.15&...&(41),(7),(7),(17)\\
060512&0.4428&0.68$\pm$0.05&...&0.11&5.93&2.0$\pm$0.4&23$\pm$20&...&...&(25),(21),(-),(20)\\
060526&3.21&0.51$\pm$0.32&0.05$\pm$0.11&0.06&893.55&606.4$\pm$303.2&105.2$\pm$21.1&-1.1$\pm$0.4&-2.2$\pm$0.4&(1),(7),(7),(17)\\
060605&3.78&1.06&0&0.07&6.32&283.0$\pm$45.0&490$\pm$251&-1$\pm$0.44&...&(42),(42),(42),(17)\\
060607A&3.082&0.72$\pm$0.27&0&0.07&14.73&2341.8$\pm$148.6&575$\pm$200&-1.09$\pm$0.19&...&(37),(7),(19),(17)\\
060614&0.125&0.47$\pm$0.04&0.11$\pm$0.03&1.55&1280.00&21.0$\pm$56.2&55$\pm$45&...&...&(21),(21),(21),(17)\\
060714&2.711&0.44$\pm$0.04&...&3.86&185.46&1510.0$\pm$195.4&234$\pm$109&-1.77$\pm$0.24&...&(43),(21),(-),(17)\\
060729&0.54&0.78$\pm$0.03&0.07$\pm$0.02&0.70&662.39&64.9$\pm$4.5&67$\pm$25&-1.8$\pm$0.1&...&(21),(21),(21),(20)\\
060904B&0.703&1.11$\pm$0.1&0.08$\pm$0.08&0.02&163.13&77.0$\pm$9.9&103$\pm$26&-0.61$\pm$0.42&-1.78$\pm$0.23&(25),(7),(7),(22)\\
060906&3.686&0.56$\pm$0.02&£¼0.09&0.66&13.61&1726.7$\pm$139.3&209$\pm$43&-1.6$\pm$0.31&...&(21),(7),(7),(17)\\
060908&2.43&0.3&0.05$\pm$0.03&0.83&7.24&1334.2$\pm$127.9&124$\pm$24&-0.89$\pm$0.2&-2.24$\pm$0.34&(37),(44),(19),(22)\\
060912A&0.937&0.62&0.46$\pm$0.23&1.10&23.90&99.8$\pm$4.5&200$\pm$110&-1.7$\pm$0.09&...&(25),(37),(45),(20)\\
060926&3.2&0.82$\pm$0.01&0.32$\pm$0.02&0.06&7.16&100.0$\pm$20.0&19$\pm$18&...&-2.5$\pm$0.3&(21),(21),(21),(20)\\
060927&5.6&0.86$\pm$0.03&£¼0.12&0.02&1.17&5815.2$\pm$861.5&473$\pm$116&-0.93$\pm$0.38&...&(21),(21),(21),(17)\\
061007&1.261&0.78$\pm$0.02&0.39$\pm$0.01&0.03&14.60&42103.8$\pm$4190.4&498$\pm$30&-0.53$\pm$0.05&-2.61$\pm$0.31&(46),(21),(46),(22)\\
061121&1.314&...&...&7.14&120.81&2665$\pm$235.3&606$\pm$90&-1.32$\pm$0.05&...&(47),(-),(-),(47)\\
061126&1.1588&0.95&0.1$\pm$0.06&0.04&156.38&28467.5$\pm$3272.1&1337$\pm$410&-1.06$\pm$0.07&...&(48),(48),(7),(17)\\
070110&2.352&\textbf{1.00$\pm$0.14}&\textbf{0.08}&0.66&34.76&723$\pm$104&110$\pm$50&-1.57$\pm$0.12&...&(Swift),(-),(-),(49)\\
070125&1.547&0.55$\pm$0.04&0.11$\pm$0.04&105.86&349.05&8968.2$\pm$773.1&367$\pm$51&-1.1$\pm$0.1&-2.1$\pm$0.15&(50),(21),(46),(51)\\
070208&1.165&0.68&...&1.17&4.85&28.0$\pm$8.0&60$\pm$20&1&...&(25),(37),(-),(20)\\
070306&1.497&0.7&5.45$\pm$0.61&12.44&215.14&600.0$\pm$100.0&105&1.67$\pm$0.1&...&(52),(52),(52),(20)\\
070311&...&1.0$\pm$0.2&0.8$\pm$0.15&0.07&350.93&...&...&-1.3$\pm$0.1&...&(-),(53),(53),(53)\\
070318&0.836&0.78&0.44$\pm$0.11&0.06&87.37&134.8$\pm$32.7&365$\pm$284&-1.34$\pm$0.27&-2.15$\pm$0.36&(25),(37),(45),(54)\\
070411&2.954&...&...&0.18&516.63&1000.0$\pm$200.0&108&1.7$\pm$0.1&...&(25),(-),(-),(20)\\
070419A&0.97&0.8&0.42$\pm$0.37&0.21&62.22&18.7$\pm$2.1&30$\pm$7&0$\pm$2&...&(25),(24),(7),(20)\\
070420&...&...&...&0.12&10.84&3100.0$\pm$500.0&150$\pm$40&1$\pm$0.2&...&(-),(-),(-),(20)\\
070518&1.16&0.8&0.3&2.11&311.76&24.6&36$\pm$33&...&-2.1$\pm$0.3&(25),(24),(7),(55)\\
070611&2.04&0.73&...&0.27&8.87&44.8$\pm$6.6&67$\pm$26&1&...&(25),(37),(-),(55)\\
070707&1.0&0.75$\pm$0.13&...&39.52&3320.00&8.9$\pm$4.2&...&-1.19$\pm$0.13&...&(56),(56),(-),(57)\\
071003&1.605&1.25$\pm$0.09&0.34$\pm$0.11&0.57&5.00&1800.0$\pm$600.0&410$\pm$190&-1.31$\pm$0.07&...&(58),(58),(7),(55)\\
071010A&0.98&0.68&0.64$\pm$0.09&0.32&523.23&13.0$\pm$2.0&37$\pm$35&...&-2.1$\pm$0.4&(25),(59),(7),(55)\\
071010B&0.947&...&...&0.06&174.46&173.8$\pm$90.0&52$\pm$10&-1.25$\pm$0.49&-2.65$\pm$0.29&(60),(-),(-),(60)\\
071025&5.2&0.42$\pm$0.08&1.09$\pm$0.2&0.17&14.88&1500.0$\pm$300.0&165$\pm$59&1.67$\pm$0.06&...&(25),(37),(36),(55)\\
071031&2.692&0.64$\pm$0.01&0.14$\pm$0.13&0.07&350.93&390.0$\pm$60.0&12$\pm$11&...&-2.3$\pm$0.3&(25),(21),(7),(55)\\
071112C&0.823&0.63$\pm$0.29&0.23$\pm$0.21&0.13&69.64&...&...&-1.09$\pm$0.07&...&(7),(7),(7),(61)\\
071122&1.14&0.83&...&1.30&9.05&30.0$\pm$10.0&...&...&...&(25),(37),(-),(-)\\
080109&...&...&...&72.58&3000.00&...&...&...&...&(-),(-),(-),(-)\\
080129&4.349&...&...&0.42&500.70&...&...&...&...&(62),(-),(-),(-)\\
080310&2.4266&0.42$\pm$0.12&0.19$\pm$0.05&0.30&124.42&590.0$\pm$100.0&22$\pm$20&...&-2.4$\pm$0.2&(25),(7),(7),(55)\\
080319A&$<$2.2&0.77$\pm$0.02&...&0.15&4.46&800.0$\pm$100.0&105$\pm$35&-1.6$\pm$0.1&...&(25),(21),(-),(55)\\
080319B&0.937&...&...&0.01&4590.00&52639.3$\pm$4024.7&$>$1382&1.09$\pm$0.02&...&(55),(-),(-),(55)\\
080319C&1.949&0.77$\pm$0.02&0.59$\pm$0.12&0.08&1.43&5206.3$\pm$1041.3&307$\pm$58&-1.01$\pm$0.08&-1.87$\pm$0.39&(21),(44),(7),(22)\\
080330&1.51&0.49&0.19$\pm$0.08&0.09&116.56&41.0$\pm$6.0&20$\pm$19&...&-2.4$\pm$0.5&(25),(59),(7),(55)\\
080413A&2.433&0.67&0.13$\pm$0.07&1.17&18.34&1855.0$\pm$397.0&126$\pm$42&-1.15$\pm$0.29&-2.12$\pm$0.33&(25),(59),(7),(54)\\
080413B&1.1&0.25$\pm$0.07&...&0.08&5190.00&175.7$\pm$21.9&67$\pm$8&-1.24$\pm$0.26&-2.77$\pm$0.22&(25),(59),(-),(54)\\
080506&...&0.95$\pm$0.05&...&0.21&5.37&190.0$\pm$40.0&67$\pm$28&-1.70.2&...&(-),(63),(-),(55)\\
080603A&1.67842&...&...&0.11&350.44&...&...&...&...&(Swift),(-),(-),(-)\\
080710&0.845&0.8$\pm$0.09&0.11$\pm$0.04&0.42&266.59&80.0$\pm$40.0&300$\pm$200&-1.3$\pm$0.2&...&(25),(7),(7),(55)\\
080721&2.602&0.68$\pm$0.02&0.6&38.40&559.52&18915.6$\pm$556.3&485$\pm$37&-0.93$\pm$0.05&-2.43$\pm$0.26&(30),(21),(64),(22)\\
080804&2.2&0.43&...&1.16&26.11&1600.0$\pm$700.0&410$\pm$200&-1$\pm$0.1&...&(25),(59),(-),(55)\\
080810&3.35&0.44&0.16$\pm$0.02&0.04&7.90&3000.0$\pm$2000.0&313.5&-0.91&...&(25),(7),(7),(65)\\
080913&6.7&0.79$\pm$0.03&-0.58$\pm$0.67&0.58&870.04&710.7$\pm$88.8&135$\pm$47&-0.4$\pm$0.9&...&(25),(21),(7),(55)\\
080928&1.692&1.08$\pm$0.02&0.29$\pm$0.03&0.39&13.43&280.0$\pm$50.0&...&-1.8&...&(21),(21),(21),(65)\\
081008&1.967&...&...&0.11&184.52&630$\pm$350&117$\pm$50&-1.26$\pm$0.24&...&(66),(-),(-),(66)\\
081028&3.038&...&...&11.19&57.50&1081.4$\pm$1405.9&67$\pm$13&1.3$\pm$0.4&...&(25),(-),(-),(55)\\
081029&3.85&1$\pm$0.01&0.03$\pm$0.02&0.53&252.67&1513.6$\pm$558.6&...&...&...&(67),(67),(67),(-)\\
081109A&0.98&...&...&0.17&66.60&530.0$\pm$80.0&99$\pm$40&-1.27$\pm$0.34&-2.19$\pm$0.42&(25),(-),(-),(54)\\
081126&...&...&...&0.10&0.54&900.0$\pm$200.0&...&...&...&(-),(-),(-),(55)\\
081203A&2.1&0.596&0.09$\pm$0.04&0.08&5.76&1700.0$\pm$400.0&201$\pm$75&-1.44$\pm$0.06&...&(25),(7),(7),(55)\\
090102&1.547&0.74&0.12$\pm$0.11&0.04&264.55&1400.0$\pm$500.0&370$\pm$220&-1.36$\pm$0.1&...&(25),(7),(7),(55)\\
090313&3.375&0.71&0.34$\pm$0.15&0.20&7870.00&460.0$\pm$50.0&55$\pm$51&-1.9$\pm$0.3&...&(25),(59),(7),(55)\\
090323&3.57&0.65$\pm$0.13&0.14$\pm$0.04&162.43&768.96&33626.4&697&-0.89&...&(68),(68),(68),(65)\\
090328&0.736&1.19$\pm$0.21&0.22$\pm$0.12&57.89&1070.00&1902.6&653&-0.93&-2.2&(68),(68),(68),(65)\\
090426&2.609&0.76$\pm$0.14&...&0.09&10.75&42.04.0&45$\pm$43&...&-2$\pm$0.3&(69),(70),(-),(55)\\
090510&0.903&...&...&0.11&103.79&...&3900$\pm$280&-0.58$\pm$0.06&-2.83$\pm$0.20&(71),(-),(-),(72)\\
090618&0.54&0.5&0.3$\pm$0.1&0.08&72.58&2476.6&155.5&-1.26&-2.5&(73),(73),(73),(65)\\
090726&2.71&...&...&0.20&3.02&186.9$\pm$17.2&27$\pm$22&-1.2$\pm$1.3&...&(25),(-),(-),(55)\\
090812&2.452&0.36&...&0.03&0.14&4585.8$\pm$597.4&190$\pm$65&-1.5$\pm$0.3&...&(25),(59),(-),(55)\\
090902B&1.822&0.68$\pm$0.11&0.2$\pm$0.06&4.80&563.79&...&...&...&...&(74),(74),(74),(-)\\
090926A&2.1062&0.72$\pm$0.17&0.13$\pm$0.06&73.16&102.15&26562.4$\pm$963.4&412$\pm$20&-0.74$\pm$0.01&-2.34$\pm$0.01&(7),(7),(7),(75)\\
091029&2.752&0.57&...&14.26&43.09&849.5$\pm$35.4&61.4$\pm$17.5&-1.46$\pm$0.27&...&(25),(59),(-),(76)\\
091127&0.49&\textbf{0.43$\pm$0.10}&0.2&7.93&8820.00&152.6$\pm$7.5&21.3$\pm$3&-1.95$\pm$0.1&...&(77),(77),(77),(78)\\
100219A&4.6667&...&...&0.94&35.0&359$\pm$64.3&140$\pm$0&-1.34$\pm$0&...&(79),(-),(-),(79)\\
100316D&0.059&...&...&40.26&4960.00&...&...&...&...&(80),(-),(-),(-)\\
100418A&0.6235&...&...&1.01&1371.57&...&...&...&...&(Swift),(-),(-),(-)\\
100728B&...&...&...&0.16&5.64&...&...&1.55$\pm$0.14&...&(-),(-),(-),(81)\\
100901A&1.408&...&...&0.64&543.01&245.50&...&1.52$\pm$0.21&...&(82),(-),(-),(82)\\
100906A&1.727&...&...&0.05&10.94&...&...&...&...&(Swift),(-),(-),(-)\\
101024A&...&...&...&0.22&160.44&...&56.25$\pm$5.54&-1.4$\pm$0.8&...&(-),(-),(-),(83)\\
101225A&0.33&...&...&5.72&3499&...&...&-1.8$\pm$0.32&...&(84),(-),(-),(84)\\
110205A&...&...&...&14.34&384.19&...&230$\pm$65&-0.59$\pm$0.06&...&(-),(-),(-),(85)\\
110213A&1.46&...&...&0.10&183.37&720.0$\pm$10.0&98.4$\pm$6.9&-0.44$\pm$0.05&...&(85),(-),(-),(85)\\
110918A&...&...&...&122.43&1410&...&...&...&...&(-),(-),(-),(-)
\enddata
\tablerefs{
(1) Liang \& Zhang (2006);
(2) Firmani  et al.(2006);
(3) Amati et al.(2002);
(4) Svensson et al.(2010);
(5) Mannucci et al.(2011);
(6) Panaitescu(2005);
(7) Kann et al.(2010);
(8) Sagar  et al.(2000);
(9)  Sokolov et al.(2001);
(10) Ghirlanda  etal.(2008);
(11) Guidorzi  et al.(2011);
(12) Ghirlanda  et al.(2004);
(13) Sakamoto  et al.(2004);
(14) Kann et al.(2006);
(15) Butler  et al.(2005);
(16) Pugliese  et al.(2005);
(17) Nava et al.(2008);
(18) Gotz, D. et al (2011)
(19) de Ugarte Postigo et al.(2011);
(20) Butler  et al.(2007);
(21) Zafar et al.(2011);
(22) Ukwatta  et al.(2010);
(23) de Ugarte Postigo et al.(2007);
(24) Xin et al.(2010);
(25) Robertson, Brant et al. (2011);
(26) Grupe et al.(2006);
(27) Antonelli et al.(2006);
(28) de Pasquale et al.(2007);
(29) Kann et al.(2007);
(30) Berger et al.(2007);
(31) Price et al.(2006);
(32) Urata et al.(2007);
(33) Castro-Tirado  (2006);
(34) Yost et al.(2007);
(35) Soderberg et al.(2006)
(36) Perley et al.(2009);
(37) Fynbo et al.(2009);
(38) Klotz et al.(2006);
(39) Donaghy  et al.(2006);
(40) Misra et al.(2007);
(41) Prochaska et al.(2007);
 (42) Ferrero et al.(2009)
 (43)Jakobsson  et al.(2006);
 (44) Covino et al.(2010);
 (45) Schady et al.(2011);
 (46) Schady et al.(2008);
 (47) Golenetskii  et al(2006);
 (48) Perley et al.(2008);
 (49) Troja  et al.(2007);
 (50) De Cia  et al.(2011);
 (51) Bellm  et al.(2008);
 (52) Jaunsen et al.(2008);
 (53) Guidorzi  et al.(2007);
 (54) Krimm et al.(2009);
 (55) Butler  et al.(2010);
 (56) Piranomonte et al.(2008)
 (57) McGlynn et al.(2008);
 (58) Kr{\"u}hler et al.(2009);
 (59) Greiner et al.(2011);
 (60) Golenetskii  et al. (2007);
 (61) Krimm et al. (2007);
 (62) Greiner  et al (2009)
 (63) Uehara et al.(2010);
 (64) Starling et al.(2009);
 (65) Guetta  et al.(2011);
 (66) Yuan et al.(2010);
 (67) Nardini et al.(2011);
 (68) McBreen (2010);
 (69) Thone et al.(2011);
 (70) Nicuesa et al.(2011);
 (71) Giuliani et al.(2010);
 (72) Asano, Katsuaki et al.(2010) ;
 (73) Cano et al.(2011);
 (74) Pandey, S. B. et al.(2010)
 (75) Zhang et al.(2011);
 (76) Barthelmy  et al. (2009);
 (77) Vergani et al.(2011);
 (78) Golenetskii  et al. (2009);
 (79) Mao et al.(2011);
 (80) Bufano et al.(2011);
 (81) Barthelmy  et al. (2010);
 (82) Sakamoto et al. (2010);
 (83) McBreen  et al. (2010b);
 (84) Palmer et al. (2010);
 (85) Cucchiara  et al.(2011);
(Swift) z in the web http://www.swift.ac.uk/xrt\_spectra/
}
\end{deluxetable}

\clearpage
\begin{deluxetable}{lccccccccccccc}
\tablewidth{620pt} 
\rotate
\tabletypesize{\footnotesize}
\tablecaption{Properties of the Flares in Our Sample}
\tablenum{2}
\tablehead{\colhead{GRB(Band)}
&\colhead{$F_m$\tablenotemark{a}}
&\colhead{$\alpha_1$}
&\colhead{$\alpha_2$}
&\colhead{$t_{\rm p}$\tablenotemark{b}}
&\colhead{$L_{\rm p,R}$\tablenotemark{c}}
&\colhead{$E_{\rm R,iso}$\tablenotemark{d}}
&\colhead{$w$\tablenotemark{b}}
&\colhead{$t_{\rm r}$\tablenotemark{b}}
&\colhead{$t_{\rm d}$\tablenotemark{b}}
&\colhead{$R_{\rm rd}$}
&\colhead{$R_{\rm rp}$}
}
\startdata
970508(R)&130.0$\pm$9.7&-5.00$\pm$1.10&2.50$\pm$0.11&150.00$\pm$10.98&0.47$\pm$0.03&8.38&134.73&45.77&88.96&0.51&0.28\\
000301C(R)&90.8$\pm$2.3&-6.00&2.11&138.06$\pm$0.80&1.95$\pm$0.05&20.18&85.00&22.37&62.63&0.36&0.16\\
000301C(R)&43.1$\pm$2.4&-3.20&11.56&336.76$\pm$2.12&0.93$\pm$0.05&37.93&101.27&69.04&32.23&2.14&0.21\\
021004(R)&125.4$\pm$14.4&-3.00&3.00&68.90$\pm$2.50&2.54$\pm$0.29&30.67&43.07&18.25&24.82&0.74&0.26\\
050401(R)&61.1$\pm$14.0&-5.48&8.94&24.58$\pm$0.82&1.90$\pm$0.44&5.10&6.42&3.50&2.92&1.20&0.14\\
051109A(R)&623.8$\pm$75.9&-2.10&2.24$\pm$0.56&7.61$\pm$0.59&18.52$\pm$2.25&20.54&6.60&2.67&3.93&0.68&0.35\\
060121(R)&7.0$\pm$0.8&-5.00$\pm$2.78&4.00$\pm$0.21&34.00$\pm$0.46&0.06$\pm$0.007&0.65&14.53&5.99&8.53&0.70&0.17\\
060206(R)&2246.0$\pm$39.8&-9.74&2.21$\pm$0.11&3.45$\pm$0.01&229.56$\pm$4.07&25.37&1.75&0.35&1.40&0.25&0.10\\
060206(R)&593.7$\pm$34.5&-5.70&2.50&6.65$\pm$0.05&60.68$\pm$3.52&20.54&3.45&1.02&2.43&0.42&0.15\\
060210(R)&888.9$\pm$245.3&-3.20&5.72&0.66$\pm$0.03&47.54$\pm$13.12&4.75&0.28&0.15&0.13&1.17&0.23\\
060607A(H)&856.2&-6.08$\pm$2.67&2.41$\pm$0.28&4.26$\pm$0.39&47.66&22.81&2.49&0.73&1.76&0.42&0.17\\
060607A(H)&1061.7&-3.00&10.89$\pm$2.77&2.17$\pm$0.04&59.10&24.85&0.72&0.48&0.24&1.96&0.23\\
060926(V)&1115.3$\pm$215.9&-2.55$\pm$1.08&3.49$\pm$0.77&0.63$\pm$0.05&77.80$\pm$15.06&3.69&0.36&0.17&0.19&0.93&0.28\\
060926(V)&2629.7$\pm$481.9&-3.47$\pm$0.87&2.00&0.09$\pm$0.01&183.45$\pm$33.62&1.06&0.07&0.02&0.05&0.48&0.24\\
070311(R)&260.8$\pm$42.4&-5.10$\pm$1.19&5.31$\pm$1.20&206.89$\pm$9.40&...&...&90.50&40.72&49.78&0.82&0.20\\
071010A(R)&157.9$\pm$5.4&-2.08&2.39&80.07$\pm$1.40&0.63$\pm$0.02&10.84&66.30&27.80&38.50&0.72&0.35\\
071025(J)&1650.9$\pm$119.1&-5.67$\pm$1.67&3.00&1.71$\pm$0.04&...&...&0.86&0.29&0.56&0.52&0.17\\
071031(R)&16.8$\pm$0.1&-8.27&3.00&16.57$\pm$0.02&0.63$\pm$0.01&0.58&7.35&2.17&5.19&0.42&0.13\\
071031(R)&49.8$\pm$0.3&-3.51&3.40&6.36$\pm$0.01&1.86$\pm$0.01&1.30&3.45&1.47&1.98&0.74&0.23\\
080506(R)&405.6$\pm$98.4&-6.84$\pm$7.49&2.38$\pm$2.49&1.19$\pm$0.13&...&...&0.68&0.19&0.49&0.39&0.15\\
090313(R)&1452.5$\pm$190.5&-6.43$\pm$1.24&2.00&19.83$\pm$0.69&97.14$\pm$12.74&107.35&13.39&3.43&9.96&0.34&0.17\\
090618(R)&33652.9$\pm$1055.6&-5.00&2.23$\pm$0.07&0.14$\pm$0.00&31.34$\pm$0.98&0.68&0.09&0.03&0.06&0.44&0.19\\
090726(R)&873.1&-2.21$\pm$0.69&5.00&0.52&38.08&2.49&0.28&0.15&0.12&1.27&0.31\\
100728B(R)&64.5$\pm$16.1&-8.00&5.60$\pm$2.29&3.00$\pm$0.39&...&...&1.71&0.70&1.01&0.69&0.22\\
\enddata
\tablenotetext{a}{In units of $10^{-15}$ erg cm$^{-2}$ s$^{-1}$.}
\tablenotetext{b}{In units of kilo seconds.}
\tablenotetext{c}{In units of $10^{45}$ erg.}
\tablenotetext{d}{In units of $10^{48}$ erg.}
\end{deluxetable}
\clearpage
\begin{deluxetable}{ccccccccccccccc}
\rotate
\tablewidth{560pt}
\tabletypesize{\footnotesize}
\tablecaption{Properties of the Shallow Decay Segments in Our Sample}
\tablenum{3}
\tablehead{
\colhead{GRB}
&\colhead{$F_{\rm m}$\tablenotemark{a}}
&\colhead{$\alpha_{1}$}
&\colhead{$\alpha_2$}
&\colhead{$t_{\rm p}$\tablenotemark{b}}
&\colhead{$L_{\rm p,R}$\tablenotemark{c}}
&\colhead{$E_{\rm R,iso}$\tablenotemark{d}}
&\colhead{$E_{\rm in}$\tablenotemark{d}}
&\colhead{$q$}
&\colhead{$\alpha_{1}(\beta_{0})$}
}
\startdata
970508(R)&23.8$\pm$3.9&0.10$\pm$0.04&3.00$\pm$0.09&1000.00$\pm$58.24&85$\pm$14&70.0&114.3&-0.006&1.665\\
000301C(R)&19.0$\pm$1.1&0.20$\pm$0&3.42&602.81$\pm$16.55&408$\pm$24&172.9&365.8&0.370&1.05\\
010222(R)&323.7$\pm$39.7&0.47$\pm$0.06&1.25$\pm$0.07&32.07$\pm$4.89&4754$\pm$583&157.6&313.2&0.261&1.605\\
020813(R)&348.9$\pm$177.3&0.50$\pm$0.35&1.37$\pm$0.11&29.91$\pm$14.82&2230$\pm$1130&90.5&141.8&0.453&1.275\\
021004(R)&1841.4$\pm$88.9&0.28$\pm$0.02&1.50$\pm$0.05&9.91$\pm$0.56&37249$\pm$1798&564.0&589.7&0.742&0.585\\
030226(R)&54.0$\pm$4.7&0.70$\pm$0.04&2.92&102.35$\pm$5.17&1097$\pm$95&142.4&343.0&0.741&1.05\\
030328(R)&245.4$\pm$66.1&0.41$\pm$0.13&1.28$\pm$0.06&16.11$\pm$4.74&1999$\pm$539&65.7&60.9&0.893&0.54\\
030429(R)&18.4$\pm$1.6&0.86$\pm$0.03&3.53$\pm$0.00&218.47$\pm$8.85&606$\pm$53&111.8&582.7&0.805&1.125\\
030723(R)&41.4$\pm$1.6&0&2.09&103.16$\pm$2.52&21$\pm$1&3.5&2.6&0.256&0.99\\
040924(R)&833.4$\pm$127.0&0.05$\pm$0.32&1.29$\pm$0.03&1.72$\pm$0.32&2484$\pm$379&6.9&5.4&0.257&1.05\\
041006(R)&325.4$\pm$16.9&0.42&1.27$\pm$0.01&11.35$\pm$0.53&586$\pm$30&13.6&12.6&0.683&0.825\\
050319(R)&55.0$\pm$4.8&0.40$\pm$0.03&1.40$\pm$0.08&121.11$\pm$15.25&3516$\pm$307&381.3&798.3&0.479&1.11\\
050319(R)&680.2$\pm$54.7&0.32$\pm$0.05&2.07&0.54$\pm$0.04&43476$\pm$3495&16.7&35.9&0.424&1.11\\
050408(R)&19.8$\pm$0.8&0.52&1.39&40.39$\pm$2.62&98$\pm$4&8.6&8.8&1.091&0.42\\
050416A(R)&27.0$\pm$1.6&0.39&1.32&15.33$\pm$1.11&57$\pm$3&1.2&1.6&0.053&1.95\\
050730(R)&590.0$\pm$83.0&0.33$\pm$0.04&1.73$\pm$0.06&11.36$\pm$1.73&41271$\pm$5806&549.0&795.1&0.641&0.78\\
050801(R)&8175.1$\pm$1158.8&0.10$\pm$0.13&2.41$\pm$0.30&0.29$\pm$0.01&128276$\pm$18183&25.8&45.4&0.064&1.5\\
050801(R)&1647.6$\pm$292.7&0.20$\pm$0.28&1.36$\pm$0.05&1.51$\pm$0.30&25852$\pm$4593&40.5&56.0&0.132&1.5\\
050922C(R)&1912.5$\pm$385.0&0.18$\pm$0.07&1.58$\pm$0.04&4.08$\pm$0.91&39381$\pm$7928&260.4&230.6&0.535&0.765\\
051021(R)&356.5$\pm$329.6&0.26$\pm$0.68&1.50&4.93$\pm$3.20&...&...&...&0.371&1.125\\
051109A(R)&4945.0$\pm$2565.0&0.40$\pm$0.08&1.04$\pm$0.04&0.44$\pm$0.32&146835$\pm$76164&89.3&107.0&0.518&1.05\\
051111(R)&2560.0$\pm$469.0&0.81$\pm$0.01&2.1$\pm$0.7&3.06$\pm$0.66&25070$\pm$4590&96.1&244.6&0.761&1.14\\
060210(R)&623.8$\pm$127.8&0.07$\pm$0.09&1.21$\pm$0.05&0.96$\pm$0.20&33362$\pm$6834&46.2&40.4&0.591&0.555\\
060526(R)&334.9$\pm$19.7&0.58$\pm$0.03&1.82$\pm$0.03&24.95$\pm$1.18&15070$\pm$886&441.7&903.8&0.855&0.765\\
060605(R)&299.2$\pm$51.4&0.18$\pm$0.12&3.74$\pm$0.56&23.28$\pm$2.20&44198$\pm$7597&299.1&1486.9&0.077&1.59\\
060729(U)&4441.9$\pm$1119.1&0.13$\pm$0.17&2.65$\pm$1.66&4.13$\pm$0.97&4669$\pm$1176&14.3&26.4&0.253&1.17\\
060927(V)&21362.0$\pm$11727&0.38$\pm$0.63&2.63$\pm$0.97&0.05&$(6.8\pm3.6)\times10^{6}$&99.4&338.6&0.287&1.29\\
061126(R)&179.4$\pm$36.4&0.40$\pm$0.12&1.29$\pm$0.05&14.81$\pm$3.03&1299$\pm$263&25.2&35.9&0.308&1.425\\
070110(U)&228.1$\pm$162.3&0.16$\pm$0.27&0.98$\pm$0.78&14.29$\pm$19.66&2302$\pm$1638&40.6&46.5&0.477&1.5\\
070411(R)&32.2$\pm$1.1&0.49$\pm$0.01&1.90&108.99$\pm$2.26&1702$\pm$57&180.5&398.5&0.541&1.125\\
070518(R)&9.8$\pm$0.7&0.65&1.90&30.00&64$\pm$4&3.0&5.2&0.607&1.2\\
070707(R)&4.3$\pm$0.9&0.44$\pm$0.14&3.20$\pm$0.24&120.01$\pm$13.81&...&...&...&0.502&1.125\\
071003(R)&34500.0$\pm$3520.0&0.89$\pm$0.04&1.86$\pm$0.02&0.18$\pm$0.01&585510&59740&88.4&0.392&1.875\\
071010A(R)&337.3$\pm$59.5&0.29$\pm$0.17&1.55&12.53$\pm$3.10&1347$\pm$238&25.6&27.6&0.459&1.02\\
080413A(R)&31509.9$\pm$4714.6&0.56$\pm$0.08&5.02$\pm$1.59&0.07$\pm$0.01&977554$\pm$146264&54.1&97.6&0.670&1.005\\
080413B(R)&88.4$\pm$5.6&0.04$\pm$0.03&2.04$\pm$0.09&159.02$\pm$9.32&335$\pm$21&91.0&67.6&0.704&0.375\\
081029(R)&536.4$\pm$39.7&0.71$\pm$0.06&10.38$\pm$0.00&2.73$\pm$0.10&47470$\pm$3510&69.0&107.3&1.710&1.5\\
090426(R)&1550.0$\pm$132.0&0.55$\pm$0.06&1.75$\pm$0.00&0.36$\pm$0.03&49920$\pm$4250&16.8&33.6&0.571&1.14\\
090426(R)&69.4$\pm$7.83&0.27$\pm$0.06&2.29$\pm$0.00&30.48$\pm$1.69&2240$\pm$252&50.2&108.4&0.367&1.14\\
090618(R)&594.0$\pm$25.0&0.62$\pm$0.00&1.63$\pm$0.06&31.12$\pm$1.46&553$\pm$23&37.9&44.8&0.899&0.75\\
091127(I)&902.4$\pm$13.6&0.45&1.48&27.47$\pm$0.57&595$\pm$9&41.1&32.4&1.161&0.645\\
101225A(R)&6.9$\pm$0.5&0.15&1.30&368.78$\pm$33.90&...&...&...&0.295&1.125
\enddata
\tablenotetext{a}{In units of $10^{-15}$erg cm$^{-2}$s$^{-1}$.}
\tablenotetext{b}{In units of ks.}
\tablenotetext{c}{In units of $10^{42}$ erg s$^{-1}$.}
\tablenotetext{d}{In units of $10^{48}$ erg.}
\end{deluxetable}

\setlength{\voffset}{0mm}

\begin{figure*}
\includegraphics[angle=0,scale=0.350,width=0.3\textwidth,height=0.25\textheight]{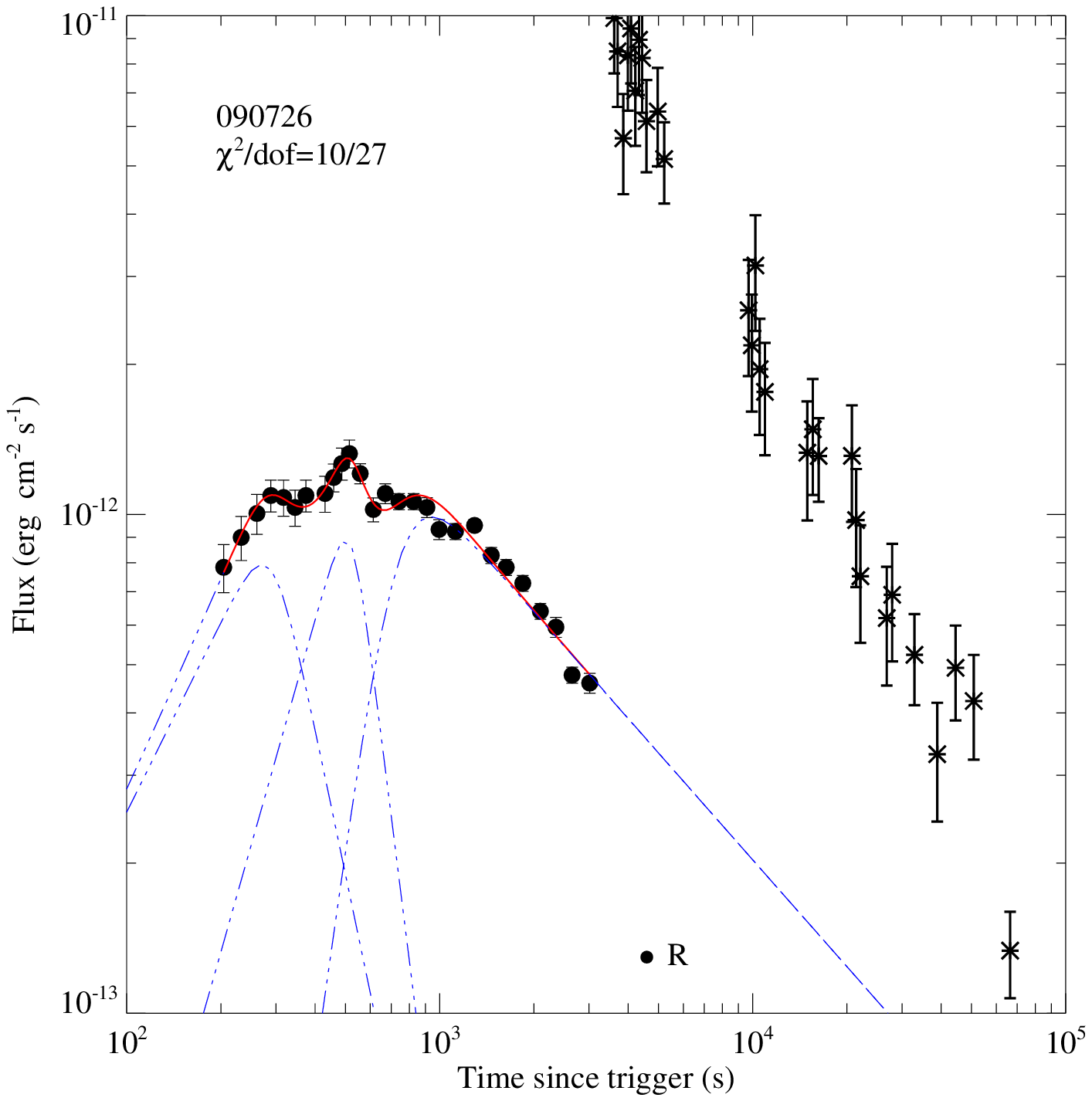}
\includegraphics[angle=0,scale=0.350,width=0.3\textwidth,height=0.25\textheight]{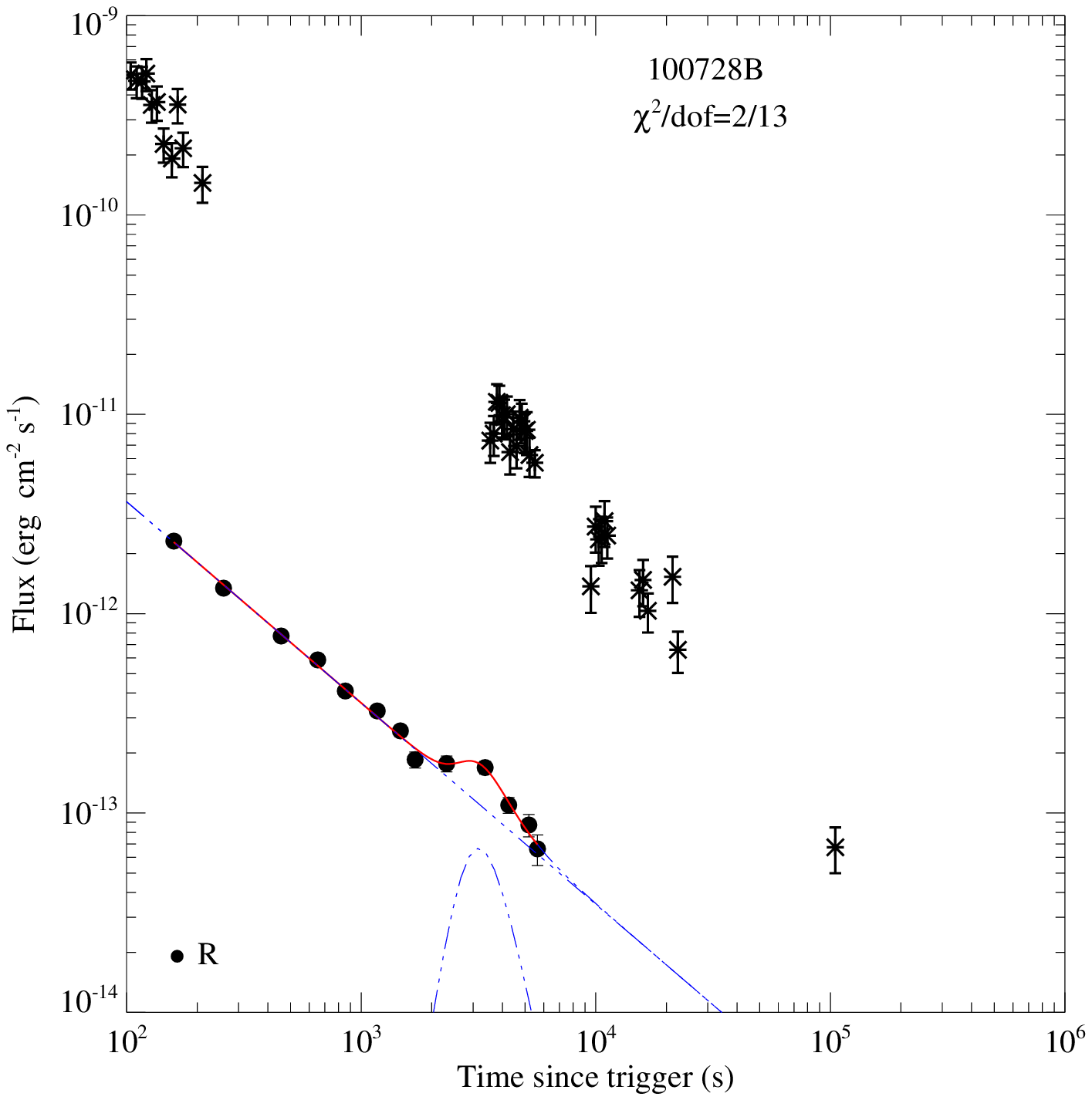}

\center{Fig. \ref{Opt_LC_Flares}--- Continued}
\end{figure*}
\newpage
\begin{figure}
\includegraphics[angle=0,scale=0.350,width=0.3\textwidth,height=0.25\textheight]{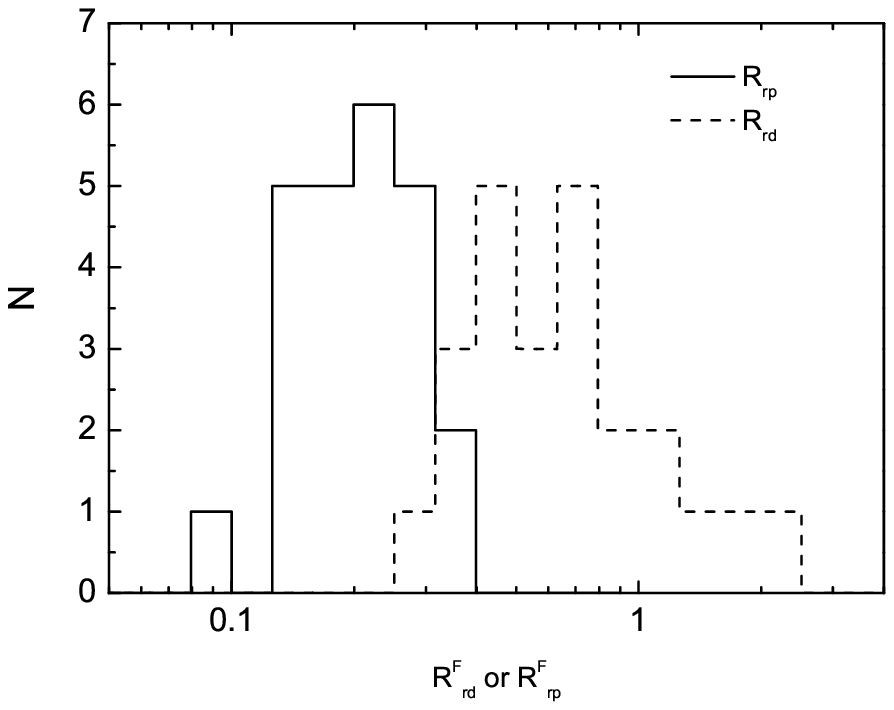}
\includegraphics[angle=0,scale=0.350,width=0.3\textwidth,height=0.25\textheight]{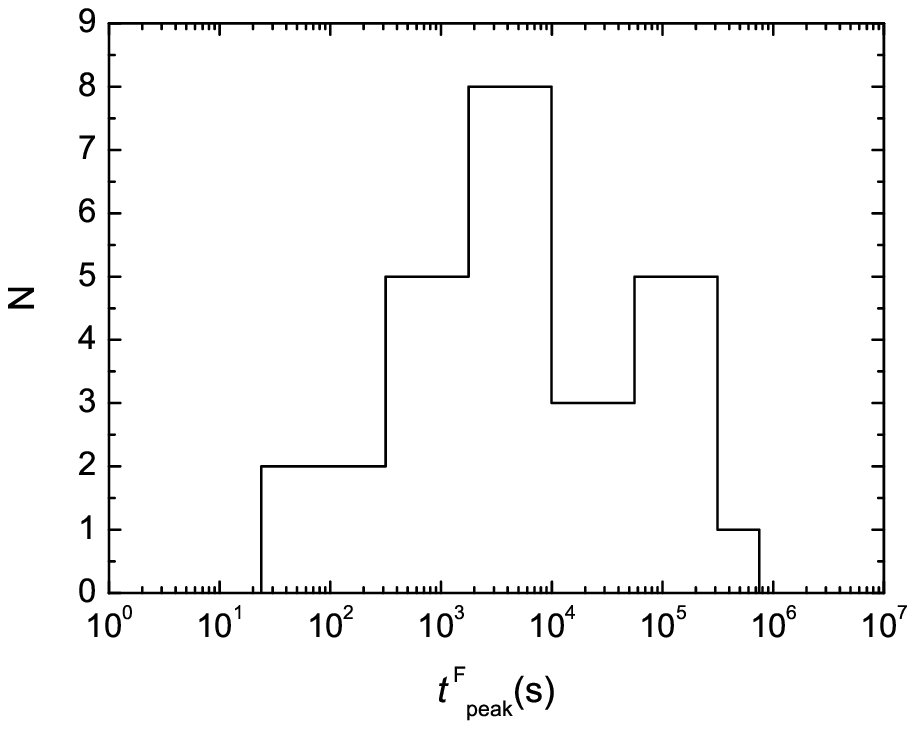}\\
\includegraphics[angle=0,scale=0.350,width=0.3\textwidth,height=0.25\textheight]{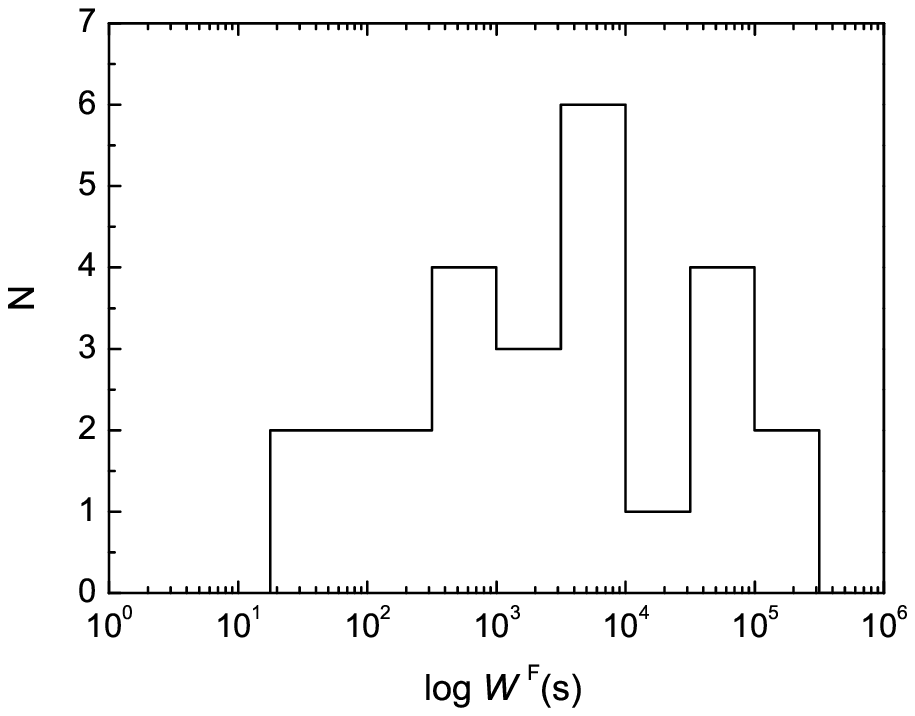}
\includegraphics[angle=0,scale=0.350,width=0.3\textwidth,height=0.25\textheight]{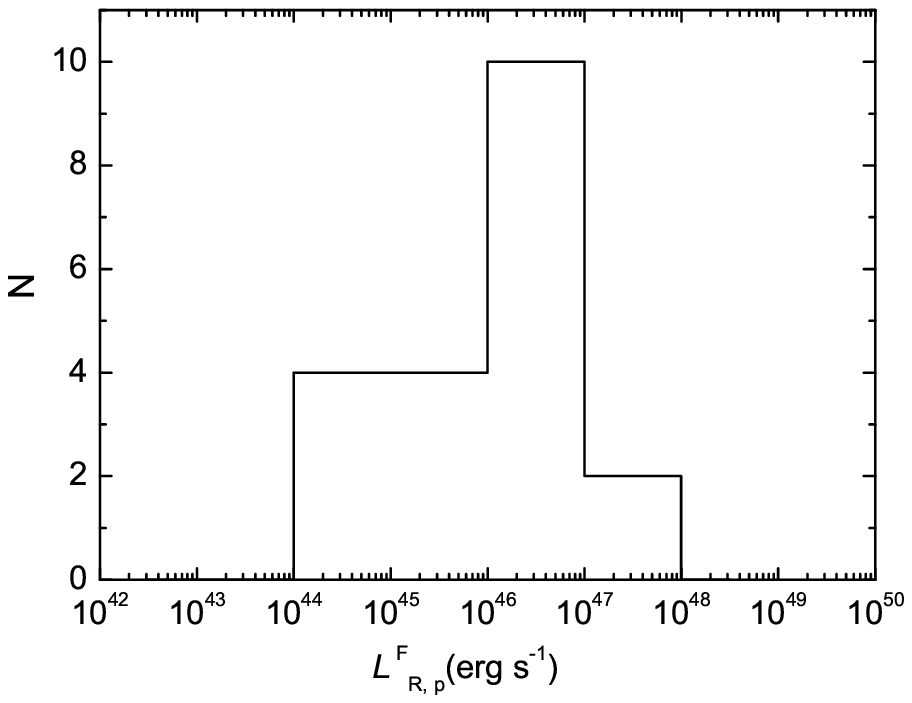}
\caption{Optical flare parameter distributions in our sample.} \label{Flare_Dis}
\end{figure}

\newpage
\begin{figure*}
\includegraphics[angle=0,scale=0.350,width=0.3\textwidth,height=0.25\textheight]{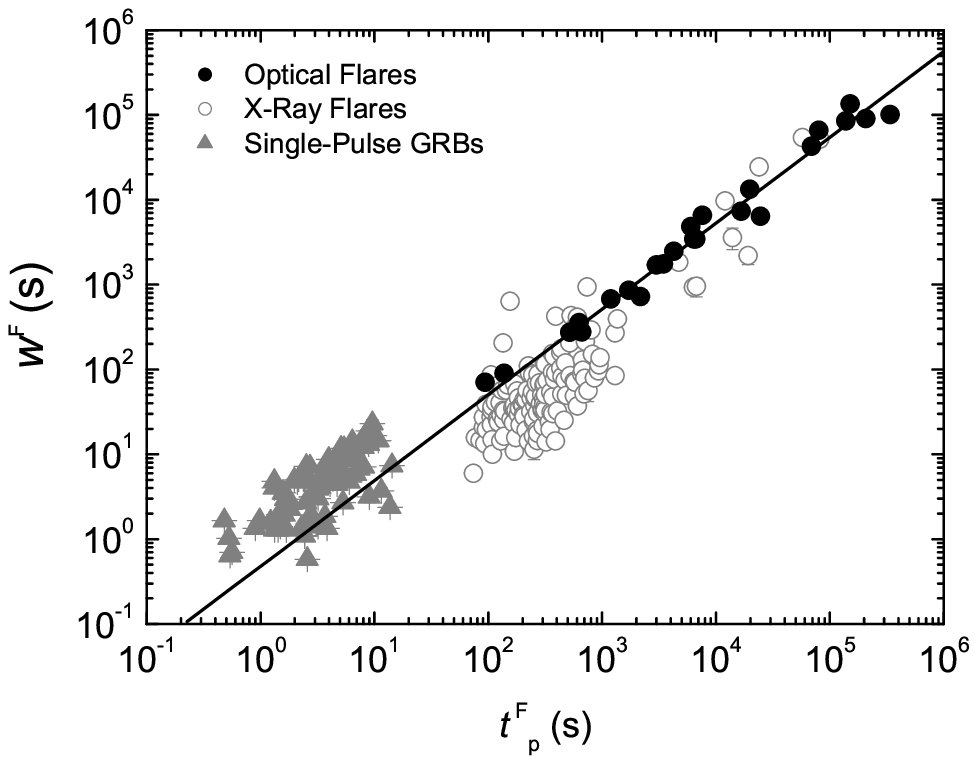}
\includegraphics[angle=0,scale=0.350,width=0.3\textwidth,height=0.25\textheight]{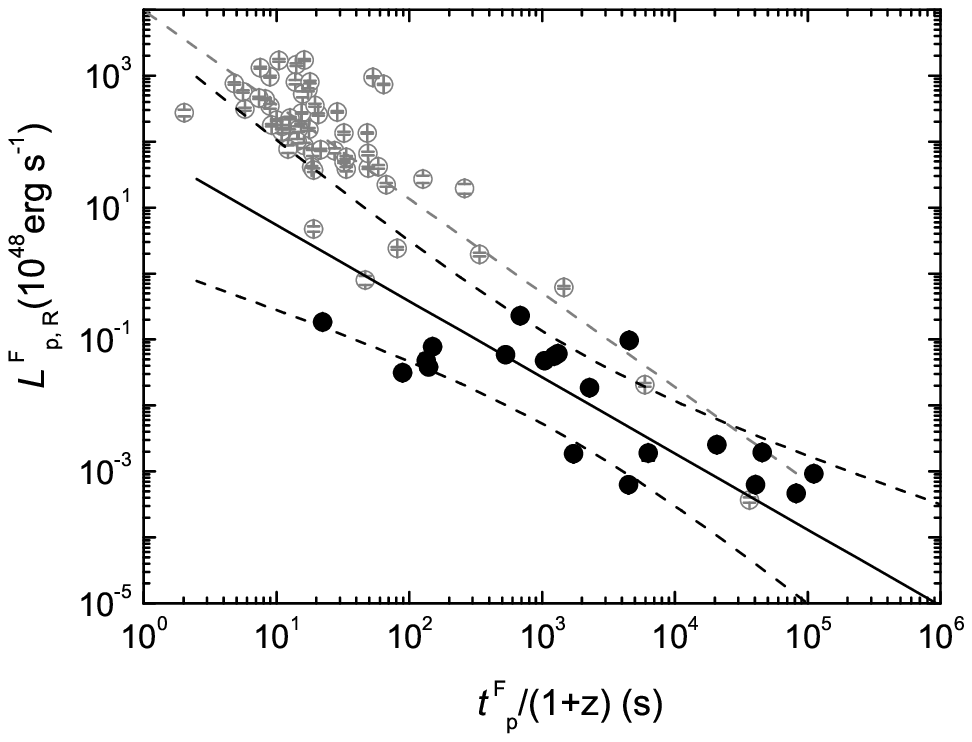}
\includegraphics[angle=0,scale=0.350,width=0.3\textwidth,height=0.25\textheight]{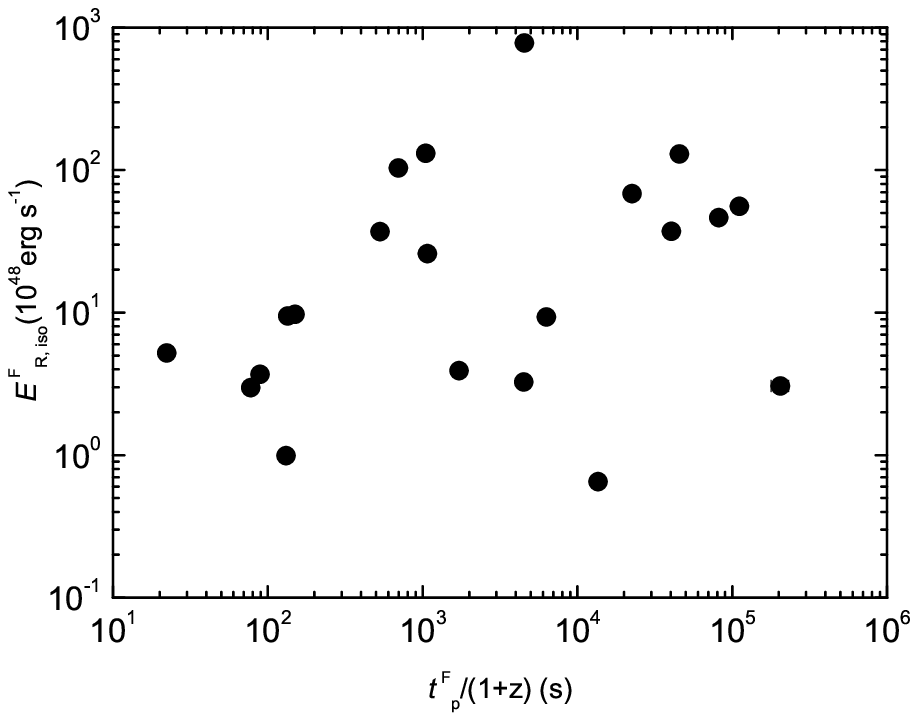}
\caption{Pair correlations between flare parameters in our sample. Lines are the regression lines.} \label{Flare_Corr}
\end{figure*}
\newpage

\begin{figure*}
\includegraphics[angle=0,scale=0.350,width=0.3\textwidth,height=0.25\textheight]{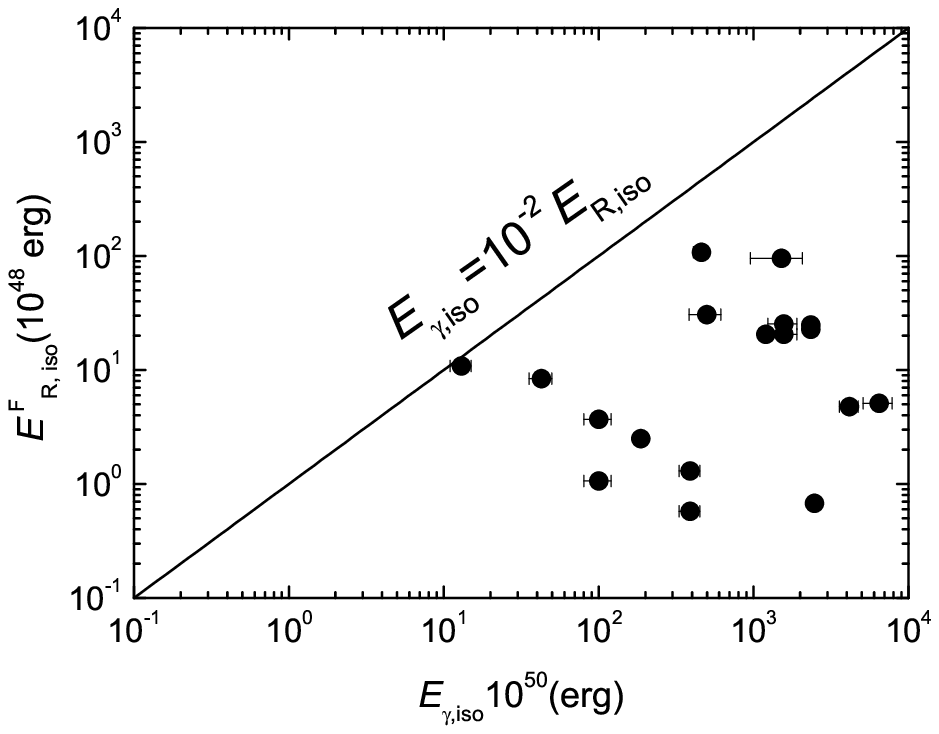}
\includegraphics[angle=0,scale=0.350,width=0.3\textwidth,height=0.25\textheight]{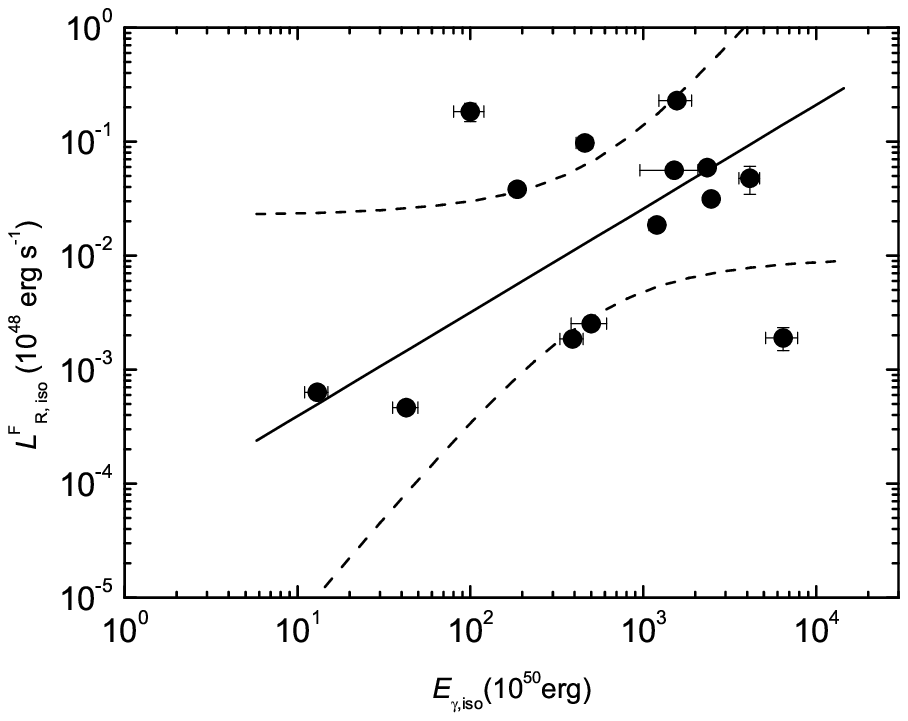}
\caption{$E^{\rm F}_{\rm R, iso}$ and $L^{\rm F}_{\rm R, iso}$ as a function  of $E_{\gamma, iso}$ for the flares in our sample. Lines in the right panel are the regression line and 3 $\sigma$ region.} \label{GRB_Flare_Corr}
\end{figure*}
\newpage

\begin{figure*}
\includegraphics[angle=0,scale=0.350,width=0.3\textwidth,height=0.25\textheight]{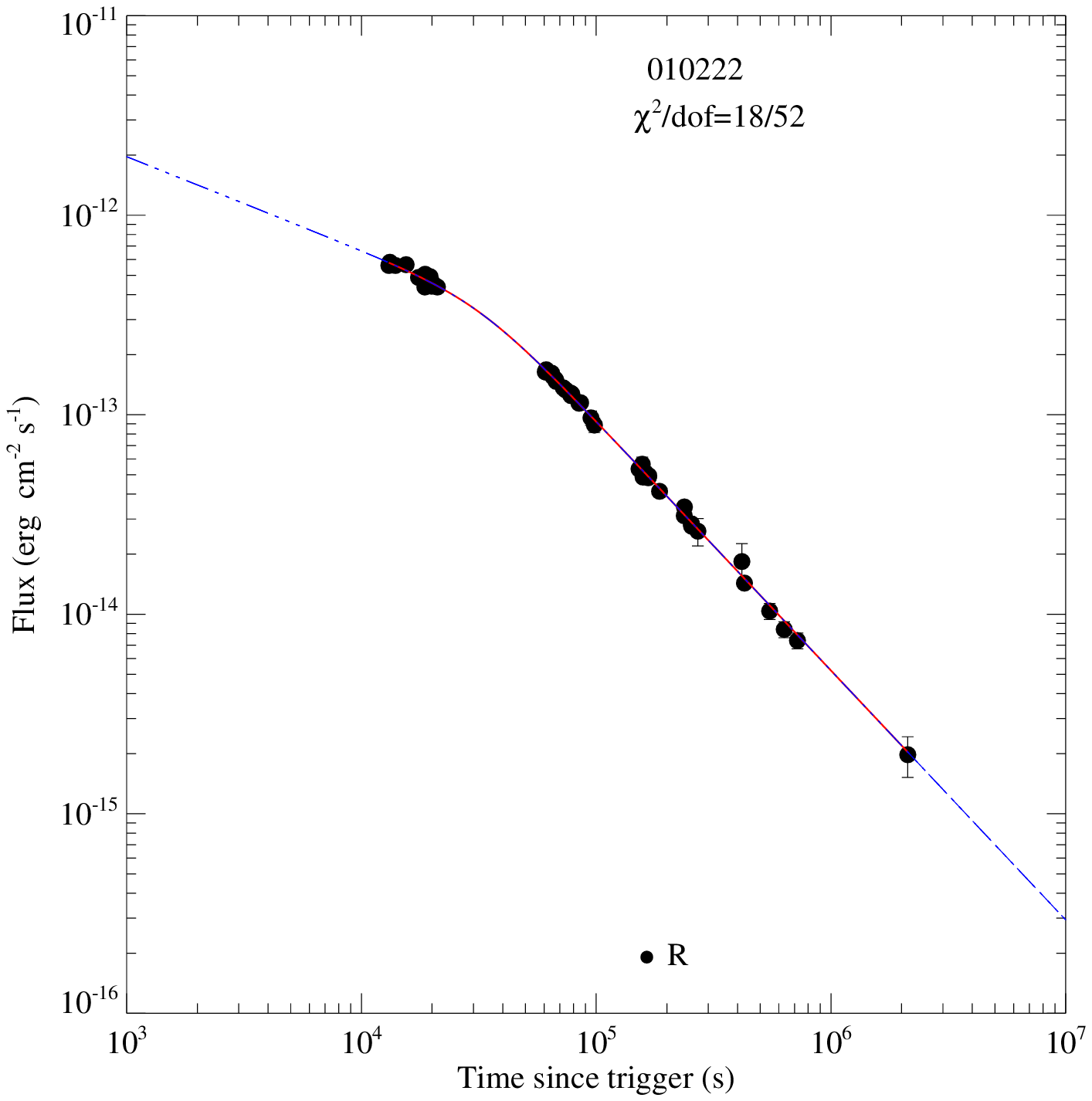}
\includegraphics[angle=0,scale=0.350,width=0.3\textwidth,height=0.25\textheight]{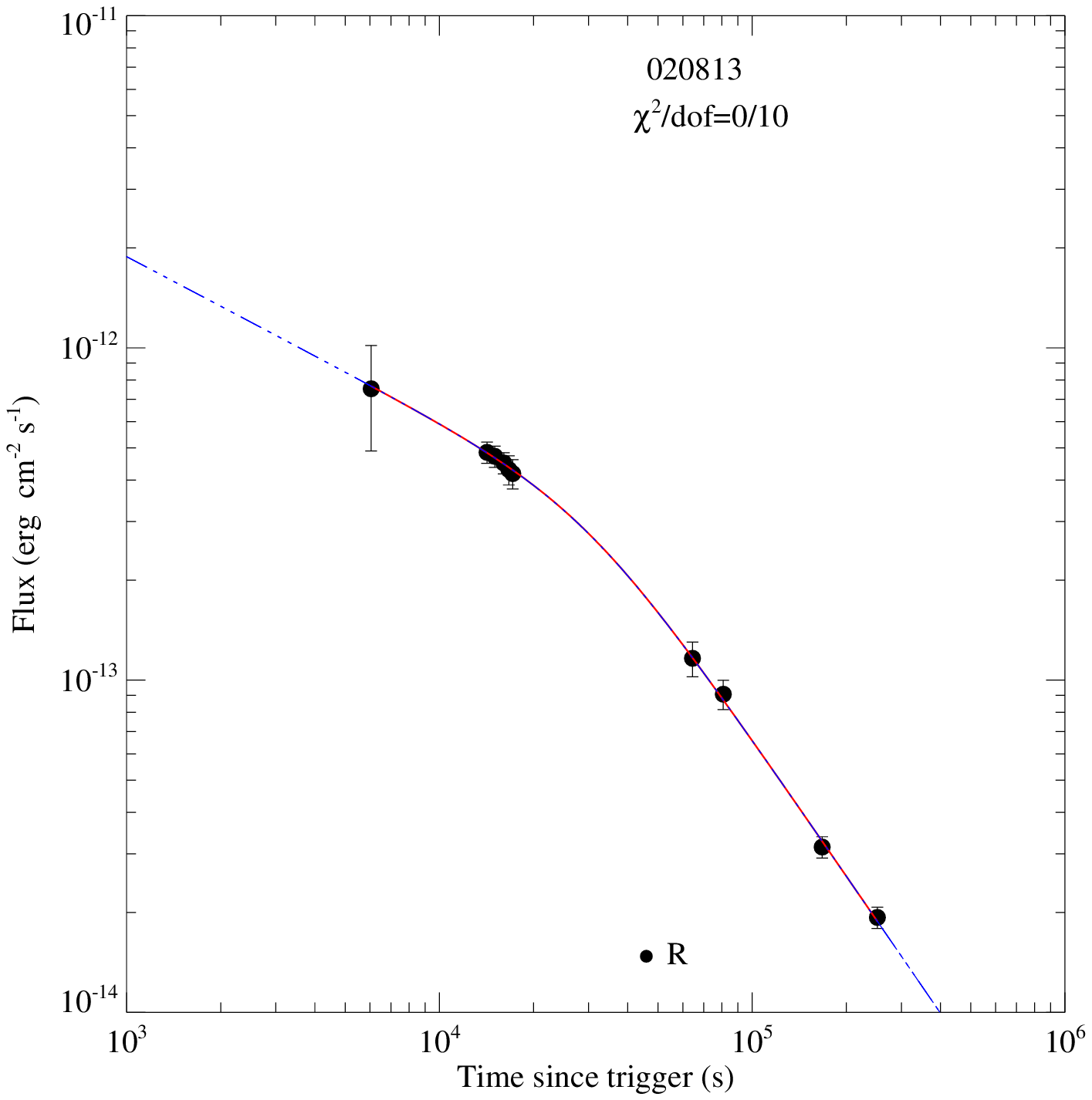}
\includegraphics[angle=0,scale=0.350,width=0.3\textwidth,height=0.25\textheight]{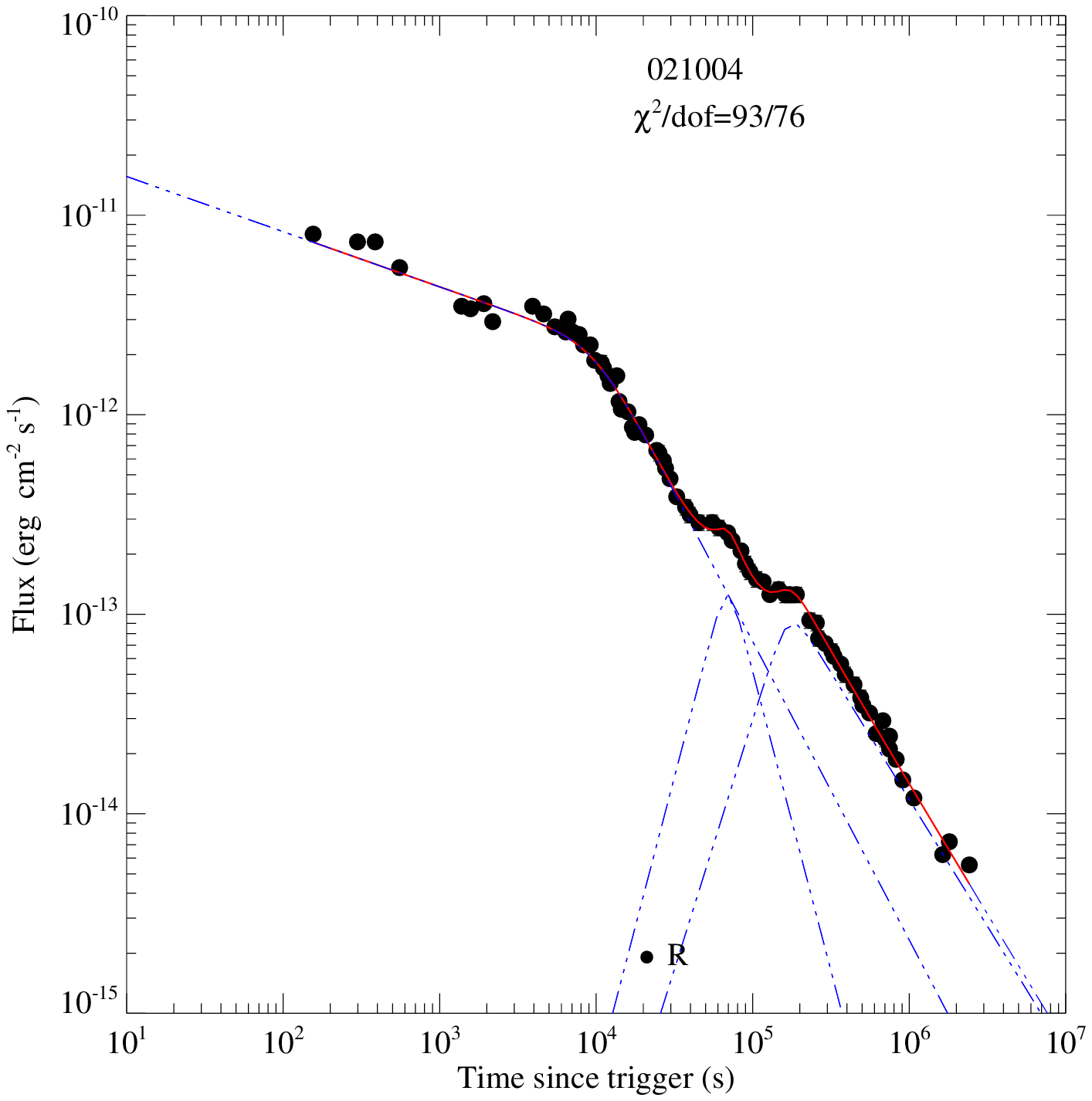}
\includegraphics[angle=0,scale=0.350,width=0.3\textwidth,height=0.25\textheight]{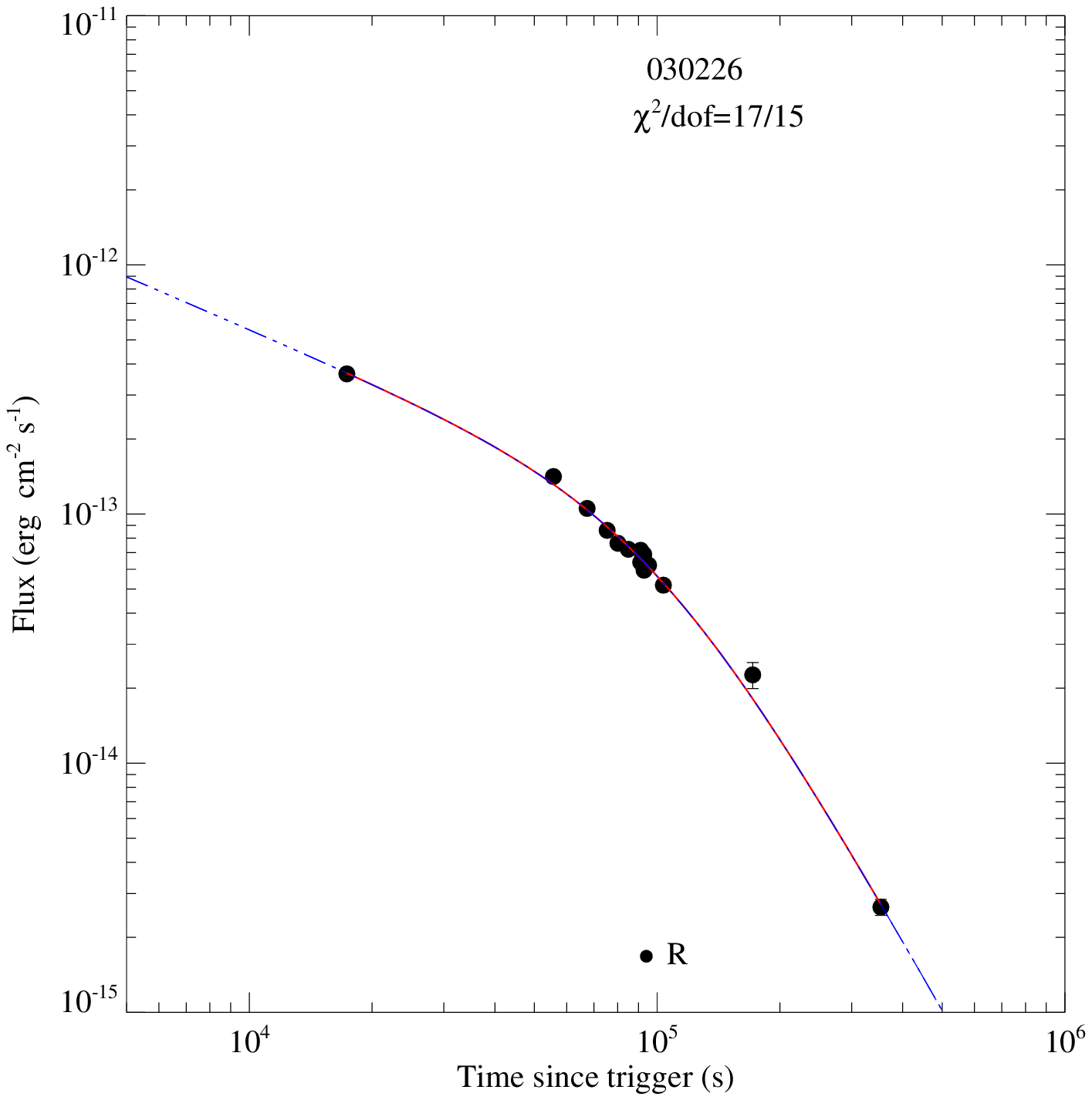}
\includegraphics[angle=0,scale=0.350,width=0.3\textwidth,height=0.25\textheight]{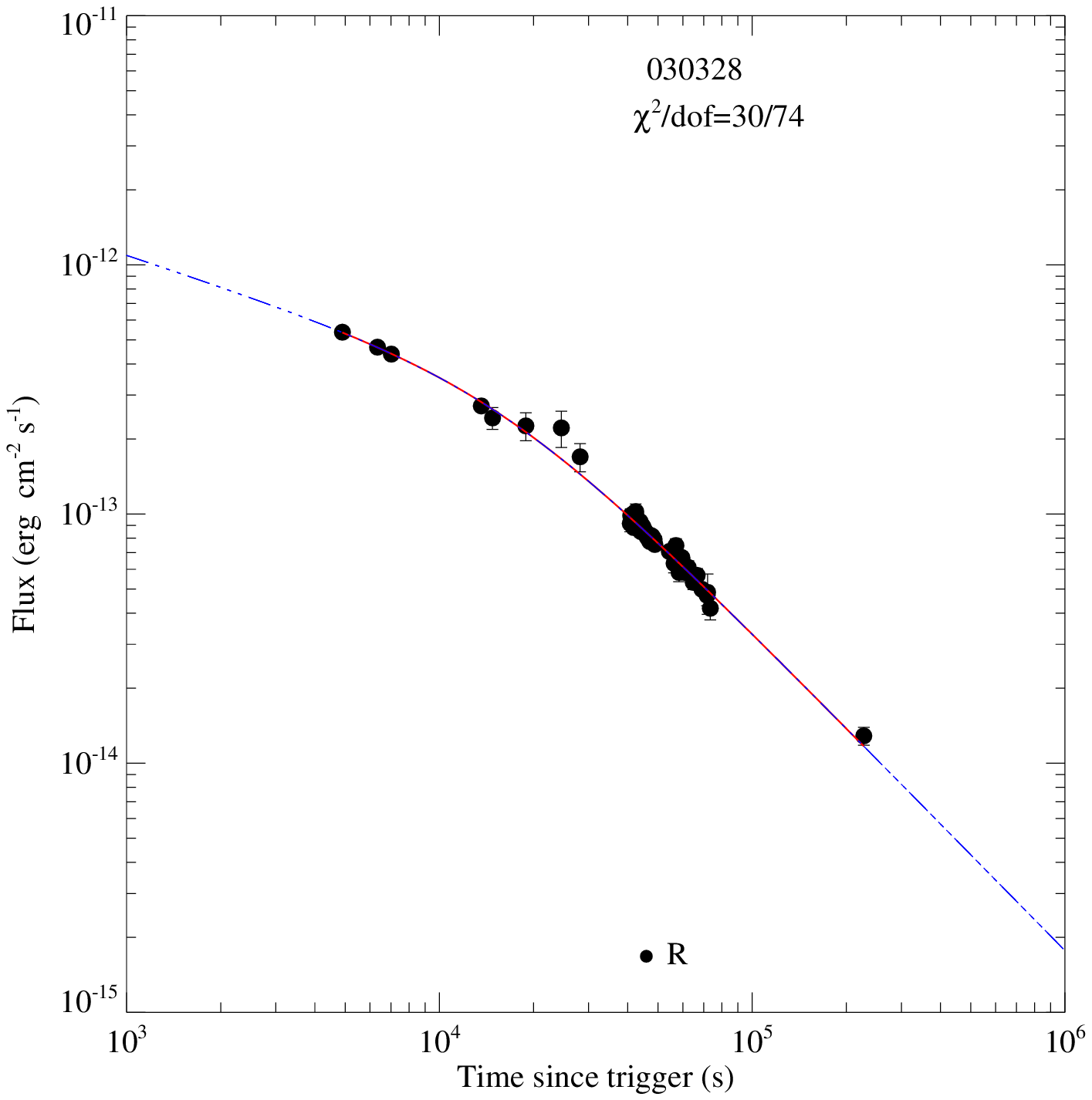}
\includegraphics[angle=0,scale=0.350,width=0.3\textwidth,height=0.25\textheight]{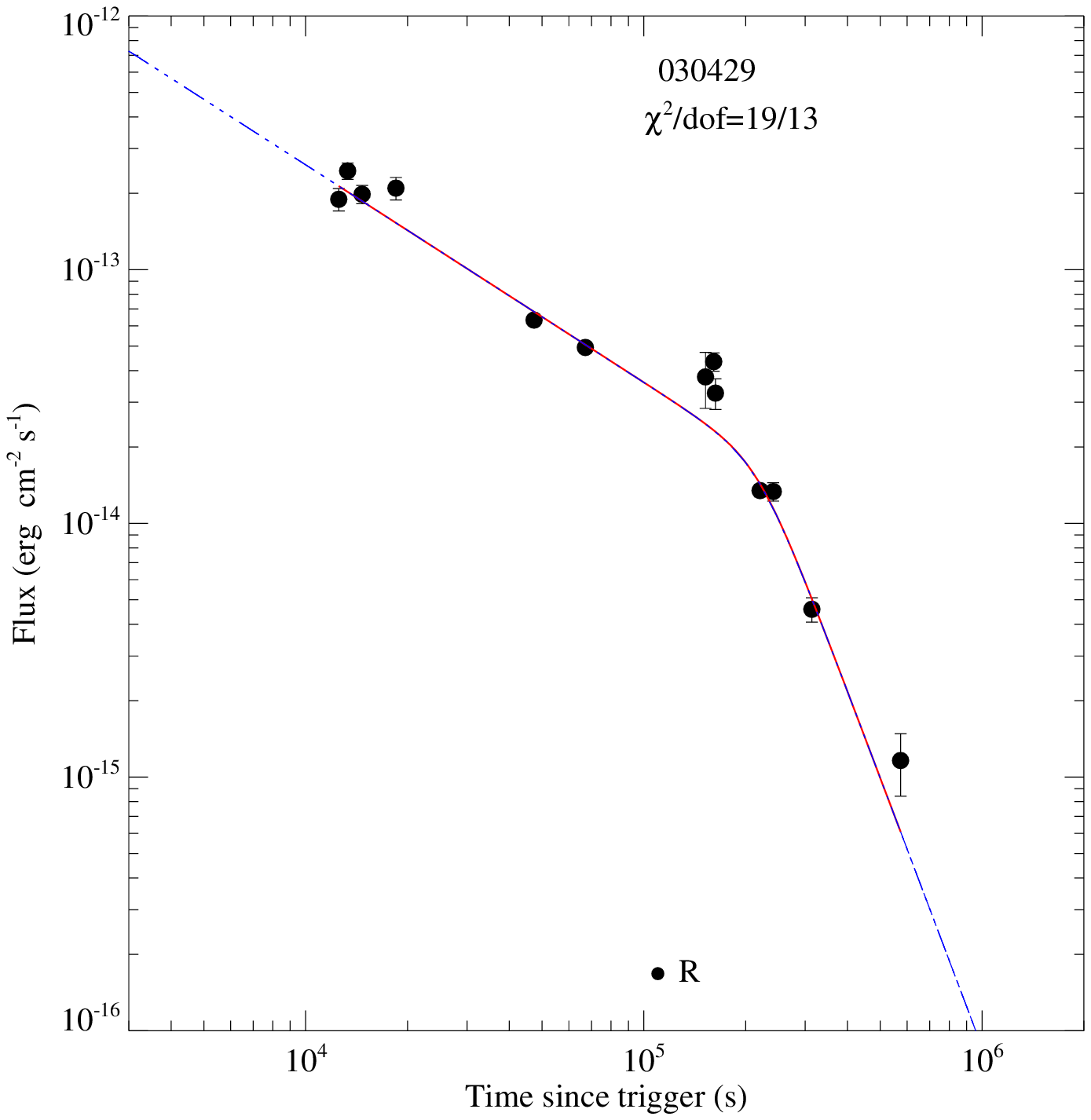}
\includegraphics[angle=0,scale=0.350,width=0.3\textwidth,height=0.25\textheight]{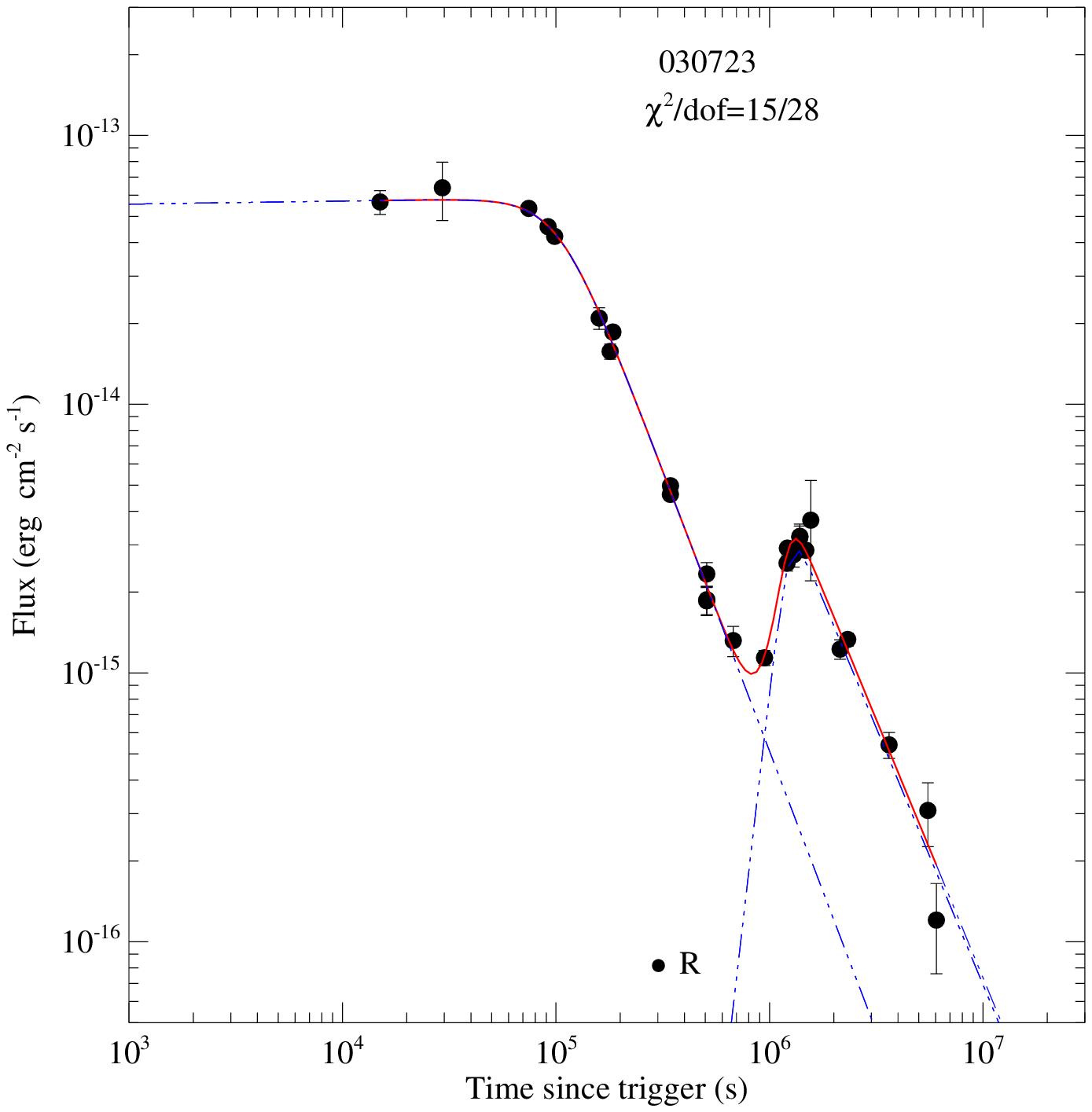}
\includegraphics[angle=0,scale=0.350,width=0.3\textwidth,height=0.25\textheight]{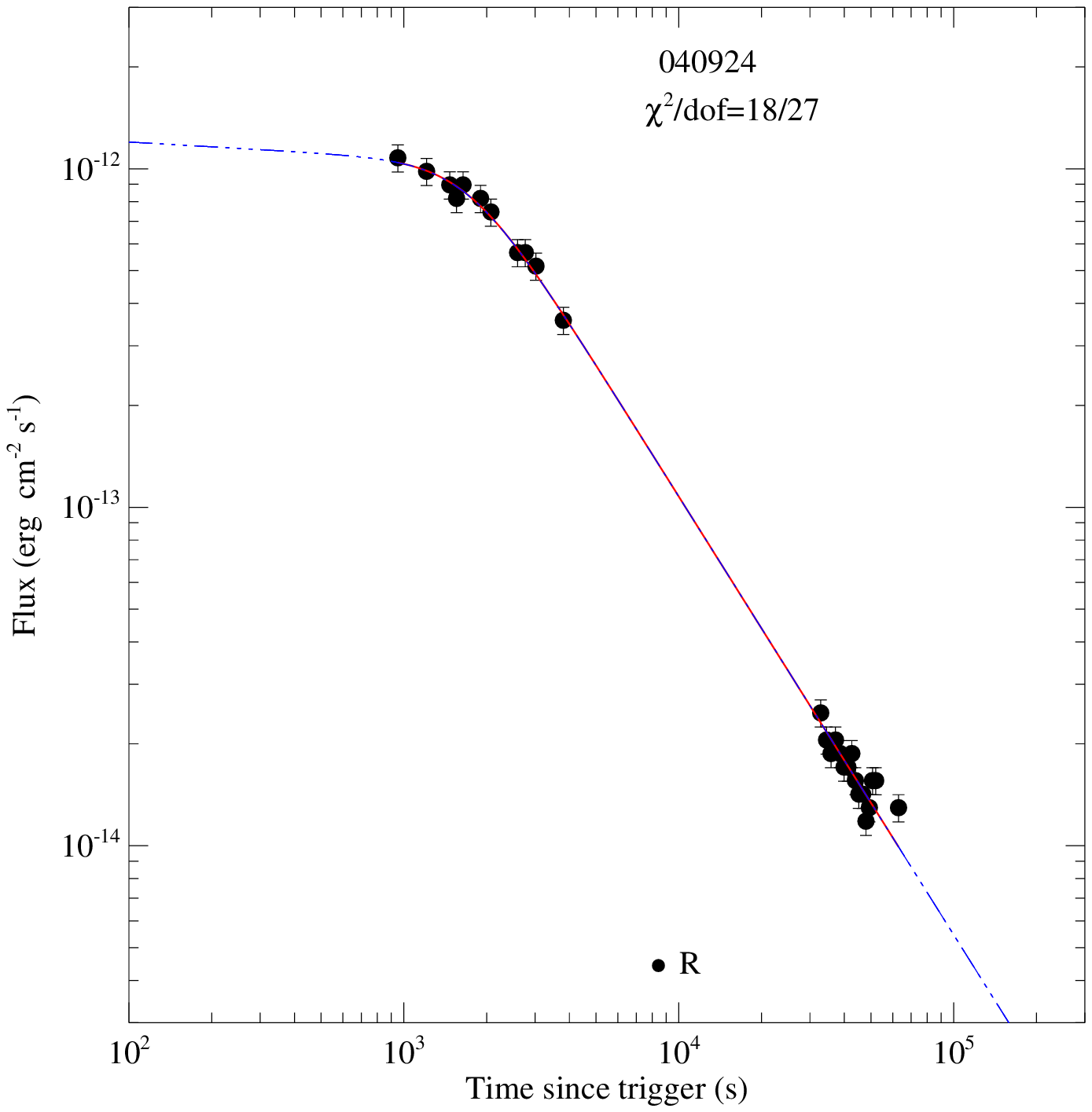}
\includegraphics[angle=0,scale=0.350,width=0.3\textwidth,height=0.25\textheight]{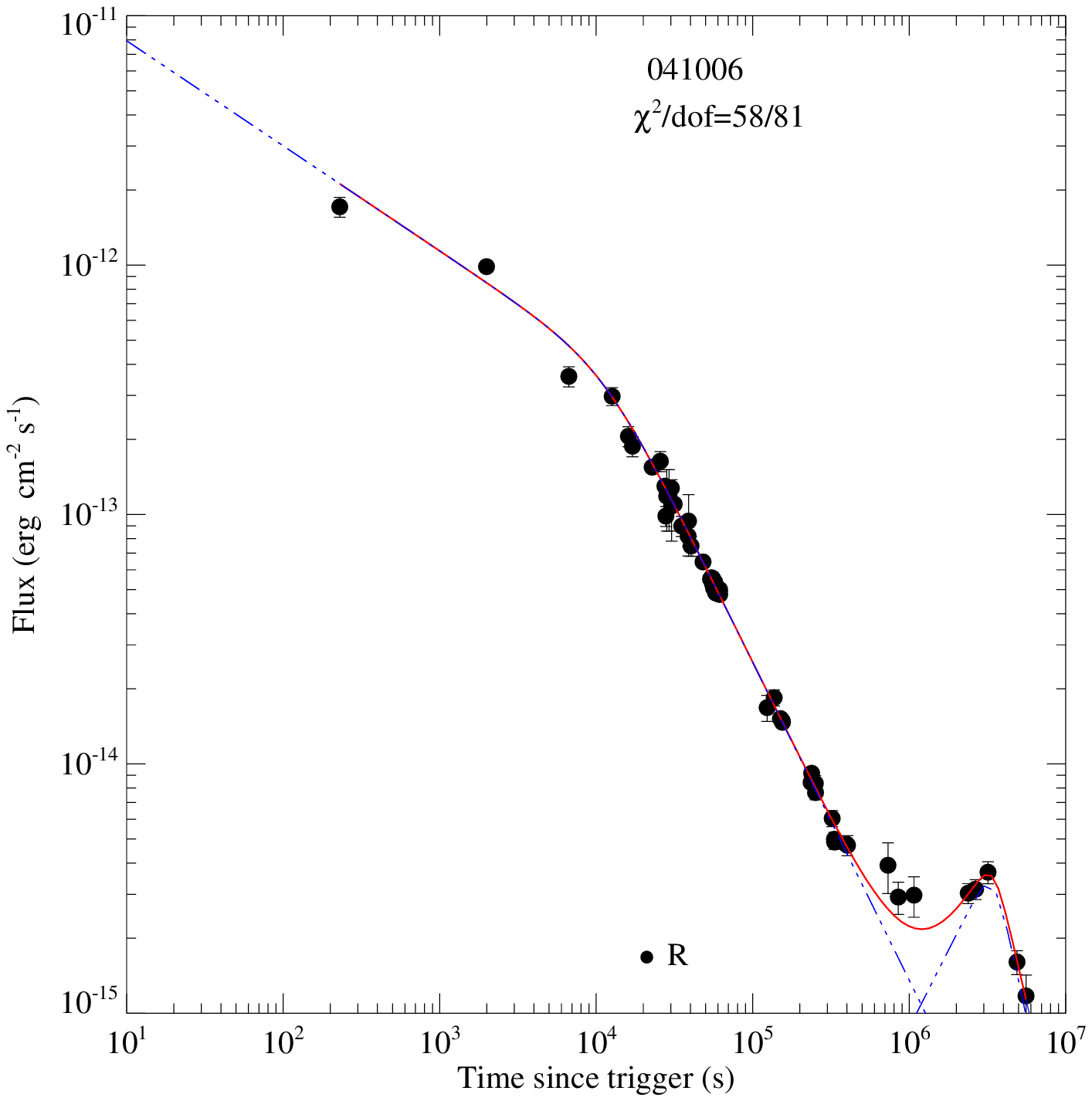}
\includegraphics[angle=0,scale=0.350,width=0.3\textwidth,height=0.25\textheight]{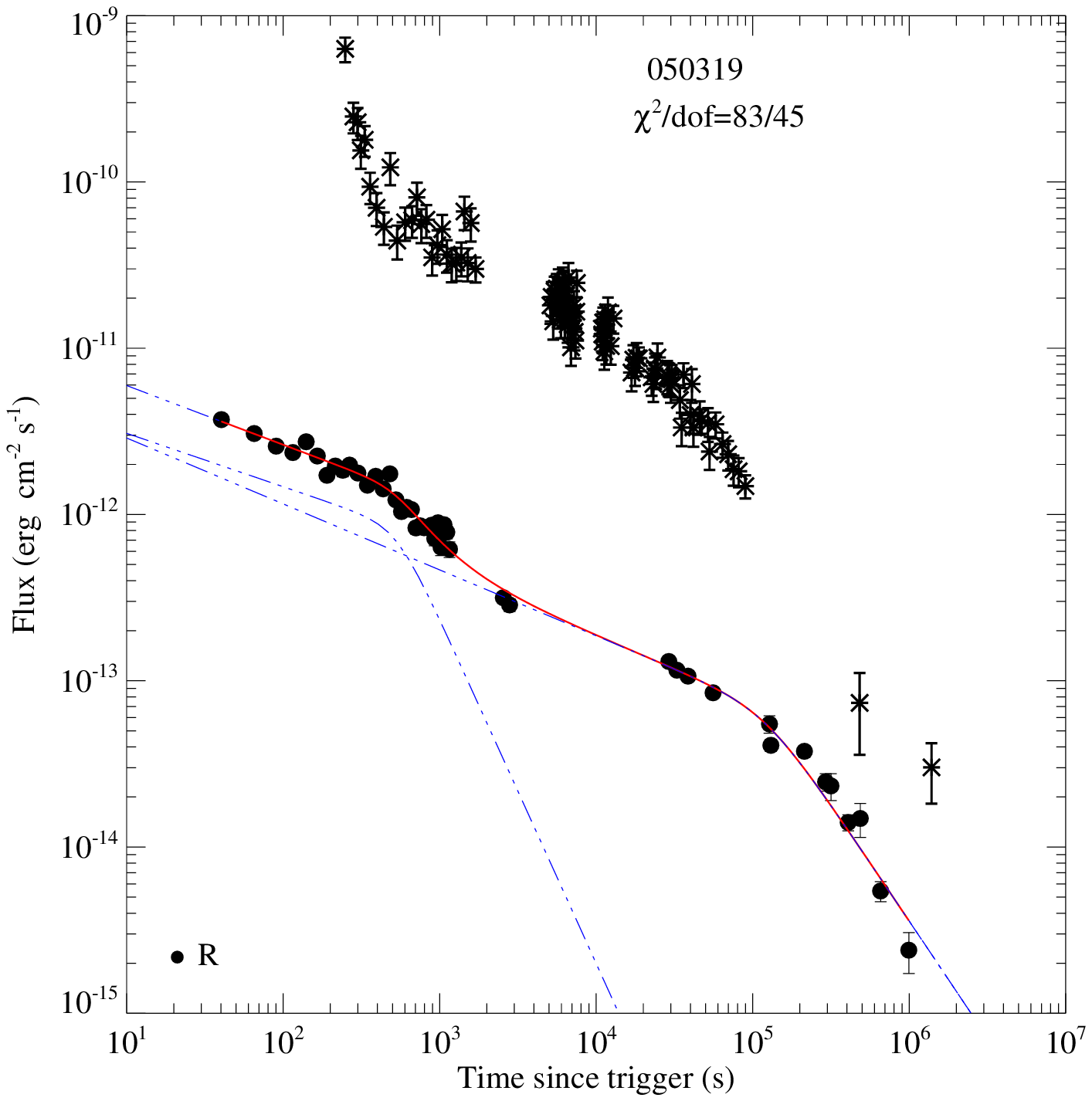}
\includegraphics[angle=0,scale=0.350,width=0.3\textwidth,height=0.25\textheight]{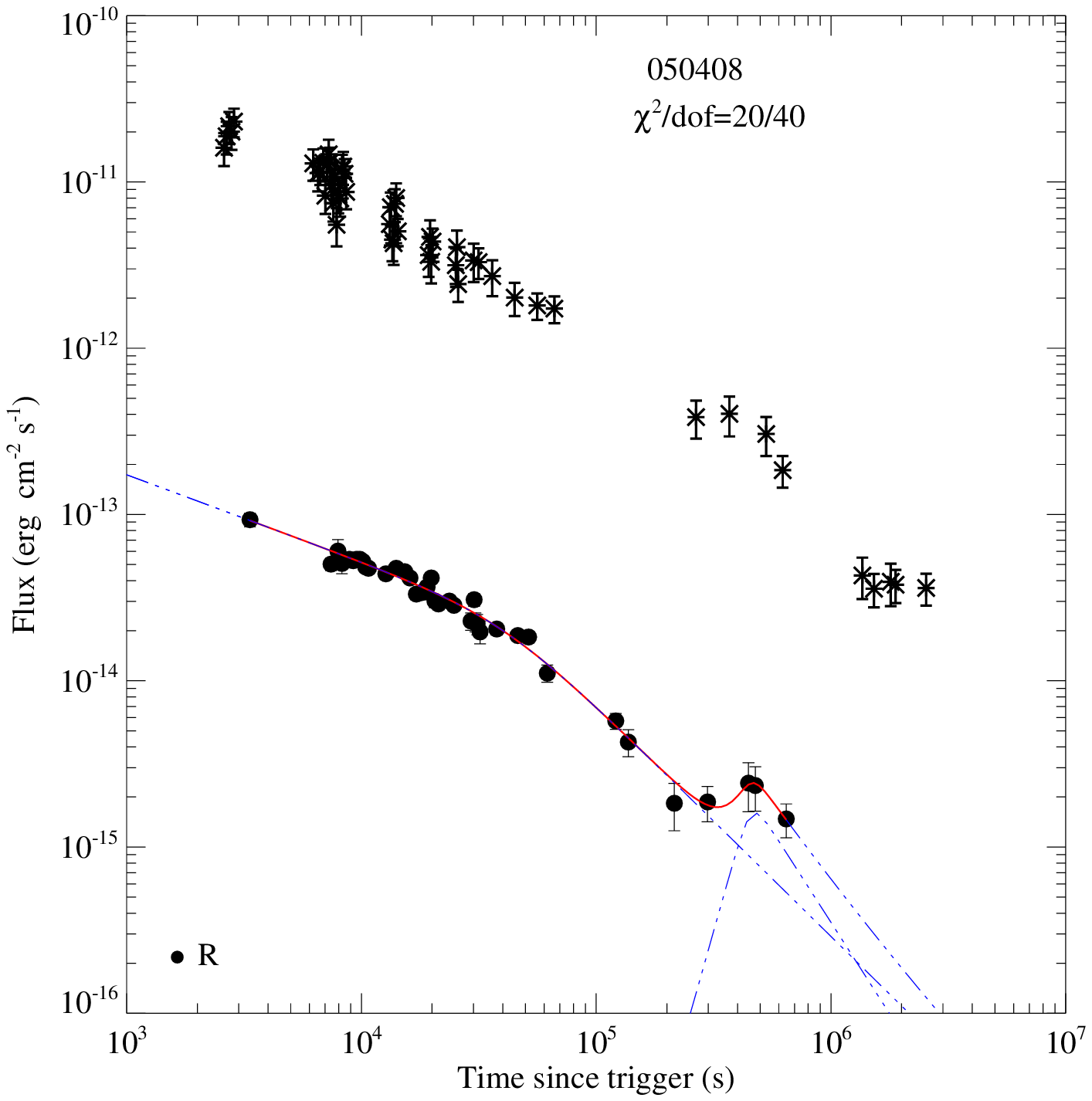}
\includegraphics[angle=0,scale=0.350,width=0.3\textwidth,height=0.25\textheight]{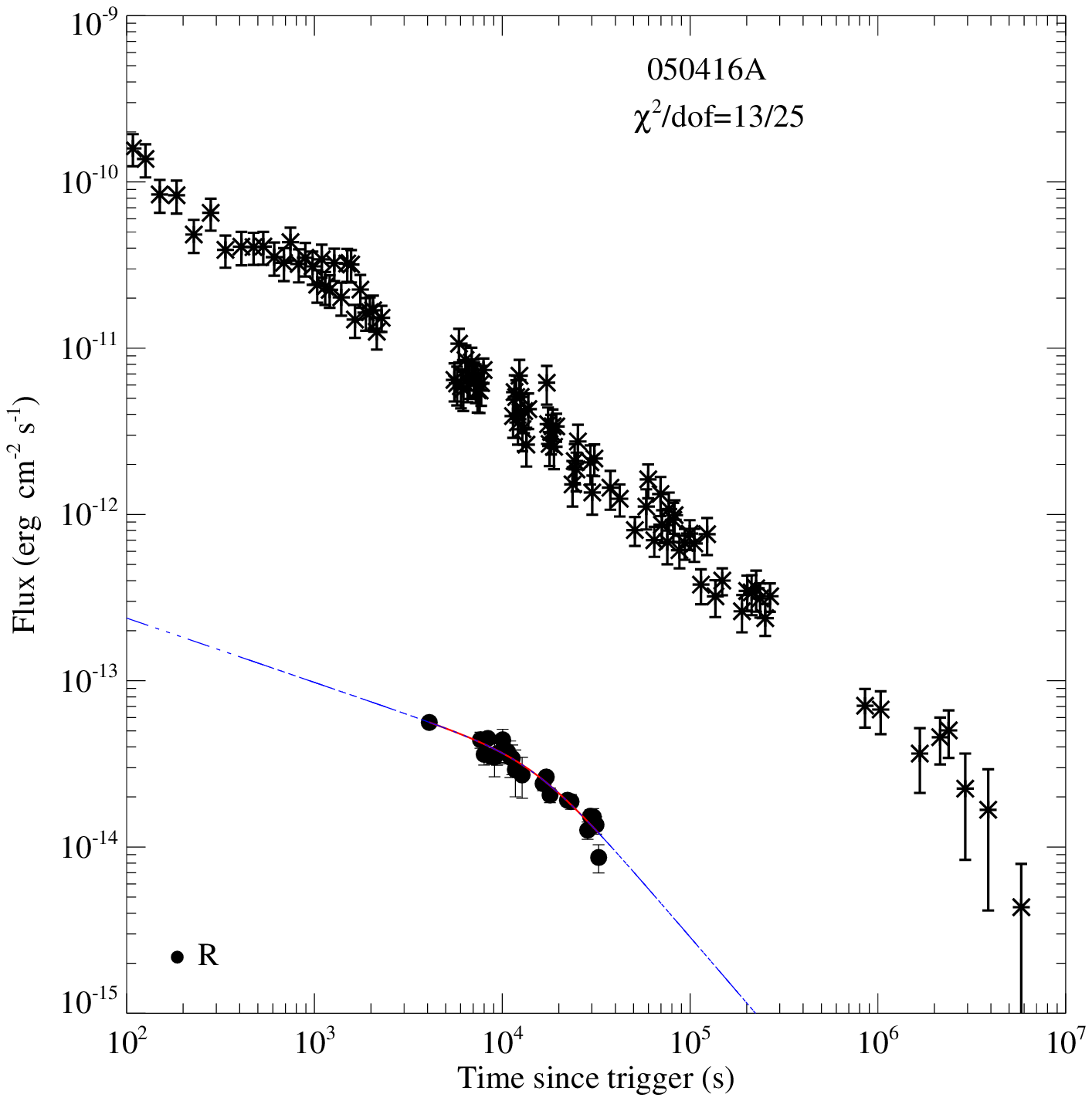}

\caption{Optical afterglow lightcurves with clear detections of the shallow decay segment(s). The symbols are the same as Figure 1.}
\label{Shallow_Opt_LC}
\end{figure*}

\begin{figure*}

\includegraphics[angle=0,scale=0.350,width=0.3\textwidth,height=0.25\textheight]{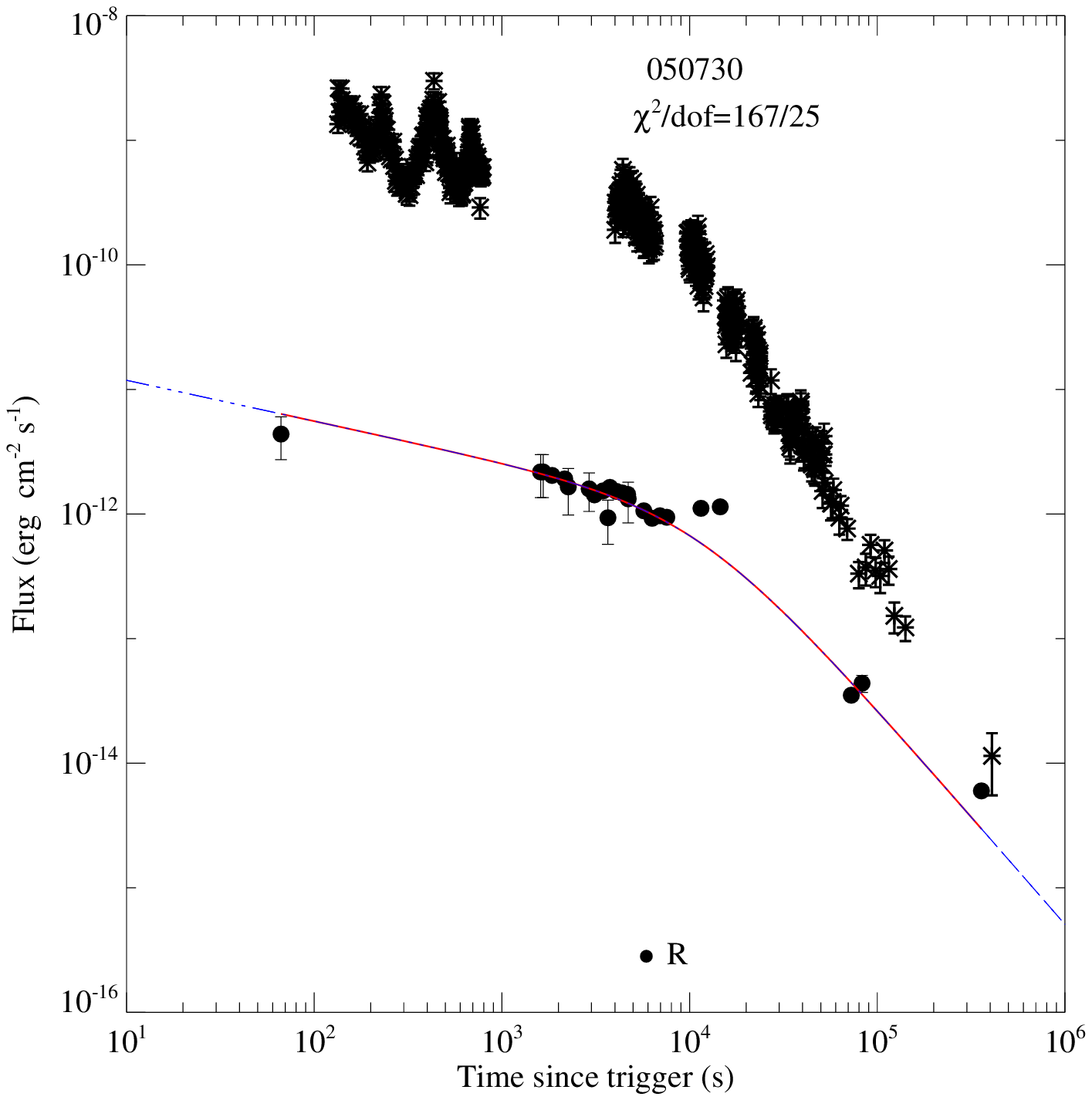}
\includegraphics[angle=0,scale=0.350,width=0.3\textwidth,height=0.25\textheight]{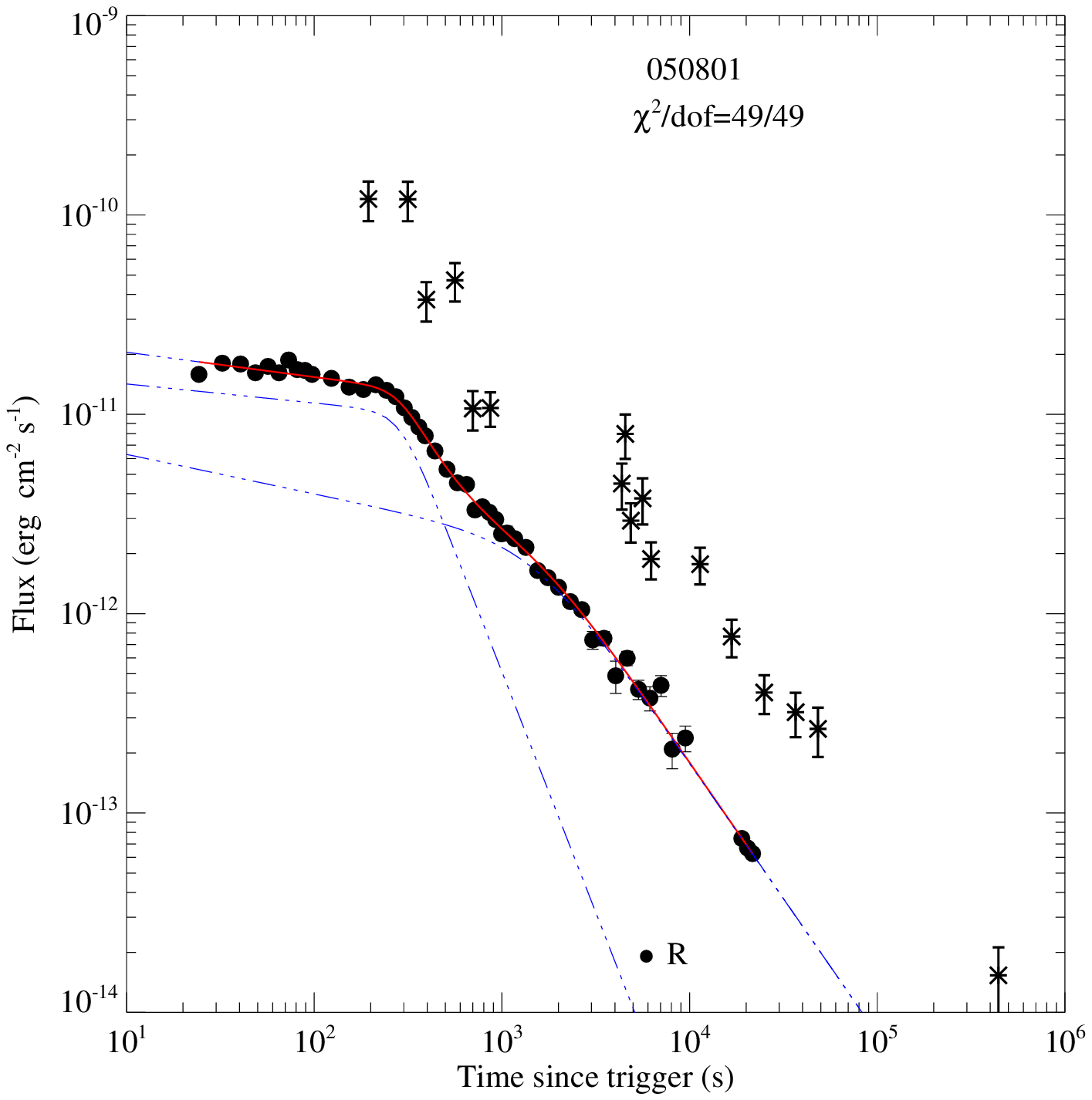}
\includegraphics[angle=0,scale=0.350,width=0.3\textwidth,height=0.25\textheight]{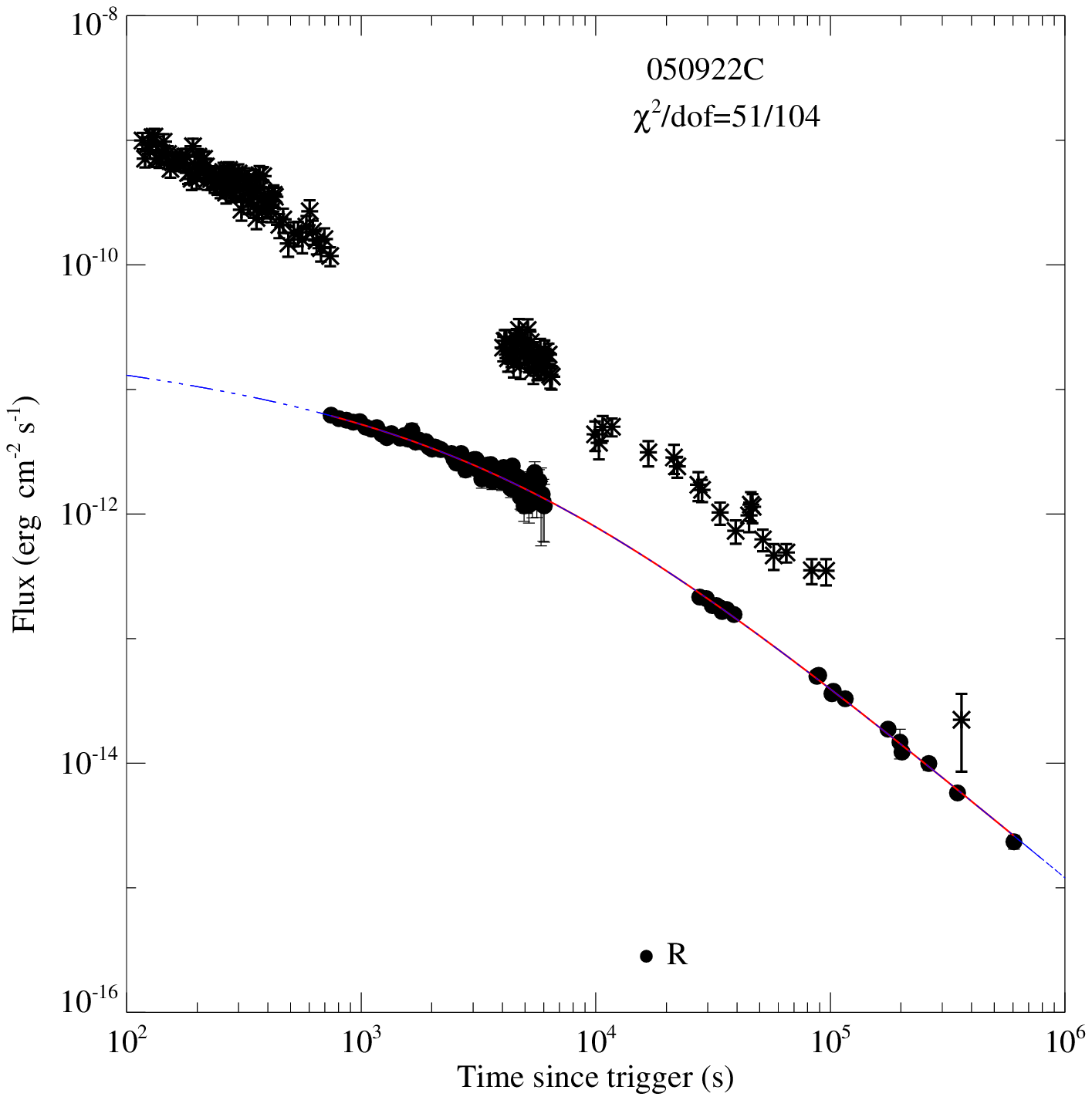}
\includegraphics[angle=0,scale=0.350,width=0.3\textwidth,height=0.25\textheight]{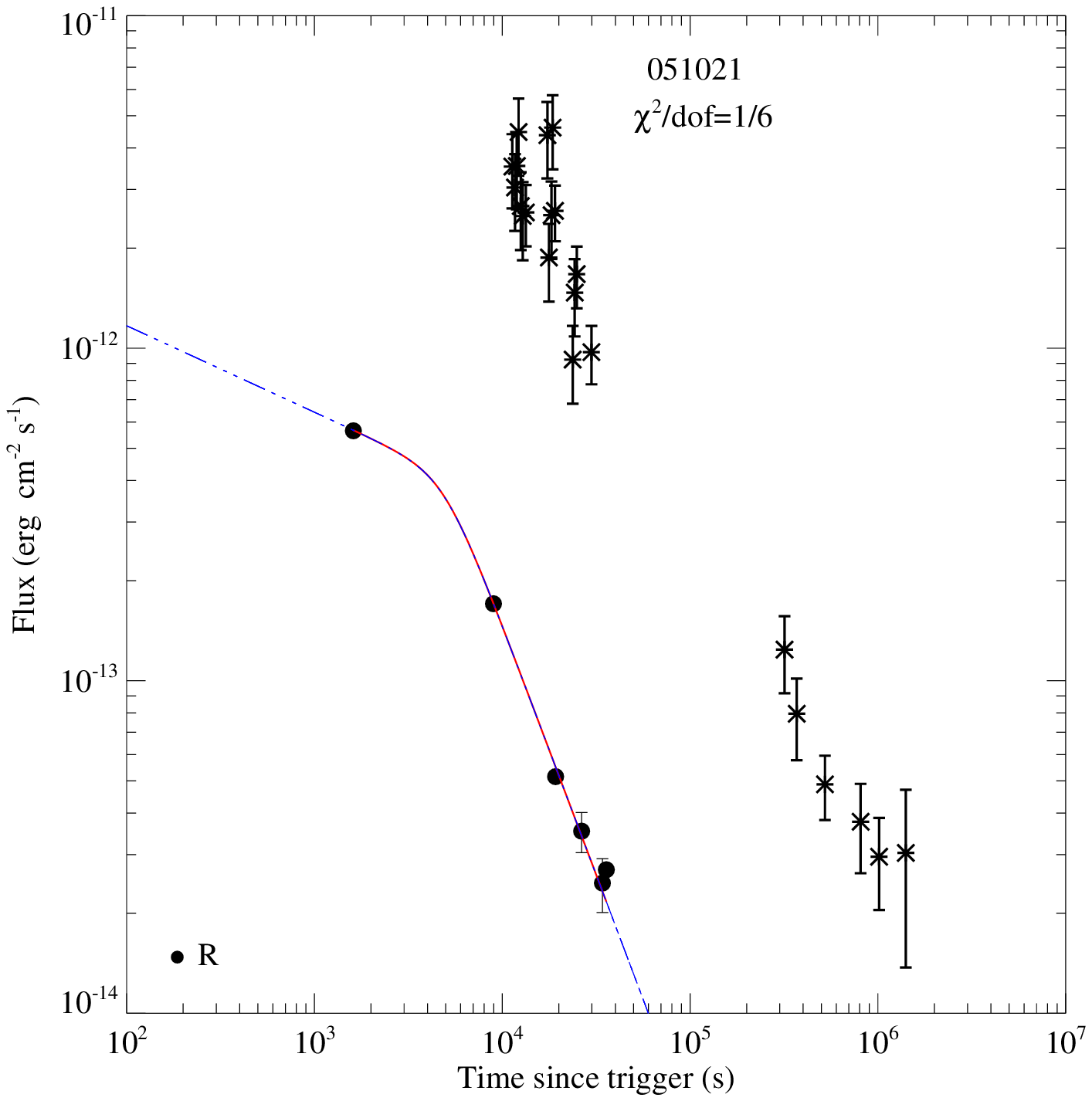}
\includegraphics[angle=0,scale=0.350,width=0.3\textwidth,height=0.25\textheight]{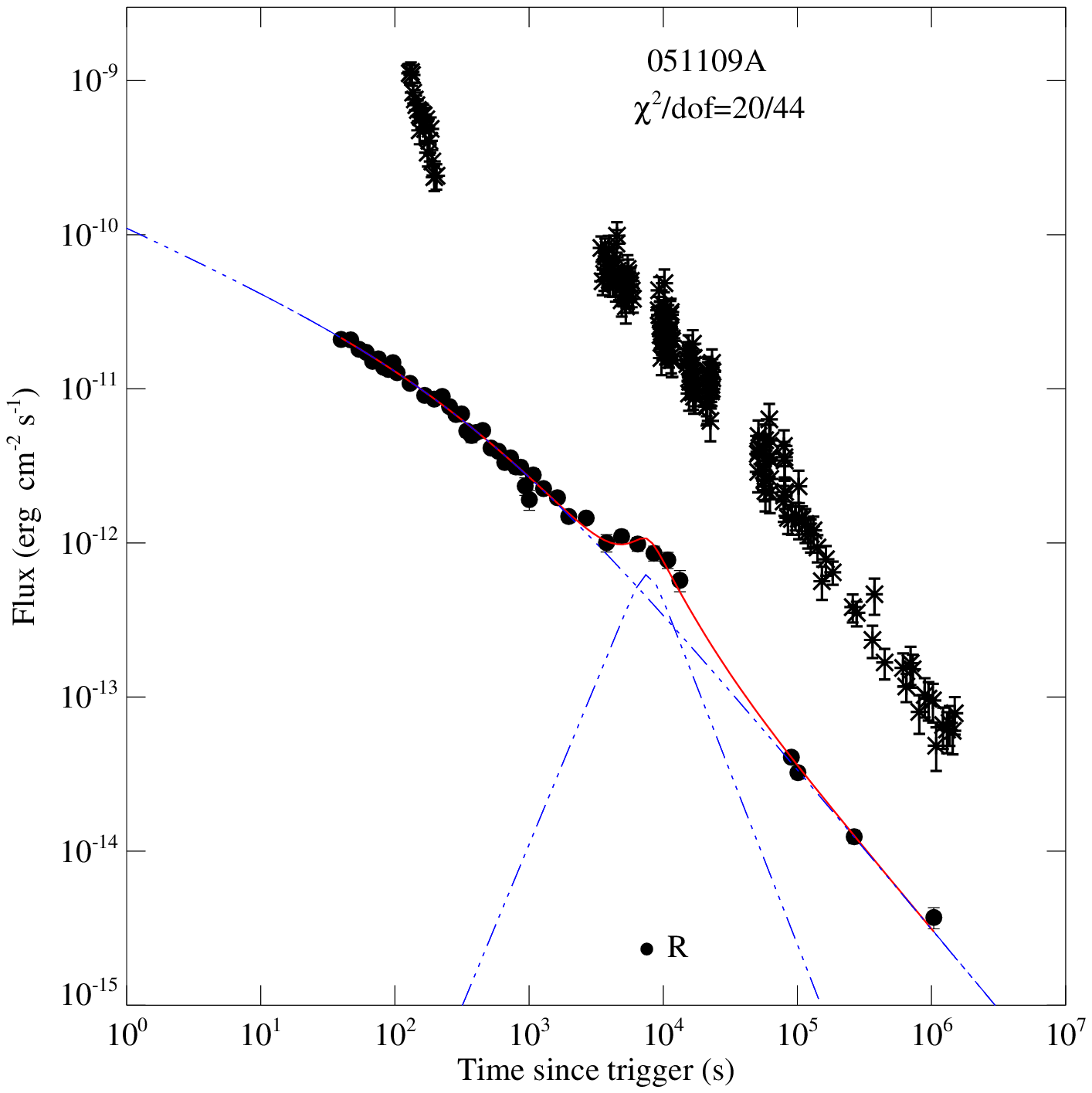}
\includegraphics[angle=0,scale=0.350,width=0.3\textwidth,height=0.25\textheight]{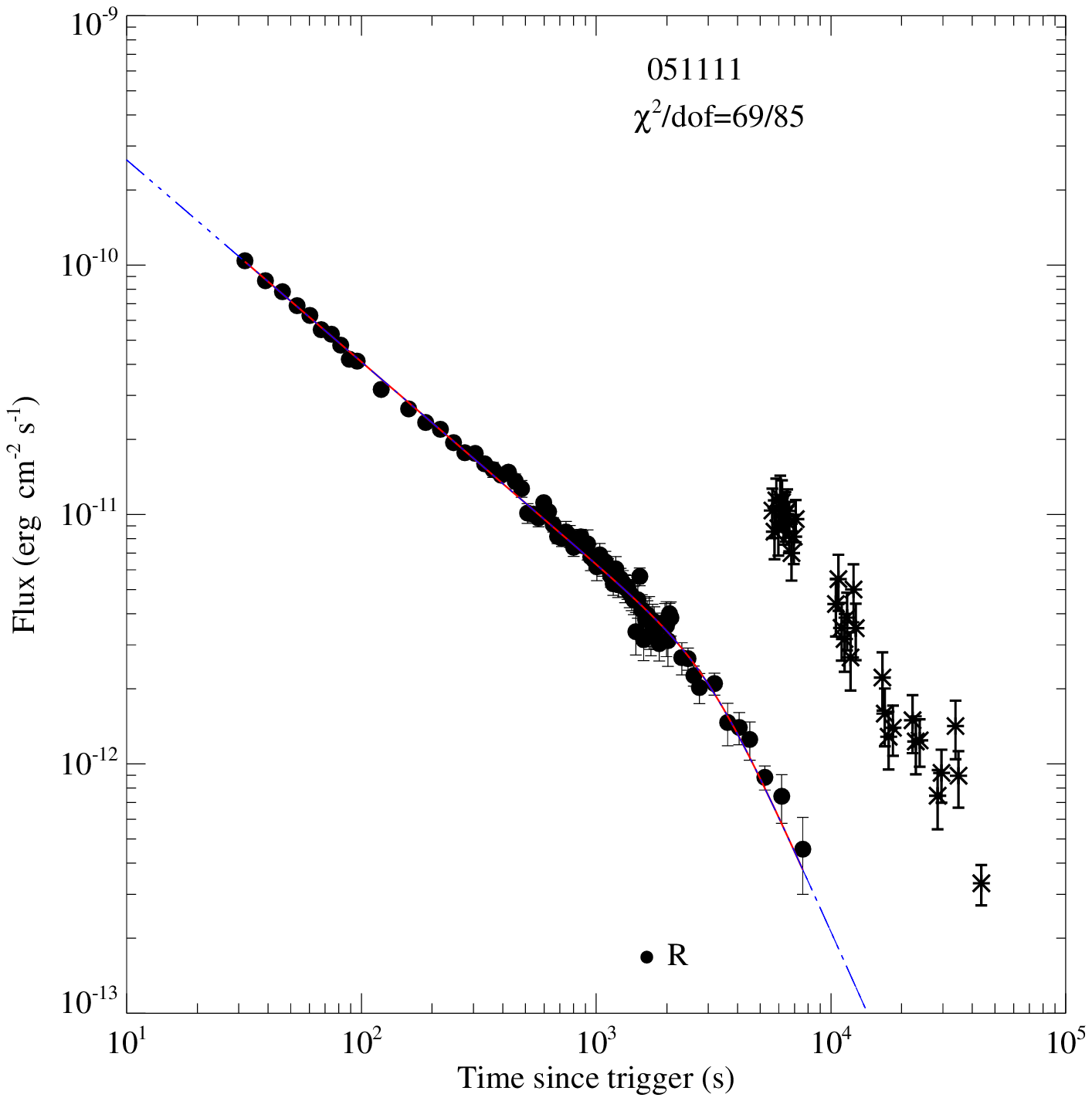}
\includegraphics[angle=0,scale=0.350,width=0.3\textwidth,height=0.25\textheight]{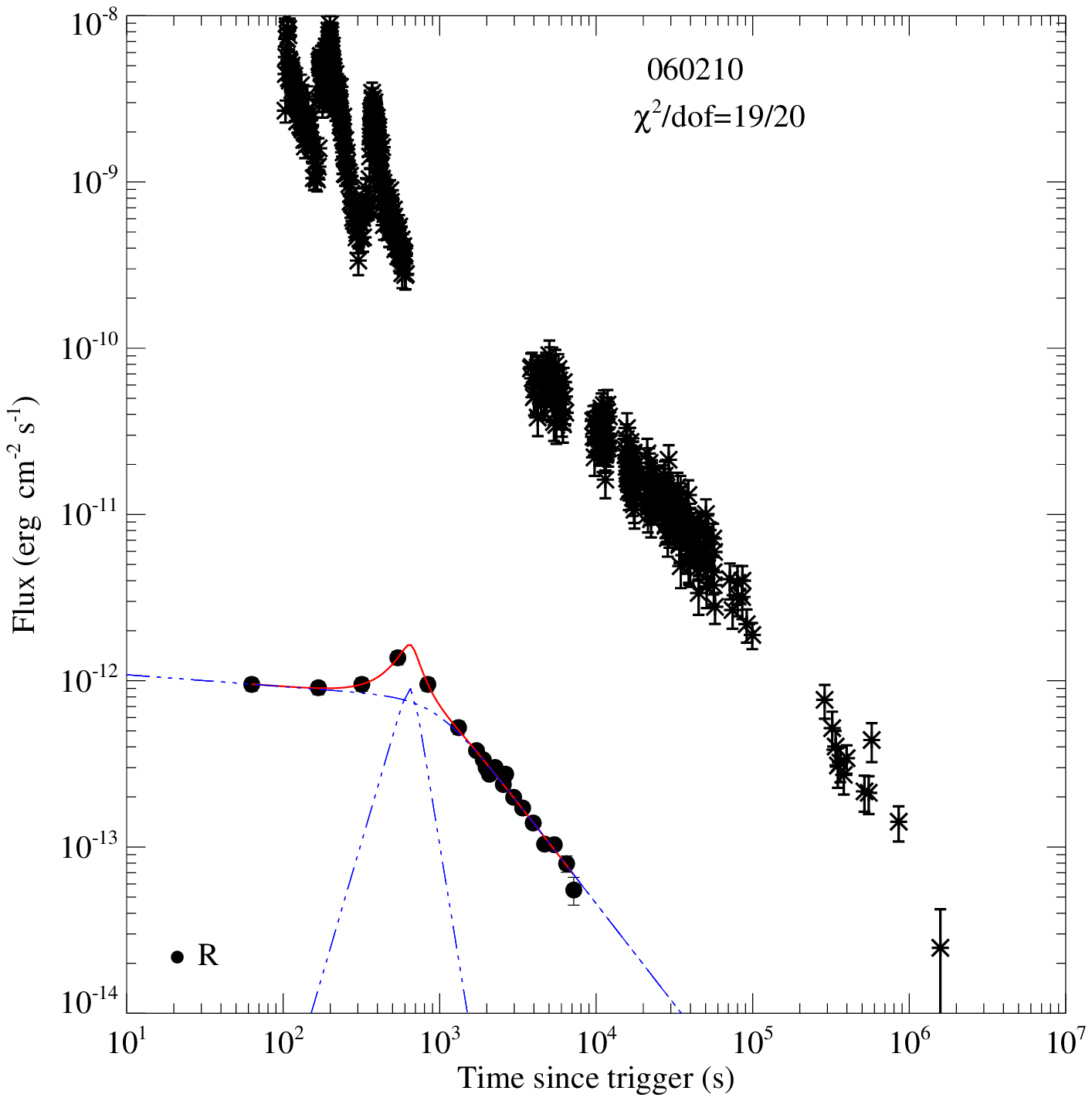}
\includegraphics[angle=0,scale=0.350,width=0.3\textwidth,height=0.25\textheight]{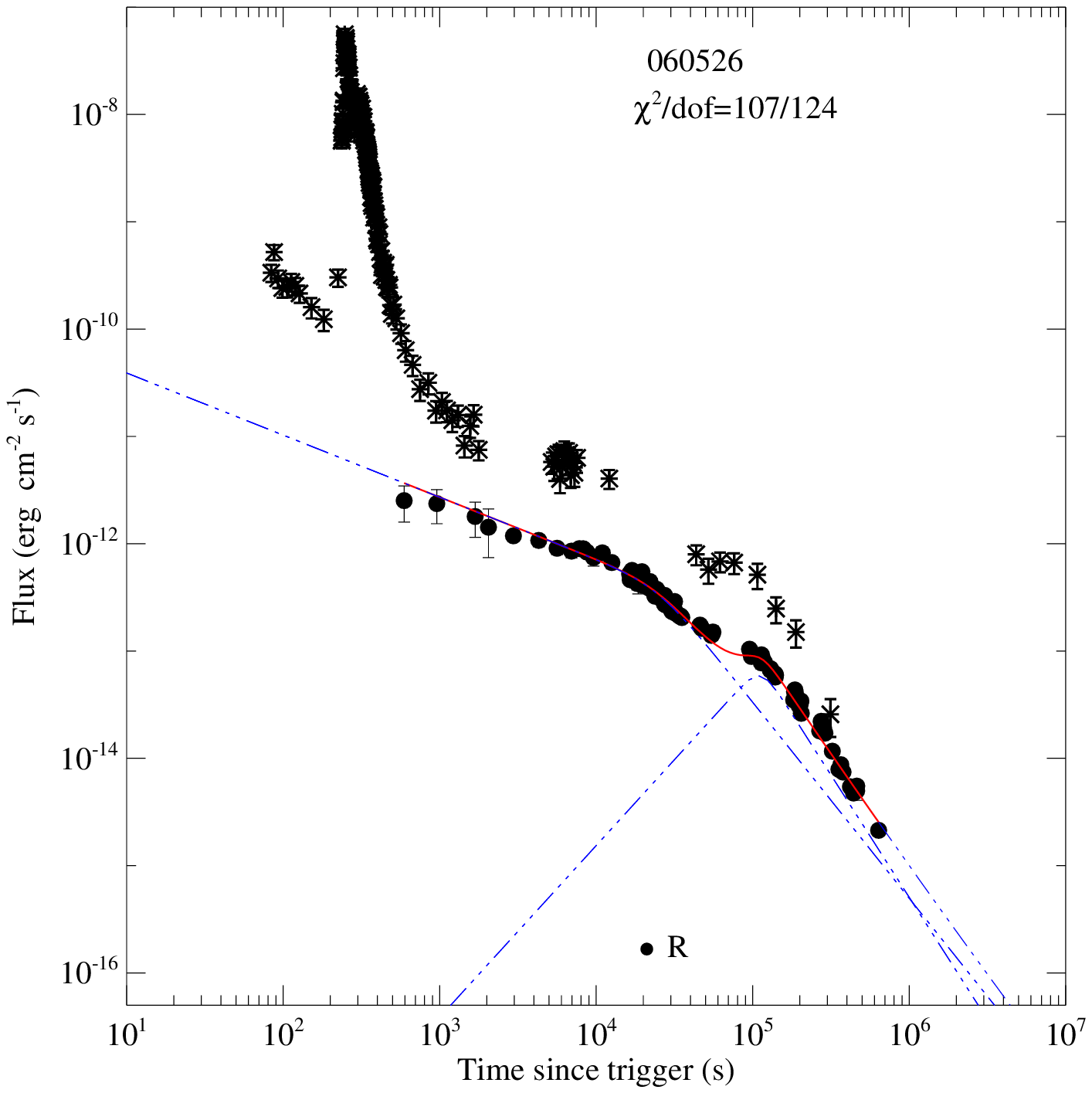}
\includegraphics[angle=0,scale=0.350,width=0.3\textwidth,height=0.25\textheight]{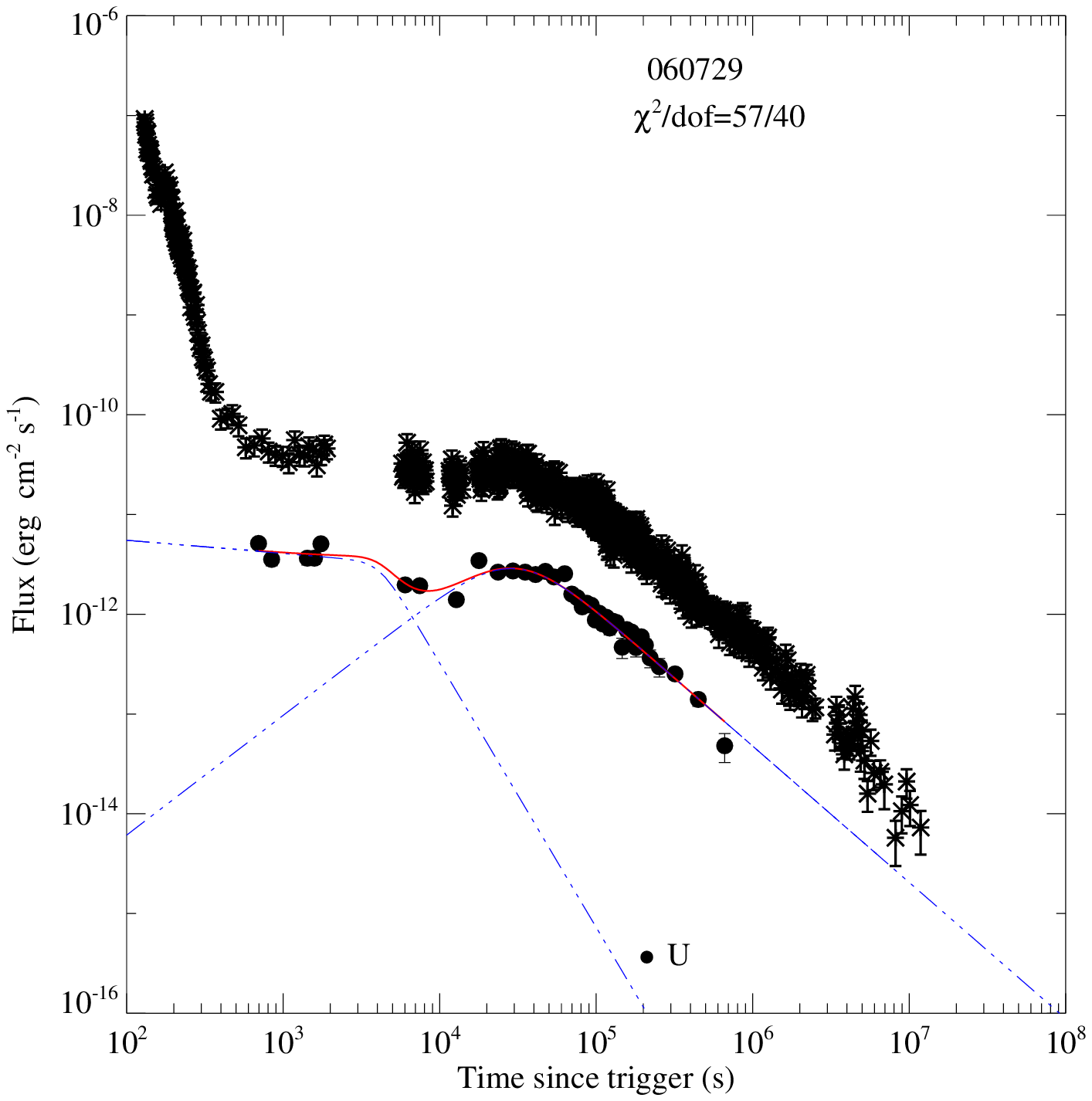}
\includegraphics[angle=0,scale=0.350,width=0.3\textwidth,height=0.25\textheight]{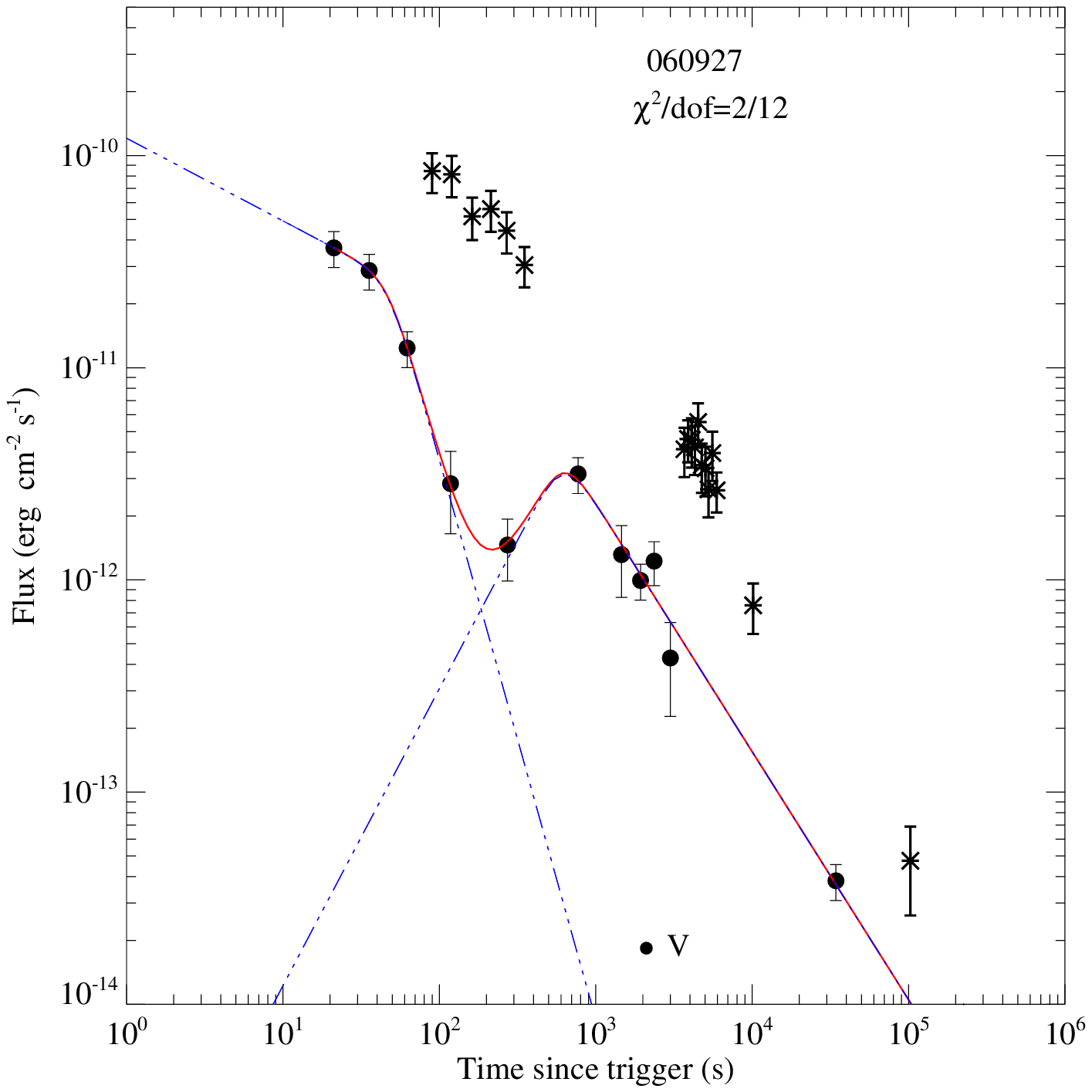}
\includegraphics[angle=0,scale=0.350,width=0.3\textwidth,height=0.25\textheight]{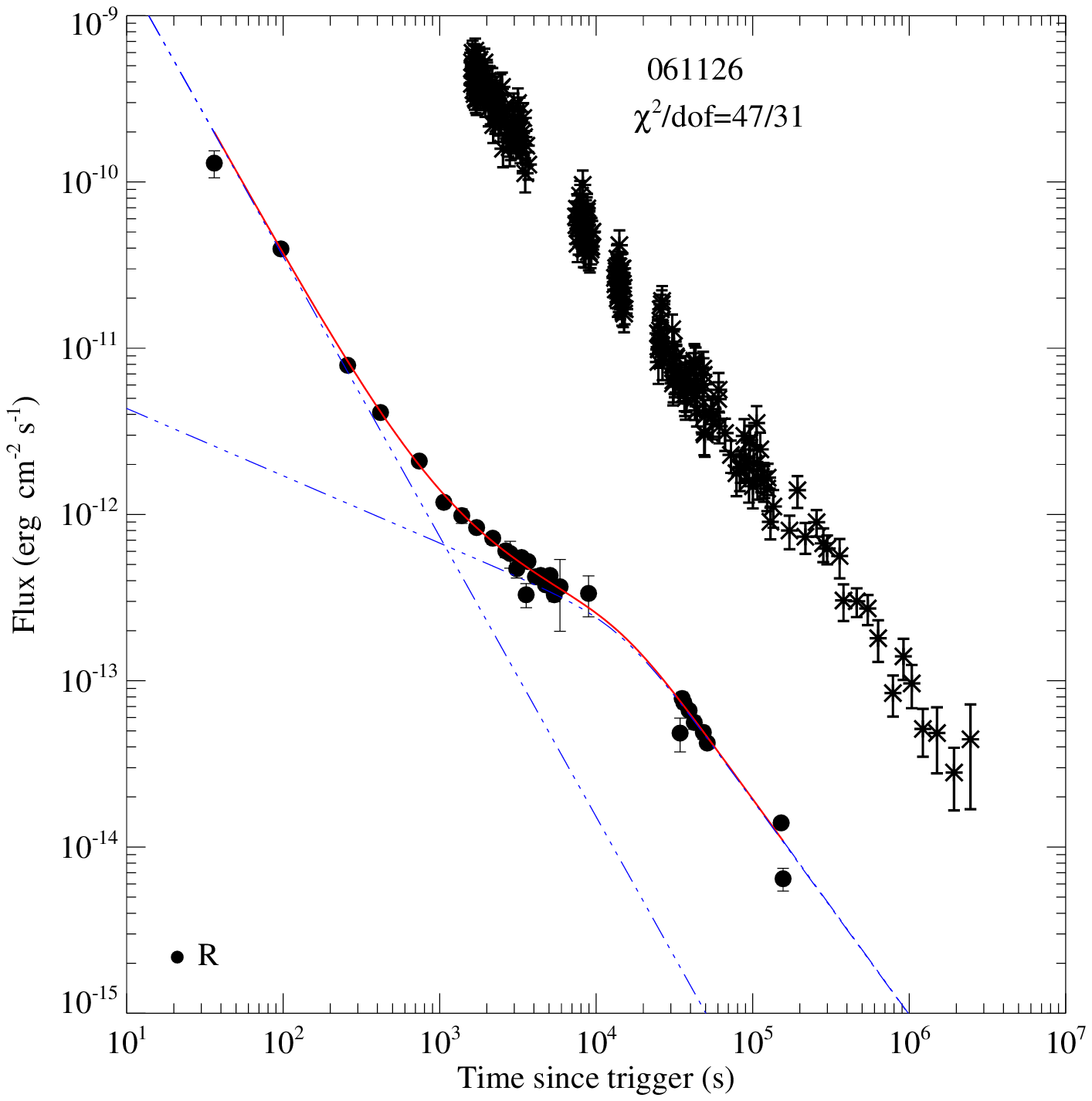}
\includegraphics[angle=0,scale=0.350,width=0.3\textwidth,height=0.25\textheight]{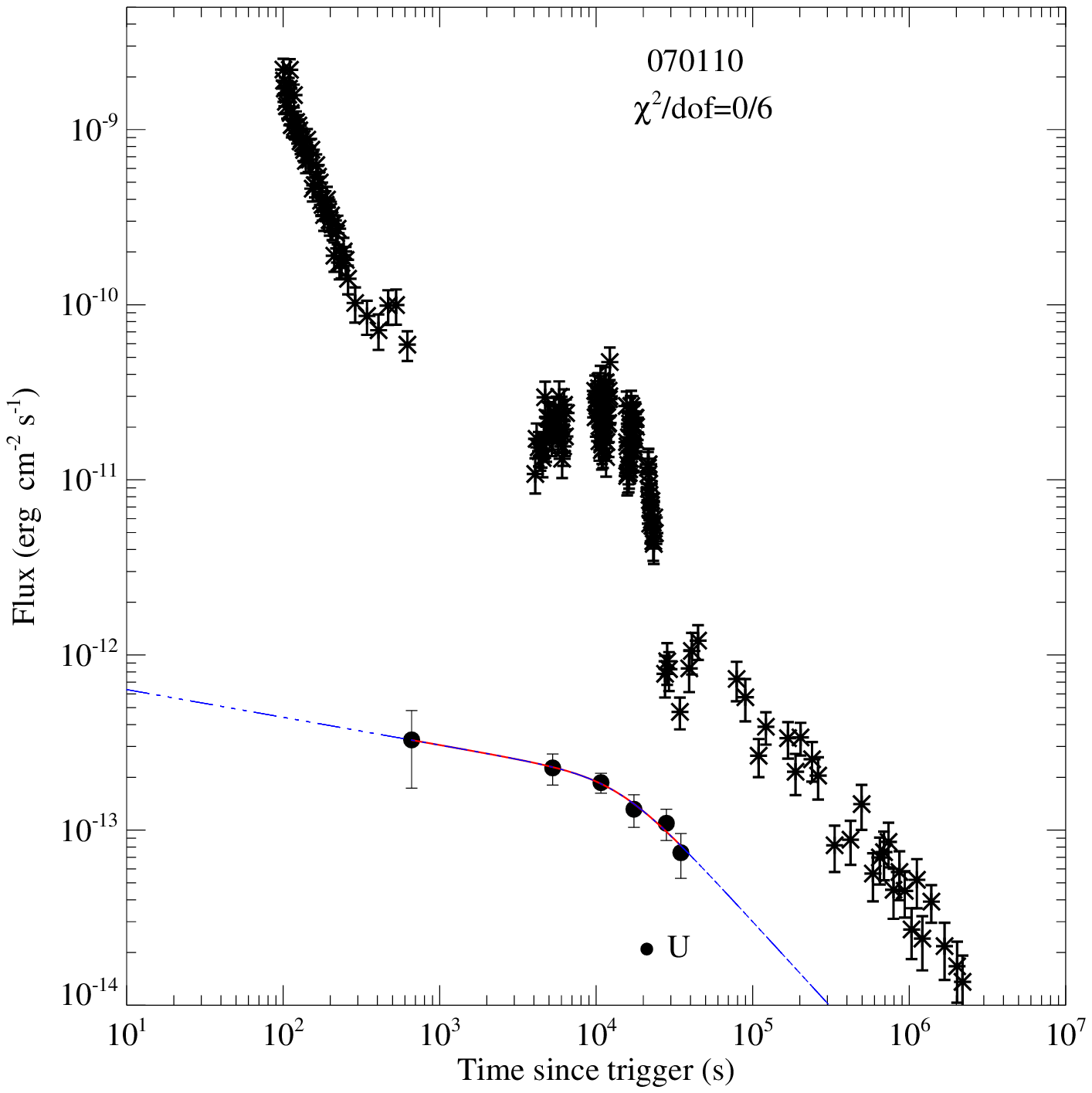}
\center{Fig. \ref{Shallow_Opt_LC}--- Continued}
\end{figure*}

\begin{figure*}

\includegraphics[angle=0,scale=0.350,width=0.3\textwidth,height=0.25\textheight]{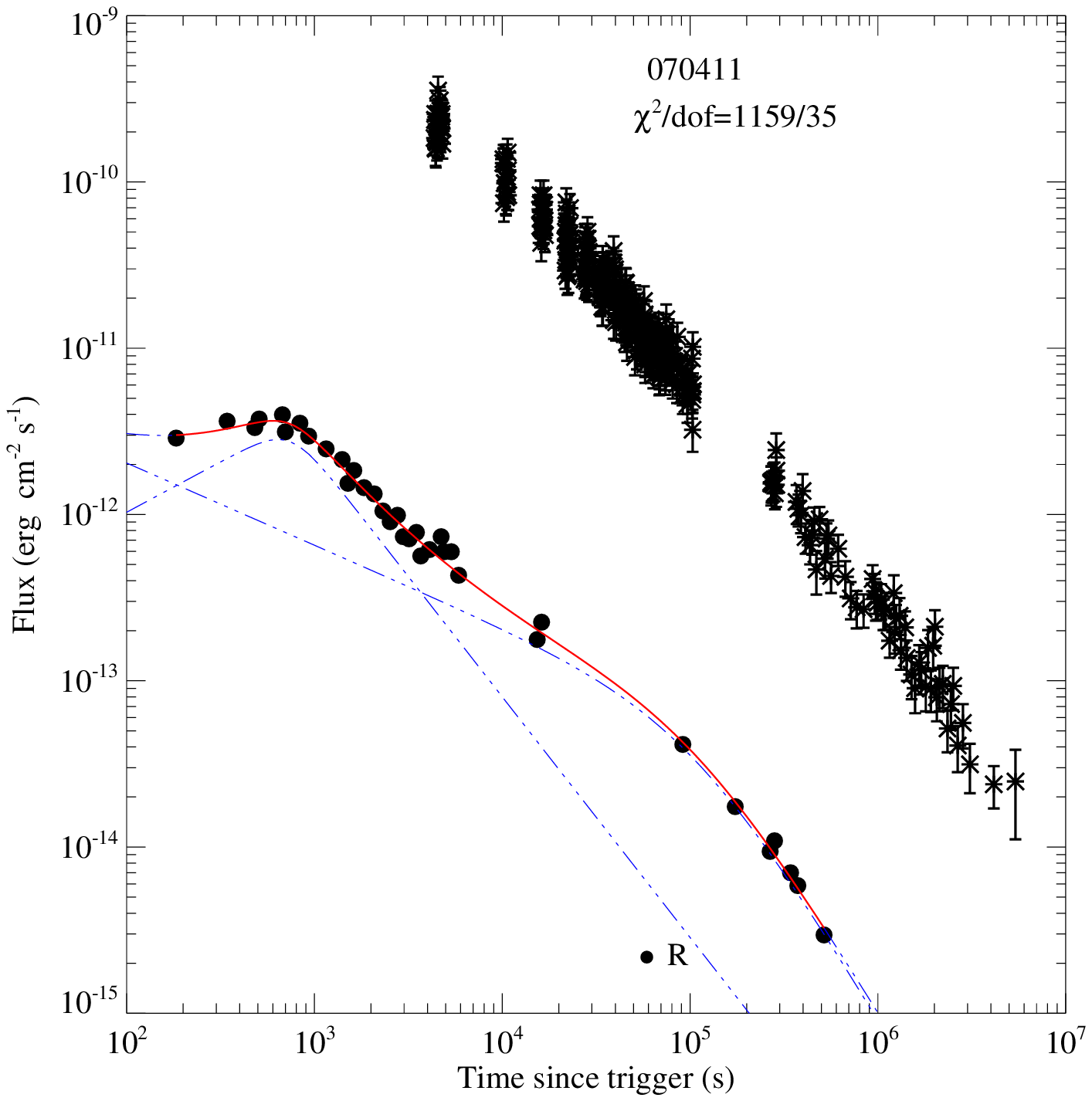}
\includegraphics[angle=0,scale=0.350,width=0.3\textwidth,height=0.25\textheight]{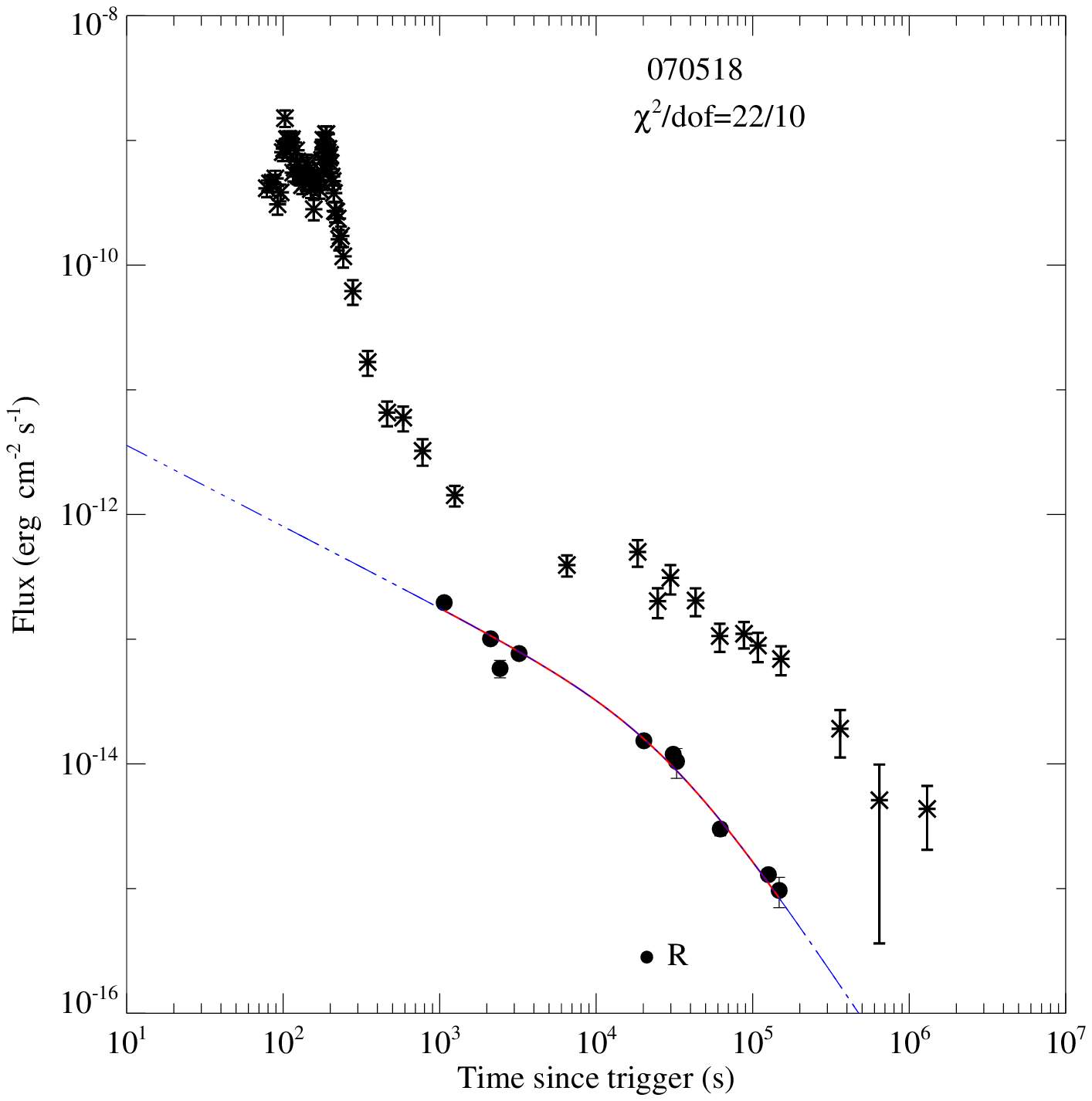}
\includegraphics[angle=0,scale=0.350,width=0.3\textwidth,height=0.25\textheight]{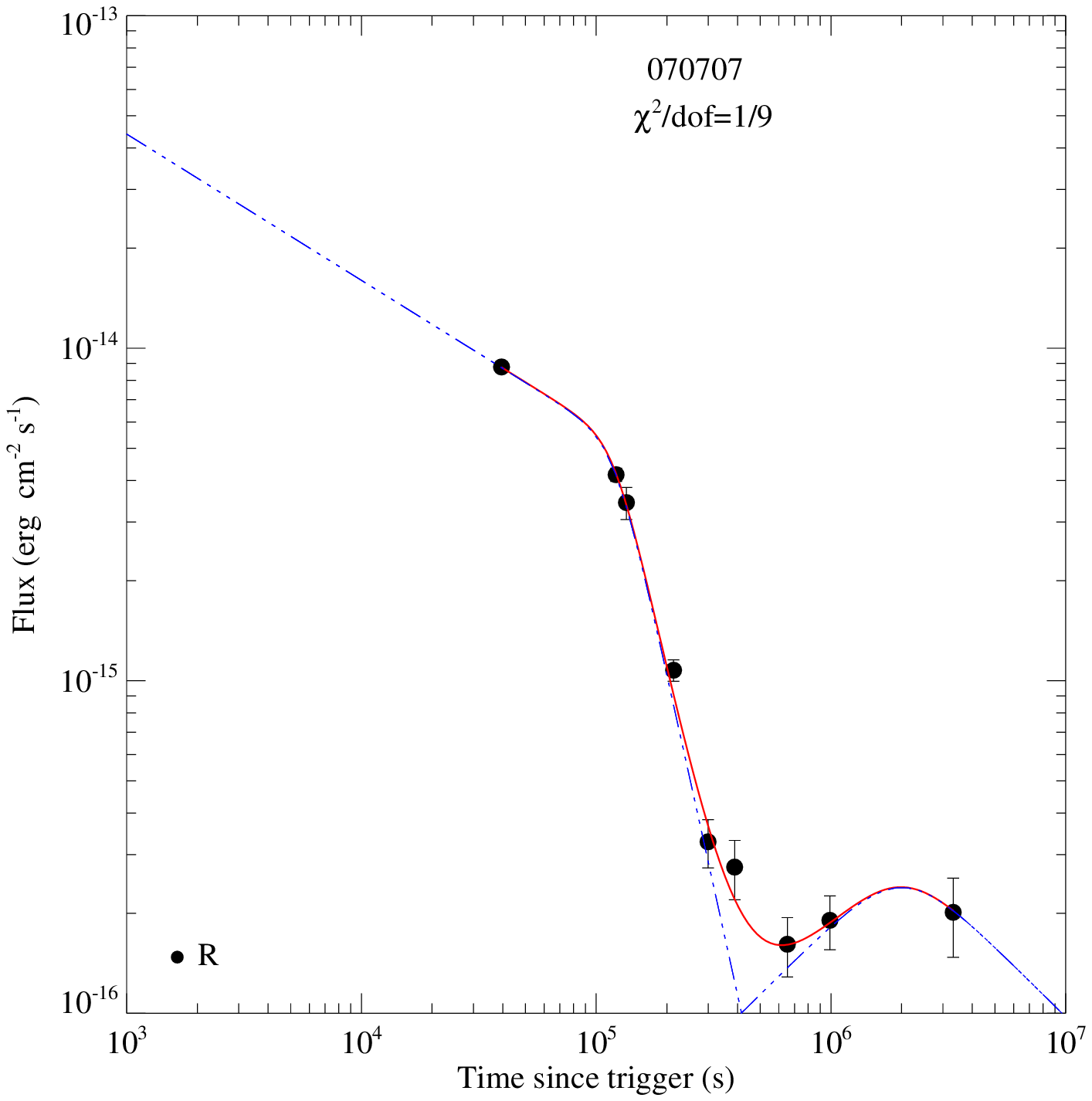}
\includegraphics[angle=0,scale=0.350,width=0.3\textwidth,height=0.25\textheight]{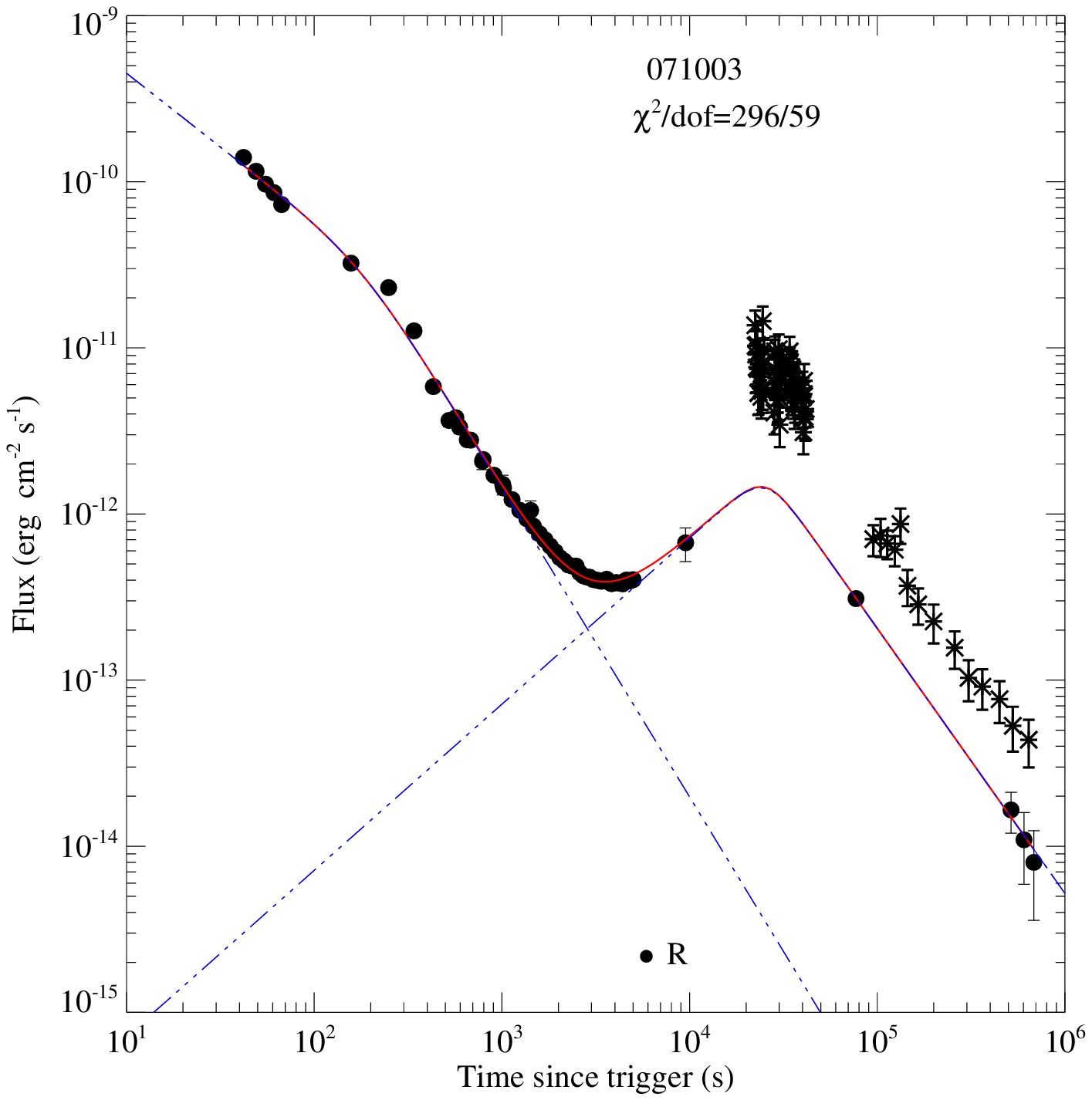}
\includegraphics[angle=0,scale=0.350,width=0.3\textwidth,height=0.25\textheight]{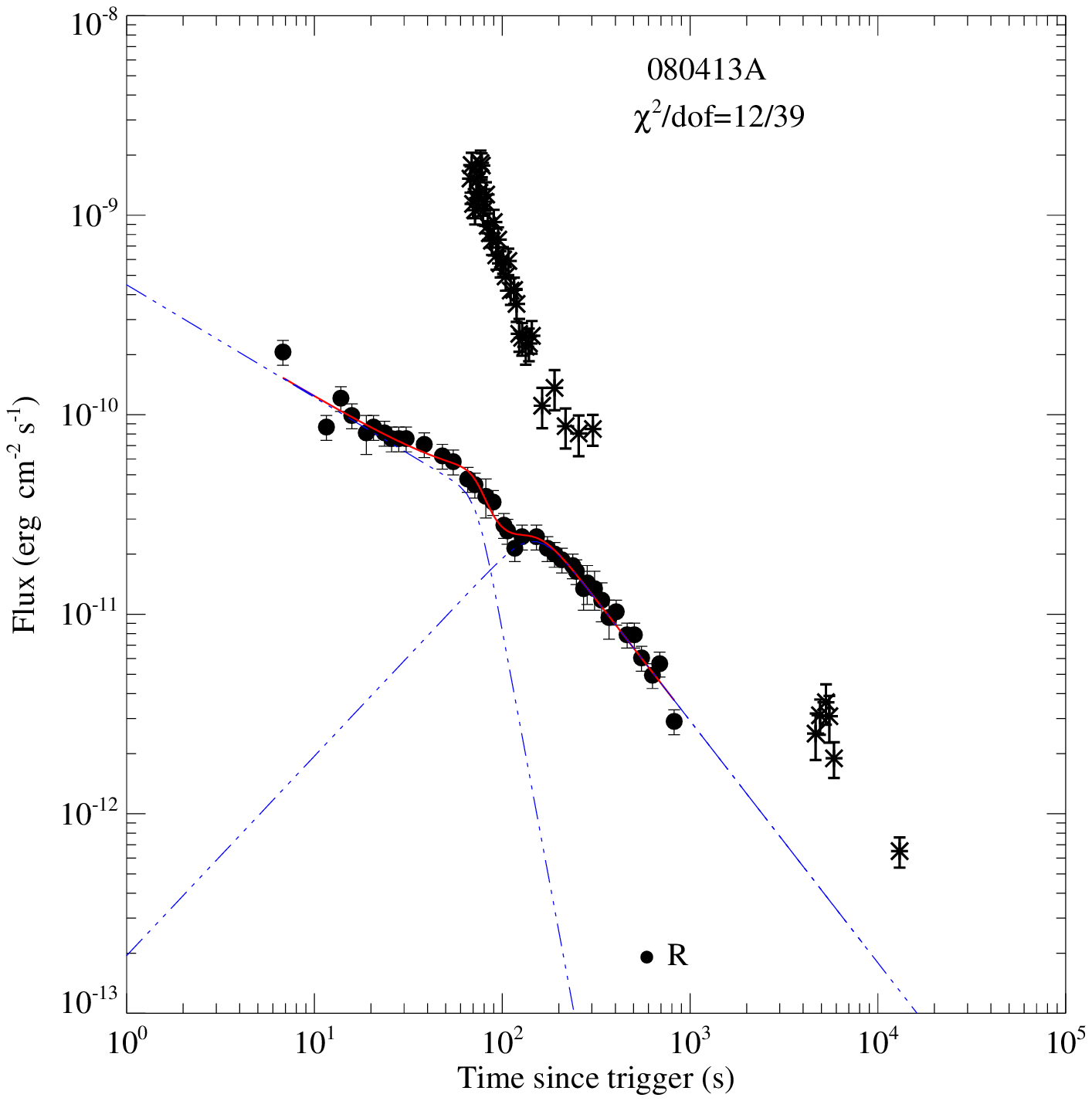}
\includegraphics[angle=0,scale=0.350,width=0.3\textwidth,height=0.25\textheight]{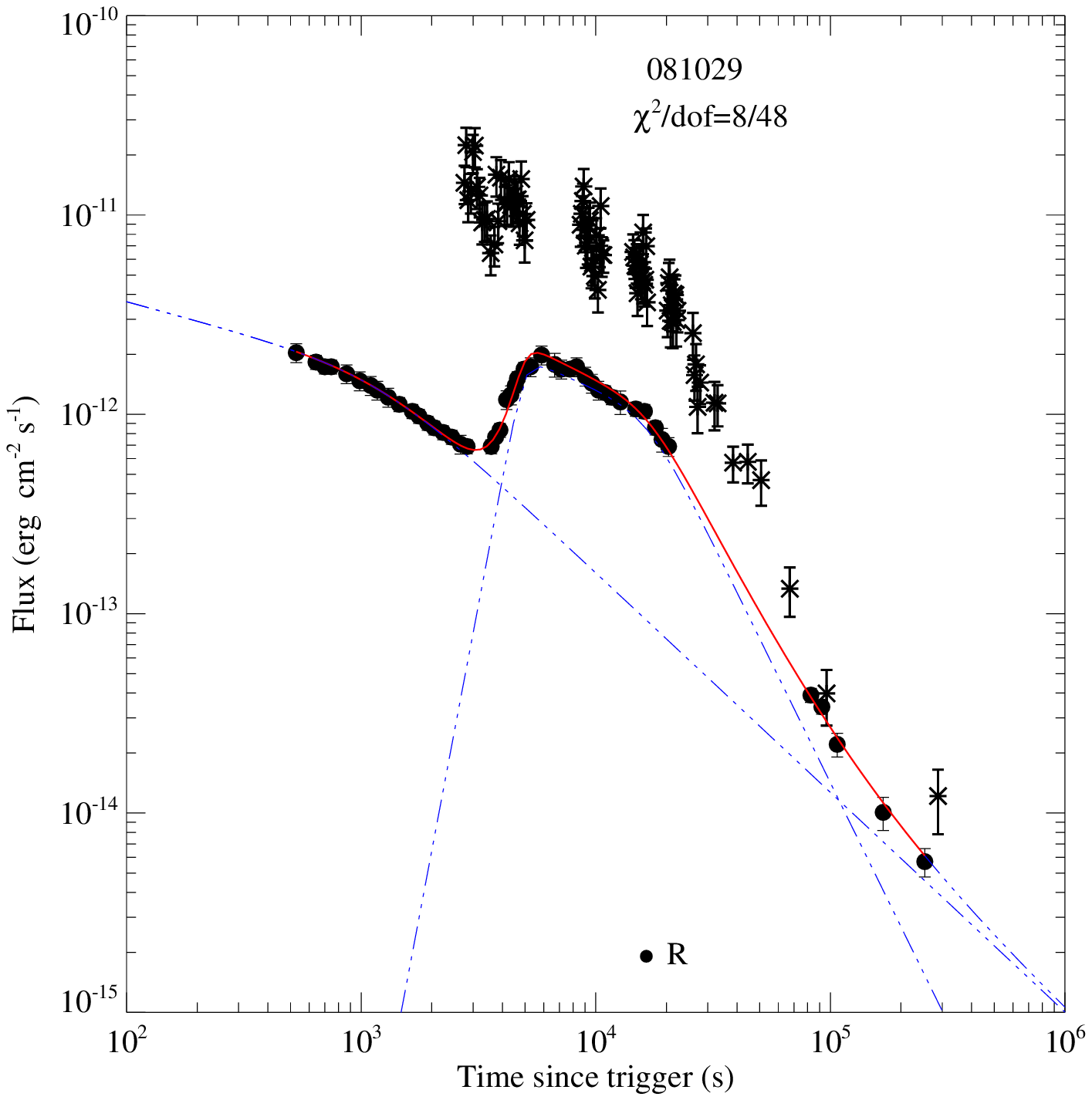}
\includegraphics[angle=0,scale=0.350,width=0.3\textwidth,height=0.25\textheight]{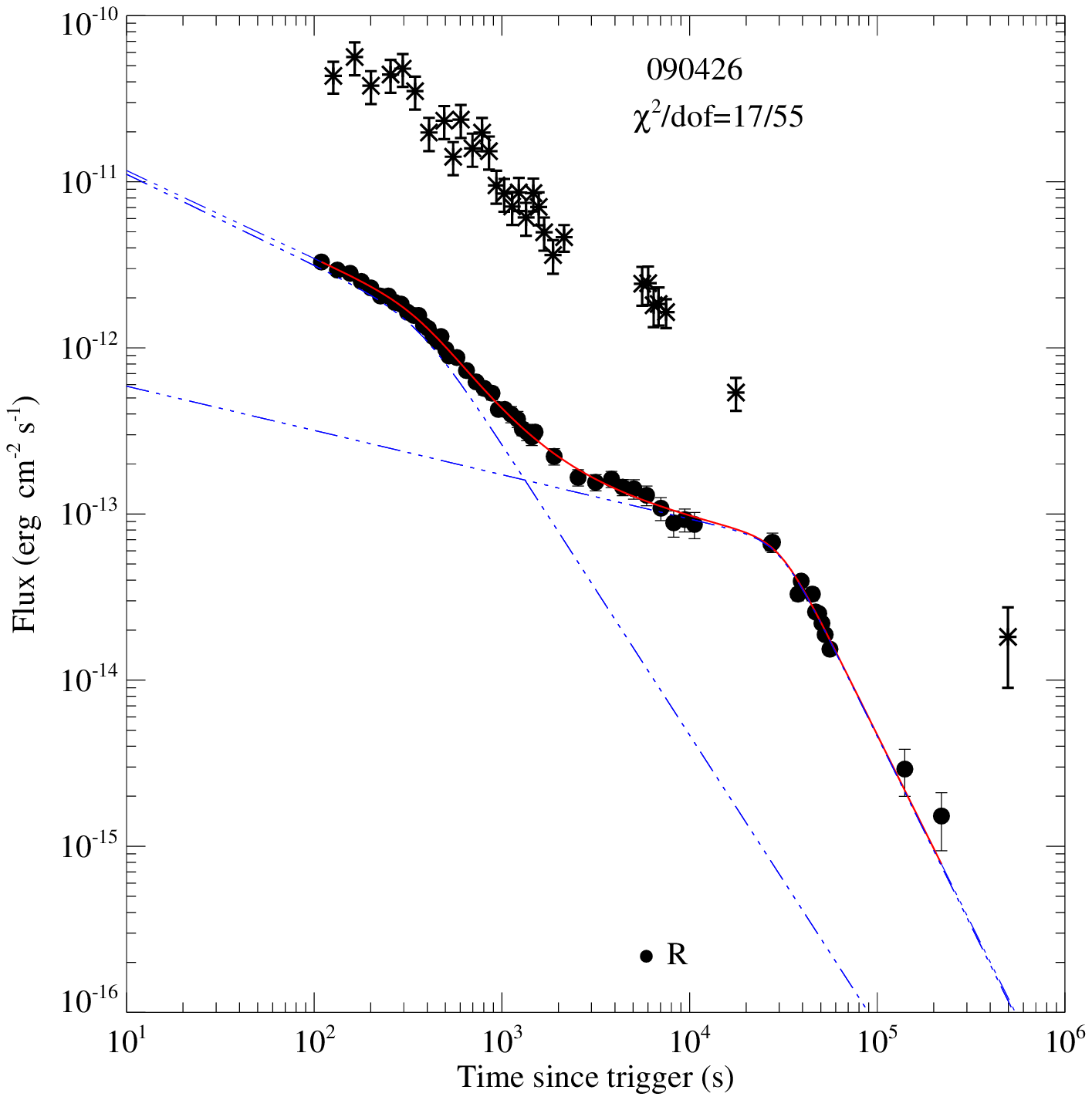}
\includegraphics[angle=0,scale=0.350,width=0.3\textwidth,height=0.25\textheight]{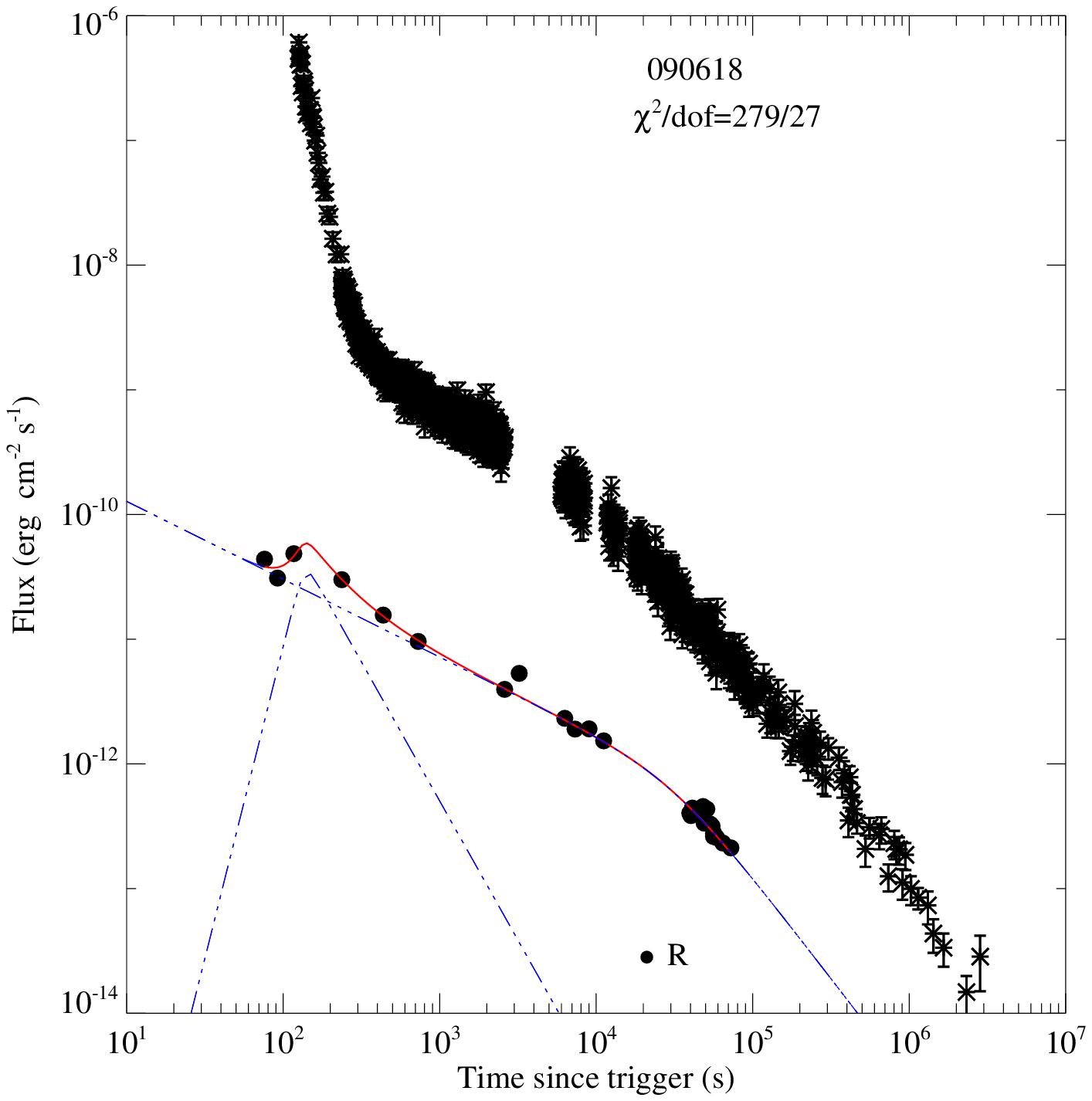}
\includegraphics[angle=0,scale=0.350,width=0.3\textwidth,height=0.25\textheight]{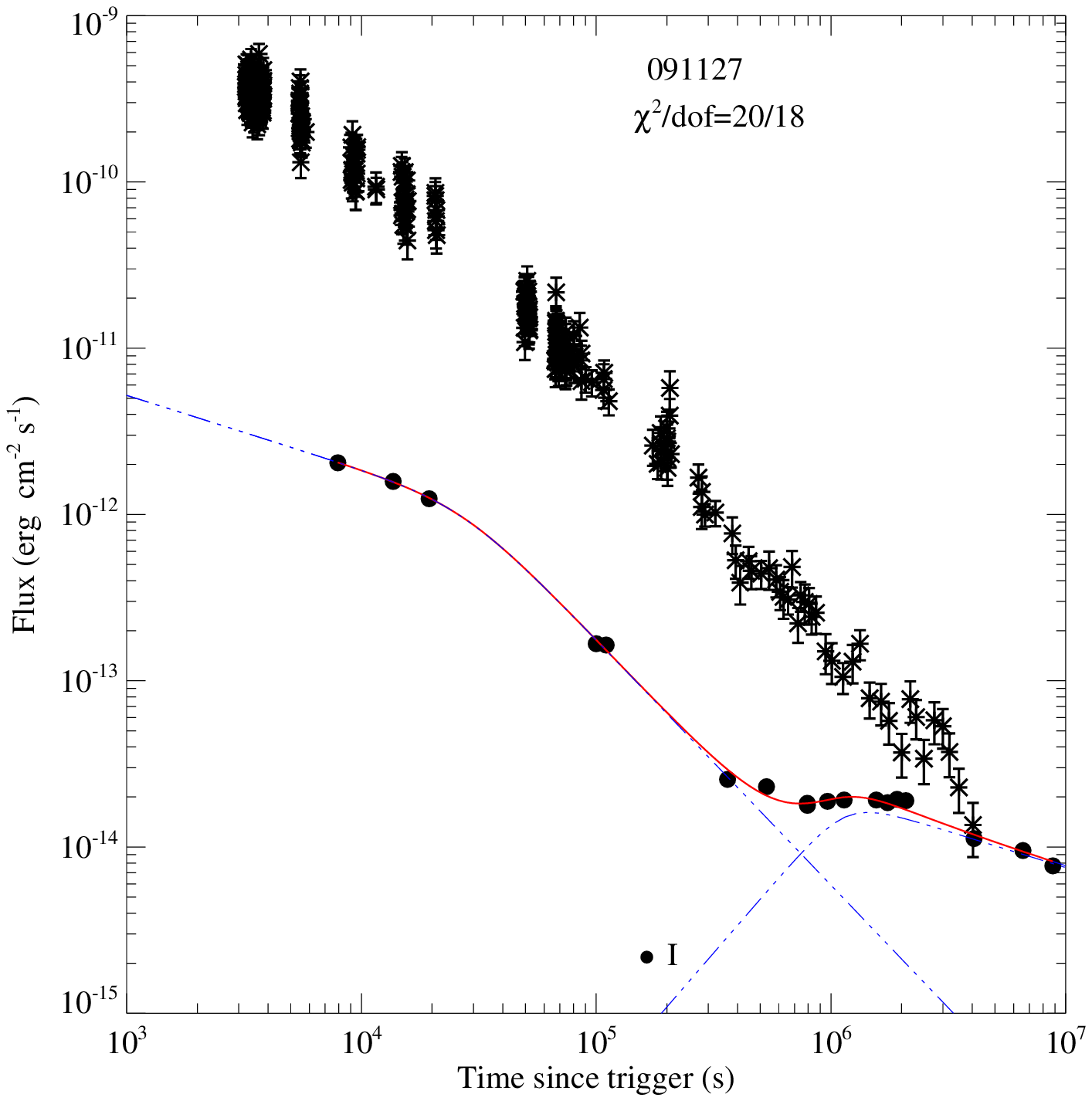}
\includegraphics[angle=0,scale=0.350,width=0.3\textwidth,height=0.25\textheight]{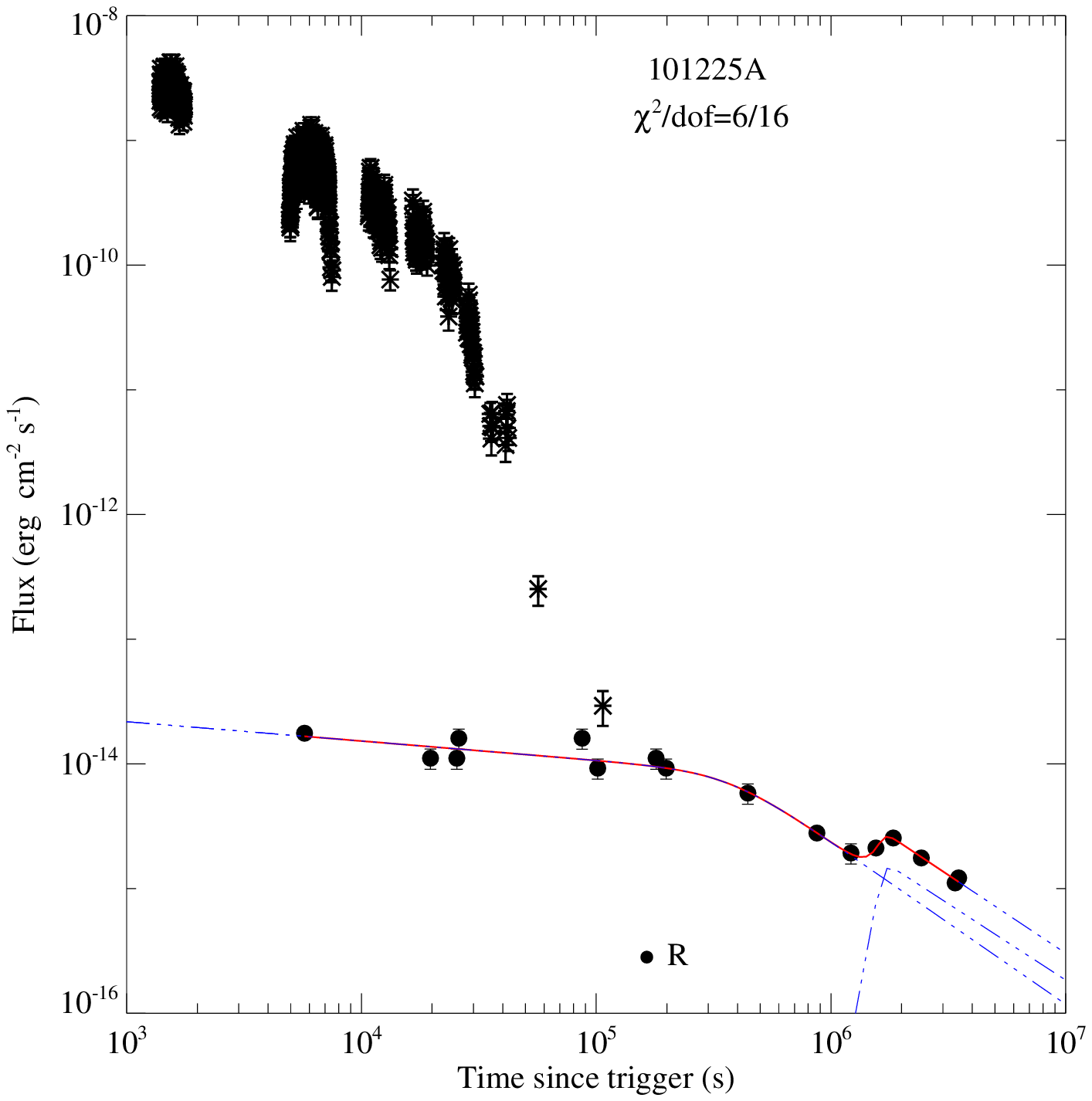}
\center{Fig. \ref{Shallow_Opt_LC}--- Continued}
\end{figure*}

\begin{figure*}
\includegraphics[angle=0,scale=0.350,width=0.3\textwidth,height=0.25\textheight]{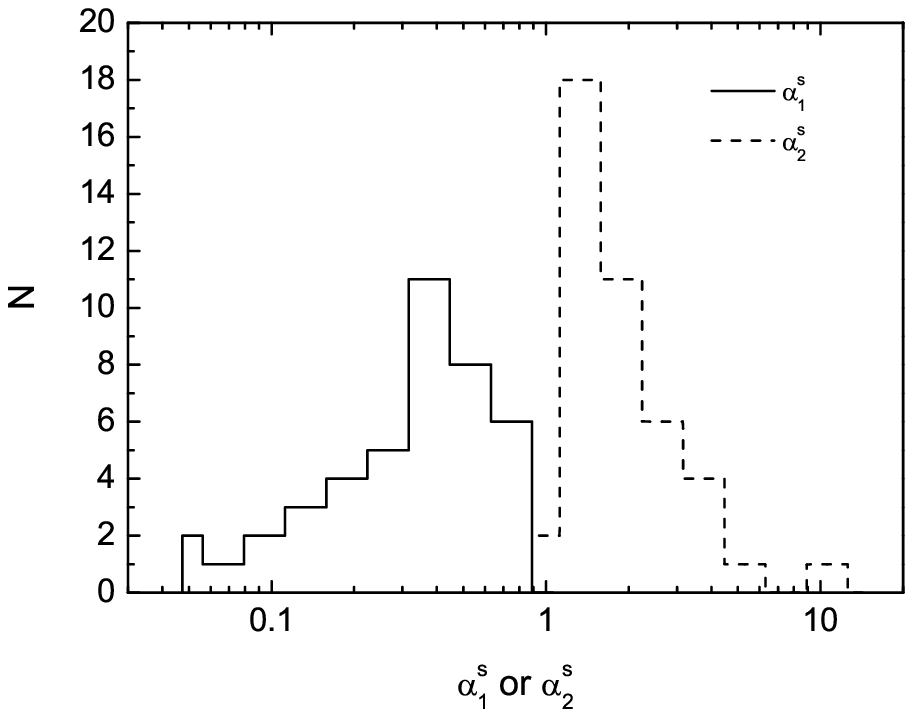}
\includegraphics[angle=0,scale=0.350,width=0.3\textwidth,height=0.25\textheight]{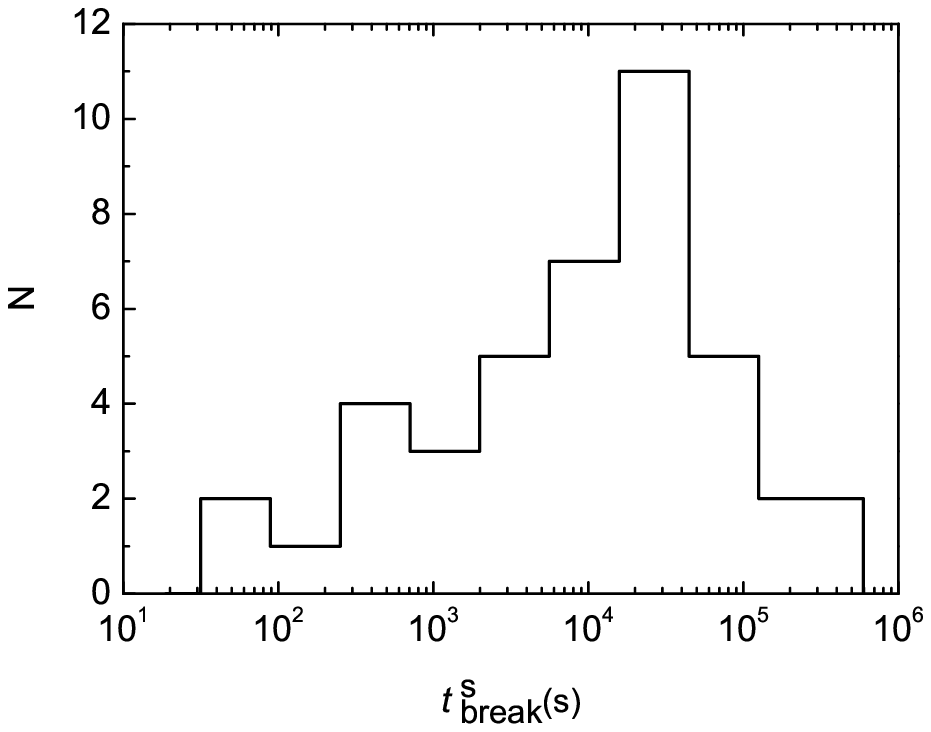}
\includegraphics[angle=0,scale=0.350,width=0.3\textwidth,height=0.25\textheight]{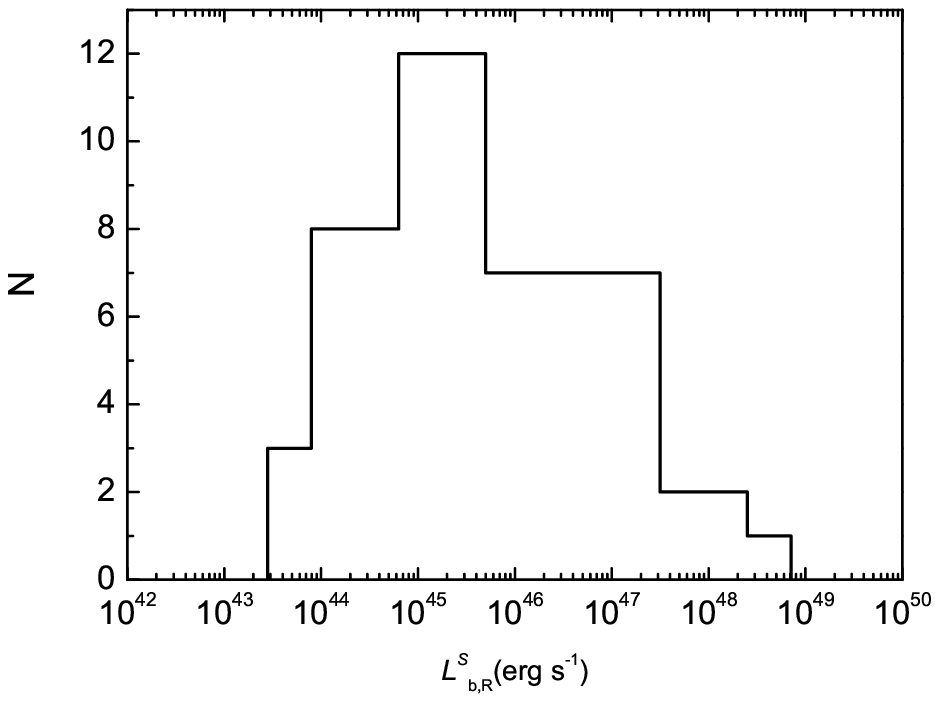}
\caption{Parameter Distributions of the shallow decay segments for the GRBs in our sample.} \label{Shallow_Dis}
\end{figure*}

\newpage
\begin{figure*}
\includegraphics[angle=0,scale=0.350,width=0.3\textwidth,height=0.25\textheight]{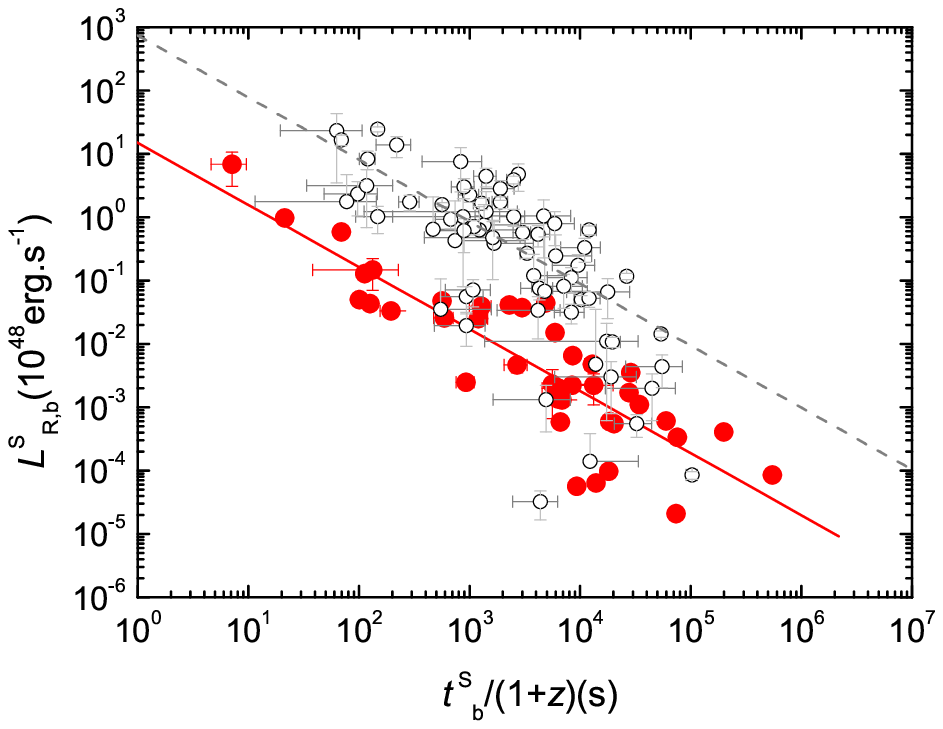}
\includegraphics[angle=0,scale=0.350,width=0.3\textwidth,height=0.25\textheight]{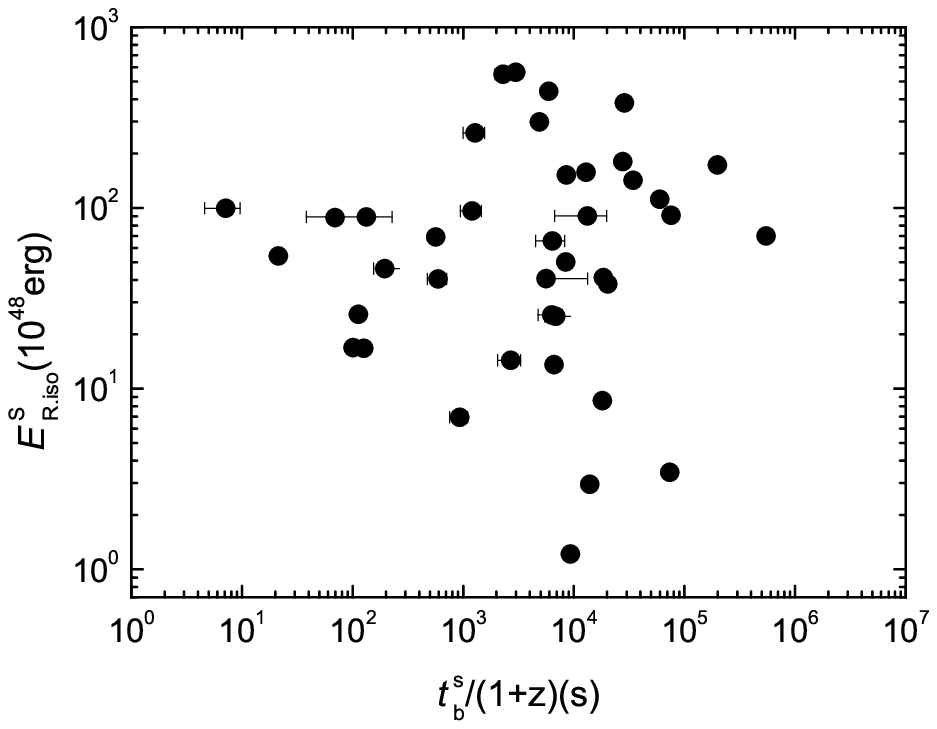}
\caption{$L^{S}_{\rm R, iso}$ ({\em left}) and $R^{\rm S}_{\rm R, iso}$ as a function of $t^{\rm S}_{b}/(1+z)$ for the GRBs with a shallow decay segment in their optical lightcurves. The grey circles are for the X-ray data from Dainotti et al. (2010). Lines are the best fit lines.} \label{Shallow_Corr}
\end{figure*}

\newpage
\begin{figure*}
\includegraphics[angle=0,scale=0.350,width=0.3\textwidth,height=0.25\textheight]{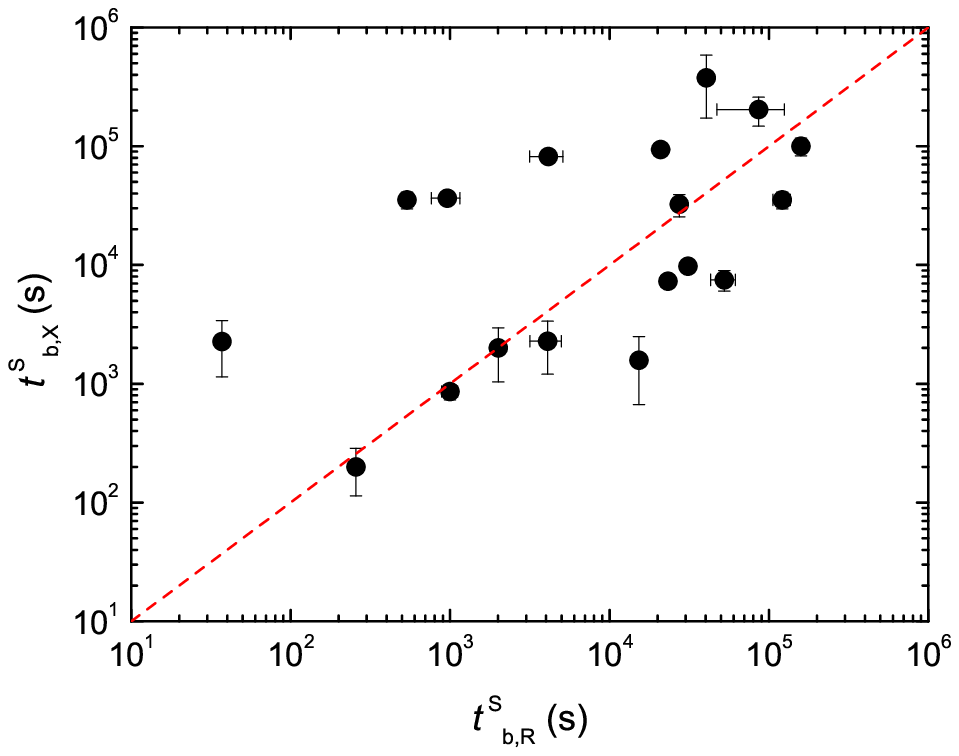}
\includegraphics[angle=0,scale=0.350,width=0.3\textwidth,height=0.25\textheight]{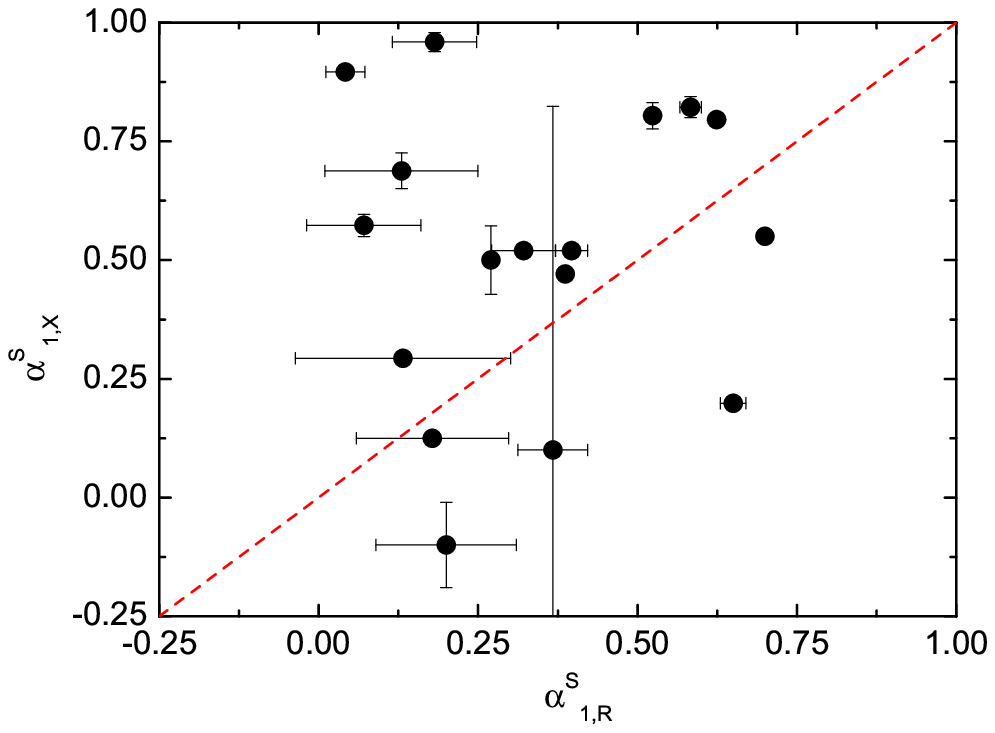}
\includegraphics[angle=0,scale=0.350,width=0.3\textwidth,height=0.25\textheight]{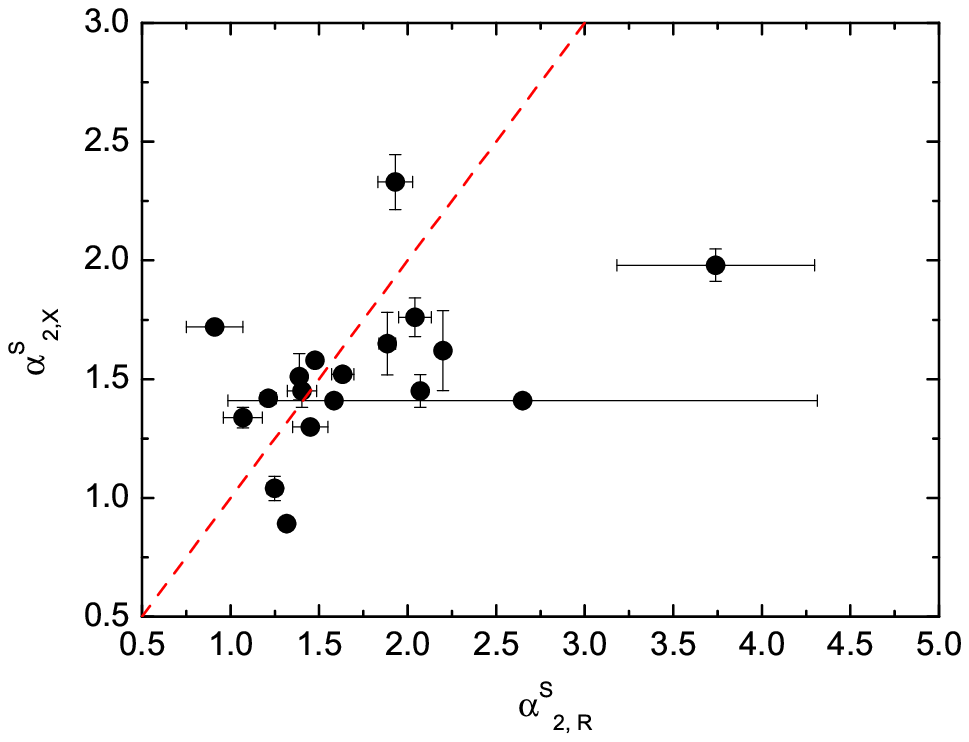}
\caption{Comparisons of the decay slopes and the break times in the optical and X-ray bands. The dashed lines are the equality lines. } \label{Shallow_Opt_Xray}
\end{figure*}

\begin{figure*}
\includegraphics[angle=0,scale=0.350,width=0.3\textwidth,height=0.25\textheight]{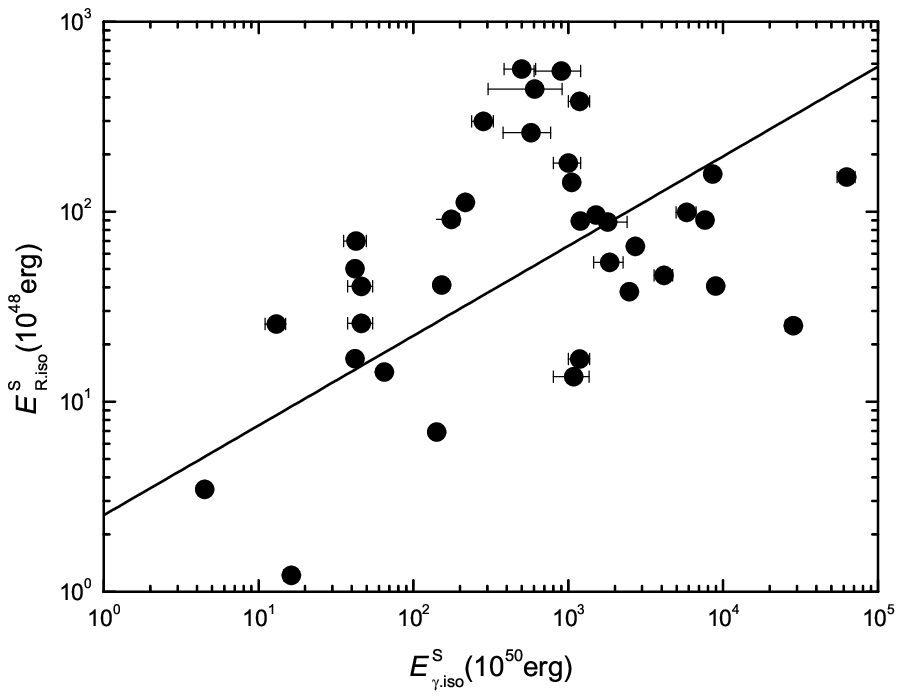}
\includegraphics[angle=0,scale=0.350,width=0.3\textwidth,height=0.25\textheight]{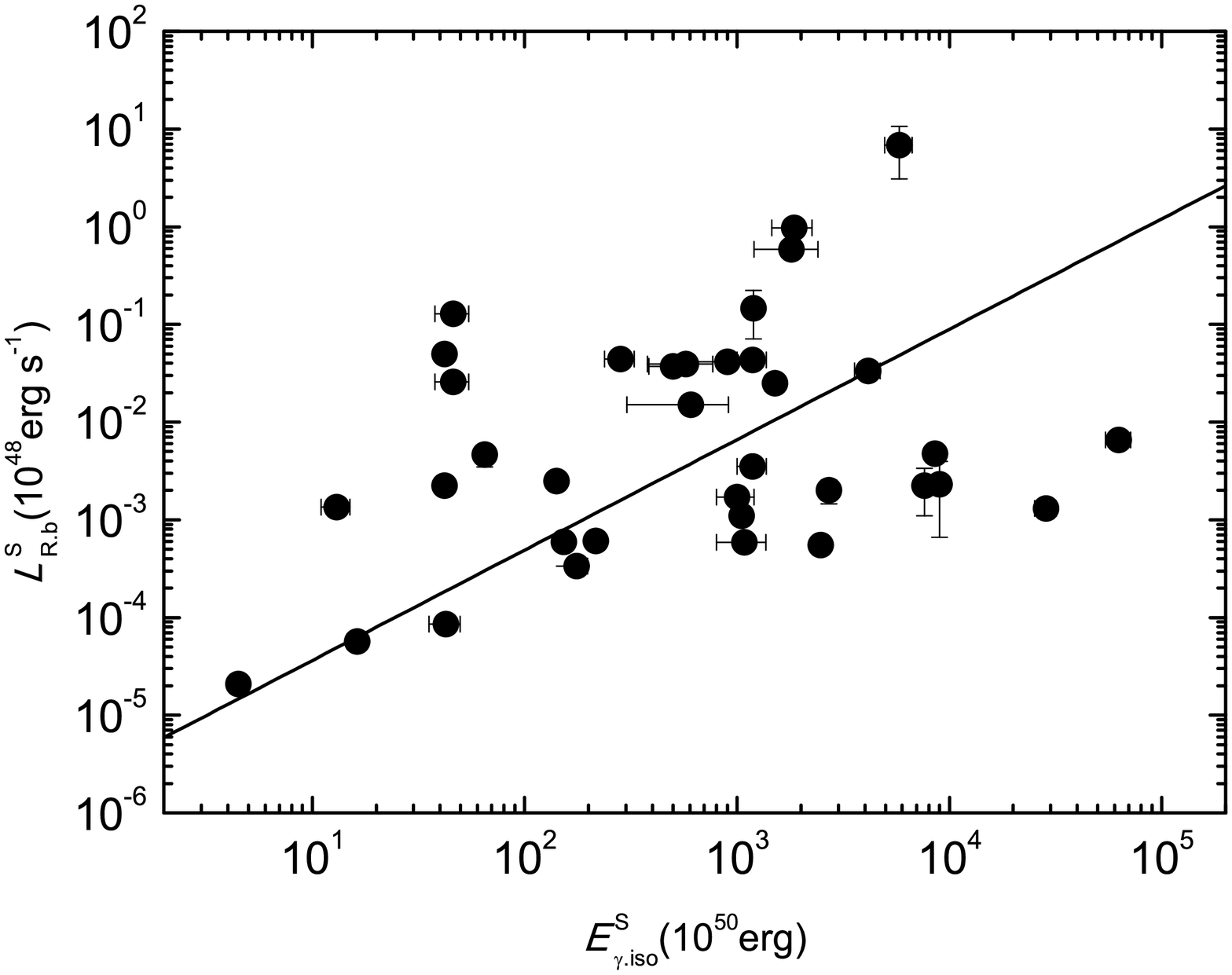}
\caption{$E_{\gamma, \rm iso}$ as a function of $E_{\rm R,iso}$, $L_{\rm R, iso}$, and $t_{p}/(1+z)$ for the GRBs with a shallow decay segment in their optical lightcurves. The lines are the robust fits to the data.} \label{Eiso_ERiso}
\end{figure*}

\begin{figure*}
\includegraphics[angle=0,scale=0.350,width=0.3\textwidth,height=0.25\textheight]{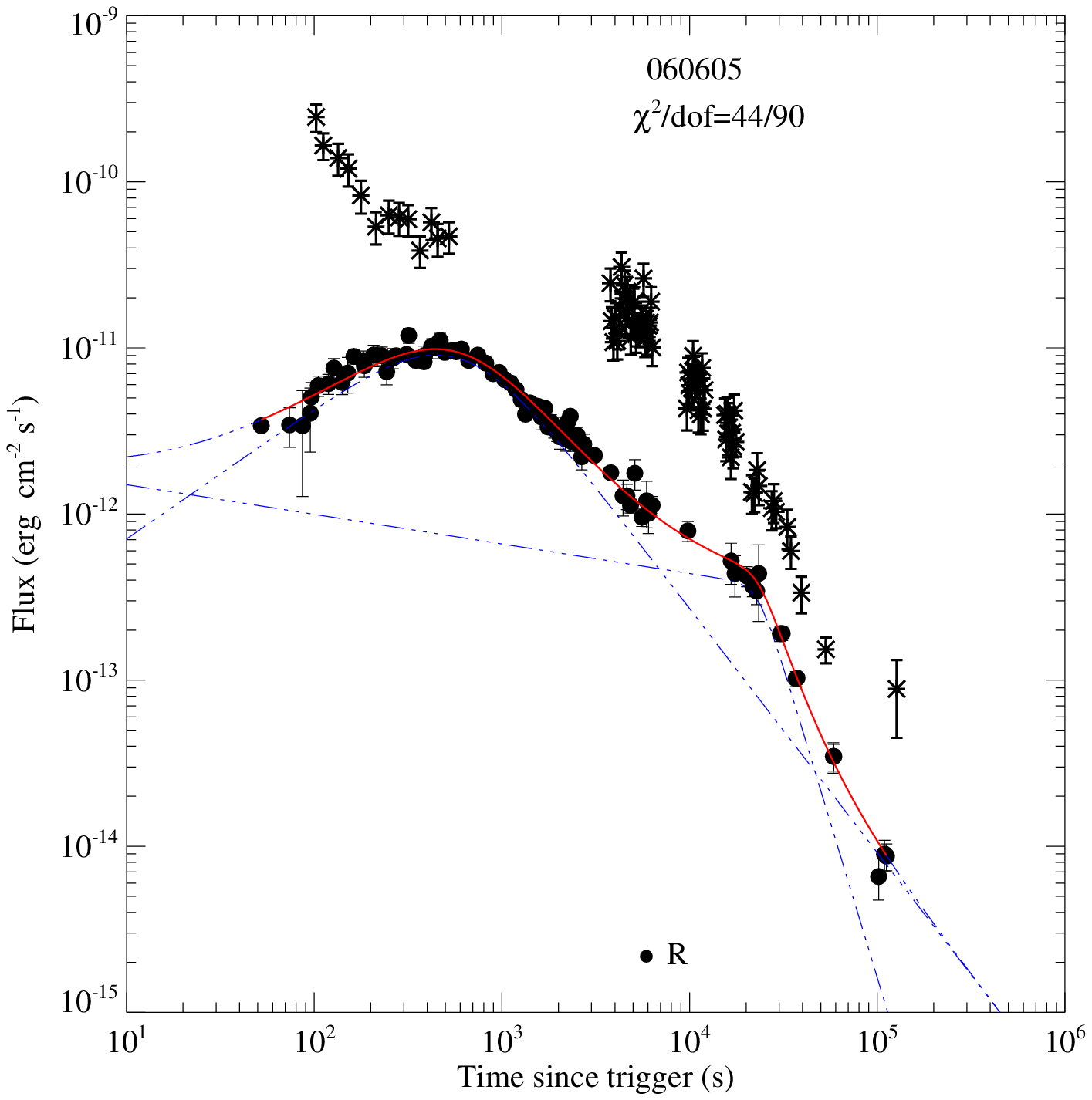}
\includegraphics[angle=0,scale=0.350,width=0.3\textwidth,height=0.25\textheight]{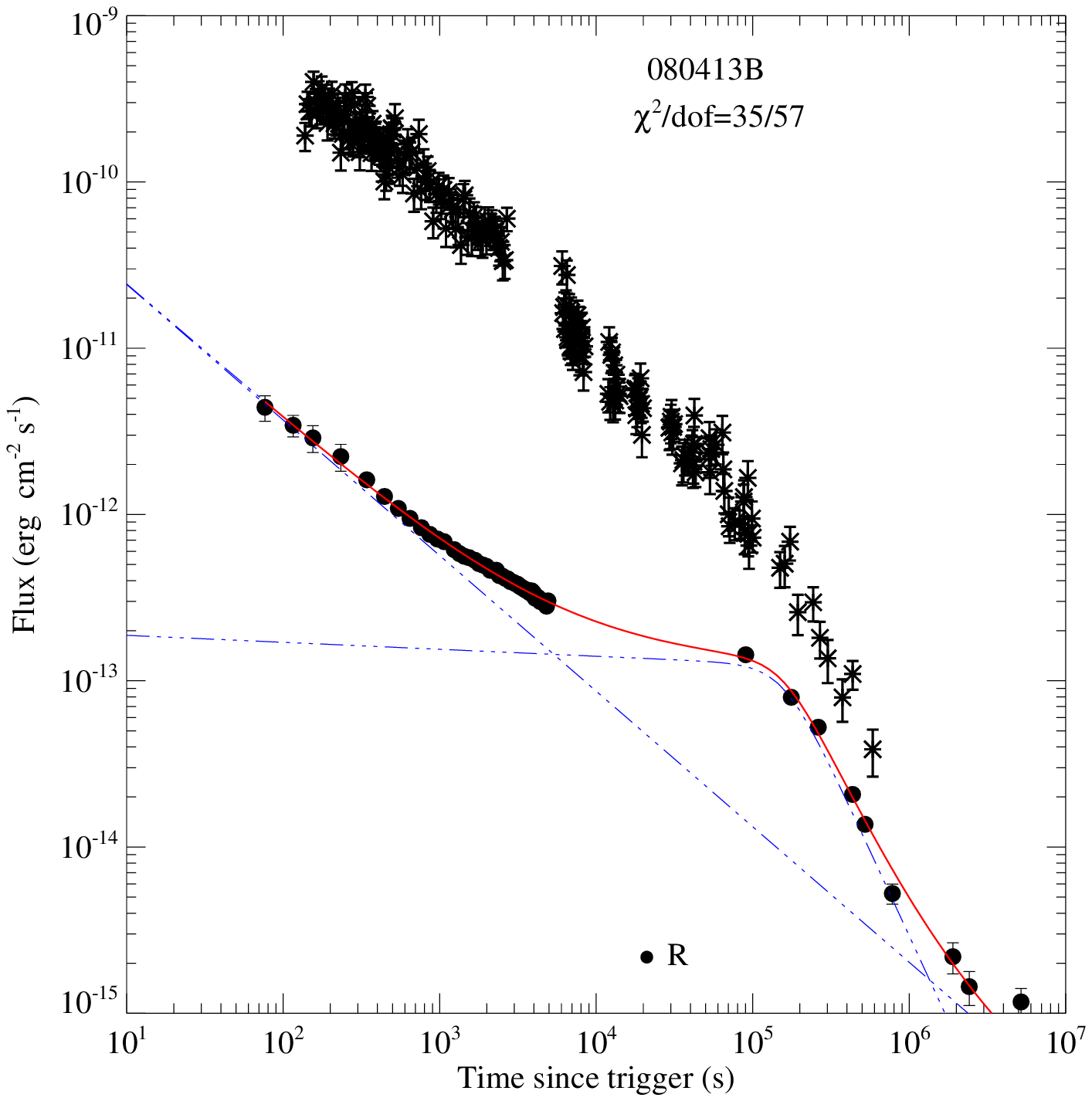}
\caption{Optical afterglow lightcurves with possible detections of an internal plateau. The symbols are the same as Figure 1.}
\label{Internal_Plateau}
\end{figure*}

\begin{figure*}
\includegraphics[angle=0,scale=0.350,width=0.3\textwidth,height=0.25\textheight]{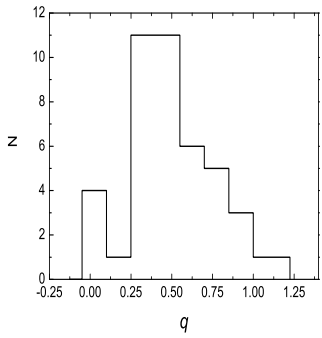}
\caption{Distribution of the $q$ parameter of our sample.} \label{Shallow_q_Dis}
\end{figure*}

\end{document}